\documentclass[11pt]{article}

\usepackage[letterpaper, hmargin=1in, top=1in, bottom=1.2in, footskip=0.6in]{geometry}
\usepackage{titlesec}
\titleformat{\section}[block]{\filcenter\normalfont\bfseries\large}{\thesection.}{.5em}{}\titlespacing*{\section}{0pt}{2\baselineskip}{1\baselineskip}
\titleformat{\subsection}[runin]{\normalfont\bfseries}{\thesubsection.}{.4em}{}[.]\titlespacing{\subsection}{0pt}{2ex plus .1ex minus .2ex}{.8em}
\titleformat{\subsubsection}[runin]{\normalfont\itshape}{\thesubsubsection.}{.3em}{}[.]\titlespacing{\subsubsection}{0pt}{1ex plus .1ex minus .2ex}{.5em}
\titleformat{\paragraph}[runin]{\normalfont\itshape}{\theparagraph.}{.3em}{}[.]\titlespacing{\paragraph}{0pt}{1ex plus .1ex minus .2ex}{.5em}

\usepackage[labelfont=bf,font=small,labelsep=period]{caption}
\usepackage[T1]{fontenc}
\usepackage[utf8]{inputenc}
\usepackage{lmodern}

\let\originalleft\left
\let\originalright\right
\renewcommand{\left}{\mathopen{}\mathclose\bgroup\originalleft}
\renewcommand{\right}{\aftergroup\egroup\originalright}

\usepackage{amsmath}
\usepackage{amssymb}
\usepackage{amsfonts}
\usepackage{latexsym}
\usepackage{amsthm}
\usepackage{amsxtra}
\usepackage{amscd}
\usepackage{bbm}
\usepackage{mathrsfs}
\usepackage{bm}
\usepackage{mathtools}

\usepackage{tikz}
\usetikzlibrary{arrows,decorations.pathmorphing,backgrounds,positioning,fit,petri}

\usepackage{graphicx, color}

\definecolor{darkred}{rgb}{0.9,0,0.3}
\definecolor{darkblue}{rgb}{0,0.3,0.9}

\definecolor{vdarkred}{rgb}{0.6,0,0.2}
\definecolor{vdarkblue}{rgb}{0,0.2,0.6}
\usepackage[pdftex, colorlinks, linkcolor=vdarkblue,citecolor=vdarkred]{hyperref}

\usepackage{booktabs}
\usepackage[nottoc,notlof,notlot]{tocbibind}
\usepackage{cite} 

\numberwithin{equation}{section}
\numberwithin{figure}{section}

\usepackage{enumitem}

\theoremstyle{plain} 
\newtheorem{theorem}{Theorem}[section]
\newtheorem*{theorem*}{Theorem}
\newtheorem{lemma}[theorem]{Lemma}
\newtheorem*{lemma*}{Lemma}
\newtheorem{corollary}[theorem]{Corollary}
\newtheorem*{corollary*}{Corollary}
\newtheorem{proposition}[theorem]{Proposition}
\newtheorem*{proposition*}{Proposition}

\newtheorem*{conjecture*}{Conjecture}

\theoremstyle{definition} 
\newtheorem{definition}[theorem]{Definition}
\newtheorem*{definition*}{Definition}

\newtheorem*{example*}{Example}
\newtheorem{remark}[theorem]{Remark}
\newtheorem*{remark*}{Remark}
\newtheorem{assumption}[theorem]{Assumption}
\newtheorem*{assumption*}{Assumption}

\renewcommand{\b}[1]{\boldsymbol{\mathrm{#1}}} 
\renewcommand{\r}{\mathrm} 
\newcommand{\bb}{\mathbb} 
\renewcommand{\cal}{\mathcal} 
 
\newcommand{\fra}{\mathfrak} 
\newcommand{\ol}[1]{\overline{#1} \!\,} 
\newcommand{\wh}{\widehat}
\newcommand{\wt}{\widetilde}

\renewcommand{\P}{\mathbb{P}}
\newcommand{\E}{\mathbb{E}}
\newcommand{\R}{\mathbb{R}}
\newcommand{\C}{\mathbb{C}}
\newcommand{\N}{\mathbb{N}}
\newcommand{\Z}{\mathbb{Z}}

\newcommand{\ee}{\r e}
\newcommand{\ii}{\r i}
\newcommand{\dd}{\r d}
\newcommand{\col}{\vcentcolon}
\newcommand*{\deq}{\mathrel{\vcenter{\baselineskip0.65ex \lineskiplimit0pt \hbox{.}\hbox{.}}}=}
\newcommand*{\eqd}{=\mathrel{\vcenter{\baselineskip0.65ex \lineskiplimit0pt \hbox{.}\hbox{.}}}}

\renewcommand{\leq}{\leqslant}
\renewcommand{\geq}{\geqslant}
\renewcommand{\epsilon}{\varepsilon}

\newcommand{\ind}[1]{\b 1_{#1}}

\newcommand{\pb}[1]{\bigl(#1\bigr)}

\newcommand{\pbb}[1]{\biggl(#1\biggr)}
\newcommand{\pBB}[1]{\Biggl(#1\Biggr)}

\newcommand{\abs}[1]{\lvert #1 \rvert}

\newcommand{\scalar}[2]{\langle#1 \mspace{2mu}, #2\rangle}

\DeclareMathOperator{\tr}{Tr}

\DeclareMathOperator{\re}{Re}

\newcommand{\cl}{{\mathrm{cl}}}
\newcommand{\lm}{{\mathrm{lm}}}

\title{Interacting loop ensembles and Bose gases}

\author{	
J\"urg Fr\"ohlich
\and Antti Knowles
\and Benjamin Schlein
\and Vedran Sohinger
}

\begin{document}

\maketitle
\begin{abstract}
We study interacting Bose gases in thermal equilibrium on a lattice.
We establish convergence of the grand-canonical Gibbs states of such gases to their mean-field (classical field) and large-mass (classical particle) limits. The former is a classical field theory for a complex scalar field with quartic self-interaction. The latter is a classical theory of point particles with two-body interactions. Our analysis is based on representations in terms of ensembles of interacting random loops, the \emph{Ginibre loop ensemble} for Bose gases and the \emph{Symanzik loop ensemble} for classical scalar field theories. For small enough interactions, our results also hold in infinite volume. The results of this paper were previously sketched in \cite{FKSS_2020_2}.
\end{abstract}

\section{Introduction and main results}

\subsection{Overview}
In this paper we study equilibrium properties of Bose gases and of systems emerging from Bose gases in various limiting regimes, by representing these systems in terms of ensembles of interacting random loops. These include the \emph{Ginibre loop ensemble} \cite{Ginibre1,Ginibre2,Ginibre3,Ginibre}, which describes an interacting Bose gas in thermal equilibrium, and the \emph{Symanzik loop ensemble} \cite{Symanzik_1968}, which describes the equilibrium state of an interacting classical field theory. The main goals of this paper are to highlight the usefulness of such random loop representations and to develop a unified approach to analyse their relationships and their behaviour in various limiting regimes.

Among different limiting regimes, we analyse the \emph{mean-field} and the \emph{large-mass} limits of interacting Bose gases using their random loop representations. These representations are particularly well-suited for proving results in infinite volume, and all our results also hold in infinite volume assuming the interaction strength is small enough. Our results on the interacting Bose gas are mostly new. We also obtain a new proof of convergence to the mean-field limit on a finite lattice, which was previsouly established using other methods.

For concreteness, we focus on the Euclidean lattice $\Z^d$, where the random loops are defined in terms of continuous-time simple random walks. With fairly straightforward modifications, our results and proofs -- with the exception of convergence to the mean-field limit in dimensions $d > 1$, treated in separate papers \cite{FKSS_2020_1, LNR_4} -- extend to continuum gases defined on $\R^d$, where the random loops are defined in terms of Brownian motion. For conciseness, we shall however not pursue this direction in the present paper.

Next, we describe the main results established in this paper and the methods used to prove them. For $d \in \N^*$ and $L \in \N^*$ we define the discrete cube $\Lambda \deq [-L/2, L/2)^d \cap \Z^d$. On $\Lambda$ we define the discrete Laplacian
\begin{equation} 
\label{discrete_laplacian}
\Delta f(x) \deq \sum_{y : \abs{y - x} = 1} (f(y) - f(x))\,, \qquad f \col \Lambda \to \C\,,
\end{equation}
with periodic boundary conditions on the cube $\Lambda$. 
For $T \geq 0$ and $x,y \in \Lambda$ we denote by $\Omega^T_{y,x}$ the set of c\`adl\`ag\footnote{Right-continuous with left limits.} paths $\omega \col [0, T] \to \Lambda$ satisfying $\omega(0) = x$ and $\omega(T) = y$. We also abbreviate $\Omega^T \deq \bigcup_{x,y \in \Lambda} \Omega^T_{y,x}$ and $\Omega \deq \bigcup_{T \geq 0} \Omega^T$. For $\omega \in \Omega^T_{y,x}$, we use the shorthands $T(\omega) \deq T$, $x(\omega) \deq x$, and $y(\omega) \deq y$.

For $x \in \Lambda$ and $T > 0$, let $\P_x^T$ denote the law on $\Omega^T$ of the continuous-time random walk starting at $x$, which is by definition the Markovian jump process with generator $\Delta/2$. 
On $\Omega^T$ we define the path measures
\begin{equation}
\label{W_path_measures}
\bb W_{y,x}^T(\dd \omega) \deq \ind{\omega(T) = y} \, \P_x^T(\dd \omega)\,, \qquad
\bb W^T \deq \int \dd x\,  \bb W_{x,x}^T\,.
\end{equation}
The measure $\bb W^{T}$ is the path measure for \emph{closed paths} (i.e.\ loops) and $\bb W^{T}_{y,x}$ the path measure for \emph{open paths} from $x$ to $y$. Here we use the abbreviation $\int \dd x \deq \sum_{x \in \Lambda}$.

By definition, a \emph{loop ensemble} is a random point process $\Phi$ on the polish space $\Omega$. That is, $\Phi$ is a random locally finite collection of elements of $\Omega$ (see e.g. \cite{Kallenberg}). To describe it in more detail, we suppose that we are given a \emph{single-loop measure} $\bb L$, which is a positive measure on $\Omega$. To simplify the presentation, we suppose here that $\bb L$ is finite (although in our proofs we shall have to abandon this assumption). As a point process, the loop ensemble is characterized by its  \emph{$p$-loop correlation functions} $\gamma_p$, $p \in \N^*$, which are defined through
\begin{equation*} 
\int f(\omega_1, \dots, \omega_p) \, \gamma_p(\omega_1, \dots, \omega_p) \, \bb L(\dd \omega_1) \cdots \bb L(\dd \omega_p) = \E \sum_{\omega_1, \dots, \omega_p \in \Phi}^* f(\omega_1, \dots, \omega_p)
\end{equation*}
for any nonnegative symmetric test function $f$, where the expectation is taken over the point process $\Phi$ and the sum is taken over all pairwise distinct $p$-tuples of loops in $\Phi$. The \emph{noninteracting loop ensemble} associated with the single-loop measure $\bb L$ is the Poisson point process on $\Omega$ with intensity measure $\bb L$. More concretely, in the noninteracting loop ensemble the loop configuration  $(\omega_1, \dots, \omega_n)$  carries the weight
\begin{equation} \label{free_loops}
\frac{1}{Z} \frac{1}{n!} \bb L(\dd \omega_1) \cdots \bb L(\dd \omega_n) \,, \qquad Z = \sum_{n \in \N} \frac{1}{n!} \int \bb L(\dd \omega_1) \cdots \bb L(\dd \omega_n)\,.
\end{equation}
Here the factor $1/n!$ compensates the overcounting from permuting the $n$ loops.
The $p$-loop correlation function of the noninteracting loop ensemble is simply $1$ for all $p \in \N^*$.

In order to define an \emph{interacting loop ensemble}, we introduce a \emph{two-loop interaction} $\cal V(\omega, \tilde \omega)$, which is a real-valued function on $\Omega \times \Omega$. This determines an $n$-loop interaction potential through
\begin{equation} \label{n_loop_interaction}
V(\omega_1, \dots, \omega_n) \deq \frac{1}{2}\sum_{i,j = 1}^{n} \cal V(\omega_i, \omega_j)\,.
\end{equation}
The interacting loop ensemble is then obtained from the corresponding noninteracting loop ensemble by weighting the contribution of each loop configuration $(\omega_1, \dots, \omega_n)$ in \eqref{free_loops} by the Boltzmann factor $\ee^{-V(\omega_1, \dots, \omega_n)}$. Recalling the definition \eqref{free_loops}, we then easily find that the $p$-loop correlation function of the interacting loop ensemble is
\begin{equation} \label{gamma_p_intro}
\gamma_p(\omega_1, \dots, \omega_p) = \frac{Z(\omega_1, \dots, \omega_p)}{Z}\,,
\end{equation}
where we defined the loop partition functions
\begin{equation}
\label{Z(omega)}
Z(\omega_1, \dots, \omega_p) \deq \sum_{n \in \N} \frac{1}{n!} \int \bb L(\dd \tilde \omega_1) \cdots \bb L(\dd \tilde \omega_n) \, \ee^{-V(\omega_1, \dots, \omega_p, \tilde \omega_1, \dots, \tilde \omega_n)}\,, \qquad Z \deq Z(\emptyset)\,.
\end{equation}

Next, we explain the three interacting loop ensembles used in this paper and explain how they are related to each other.
The \emph{Symanzik loop ensemble} has the single-loop measure
\begin{equation} \label{L_cl}
\int \bb L^\cl(\dd \omega) \deq
\int_0^\infty \dd T \, \frac{\ee^{-\kappa T}}{T} \int \bb W^T(\dd \omega)\,,
\end{equation}
where $\kappa > 0$ is a killing rate (a negative chemical potential in physics terminology).
The factor $1/T$ has the interpretation of compensating the overcounting arising from the arbitrary choice of the origin in the time interval $[0,T]$ parametrizing the closed loop. The factor $\ee^{-\kappa T}$ entails an exponential suppression of long loops. We note that $\bb L^\cl$ is not finite, owing to the contribution of very short-lived loops; we temporarily ignore this issue here. In our proofs, we shall regularize $\bb L^\cl$ by truncating it at small values of $T$, and then show that the truncation can be removed in the quotient \eqref{gamma_p_intro}. The two-loop interaction of the Symanzik loop ensemble is
\begin{equation} \label{V_symanzik}
\cal V^\cl(\omega, \tilde \omega) \deq \int_0^{T(\omega)} \dd t \int_0^{T(\tilde \omega)} \dd \tilde t \, v(\omega(t) - \tilde \omega(\tilde t))\,,
\end{equation}
where $v \col \Lambda \to \R$ is a \emph{two-body interaction potential}. We denote the associated $n$-loop interaction potential in \eqref{n_loop_interaction} by $V^{\cl}(\omega_1,\ldots,\omega_p)$

The Symanzik loop ensemble was introduced in \cite{Symanzik_1968} to describe interacting Euclidean field theories. To define a field theory of a complex scalar field $\phi$ on $\Lambda$ with interaction potential $v$, we define the complex Gaussian measure with mean 0 and covariance $(-\Delta/2+\kappa)^{-1}$ through
\begin{equation}
\label{Gaussian_measure_Phi}
\mu_{(-\Delta/2+\kappa)^{-1}}(\dd \phi) \deq \frac{1}{\pi^{|\Lambda|} \det \,(-\Delta/2+\kappa)^{-1}} \, \ee^{\scalar{\phi}{(\Delta/2-\kappa) \phi}} \, \dd \phi\,,
\end{equation}
where $\dd \phi$ denotes Lebesgue measure on $\C^\Lambda$.
The \emph{(relative) classical partition function} is
\begin{equation}
\label{Z_hat}
\cal Z^\cl \deq \int  \mu_{(-\Delta/2+\kappa)^{-1}} (\dd \phi)\,\ee^{-\frac{1}{2} \int_{\Lambda} \dd x\, \int_{\Lambda} \dd y\, |\phi(x)|^2 \,v(x-y)\, |\phi(y)|^2}\,.
\end{equation}
The \emph{classical $p$-point correlation function} is
\begin{equation}
\label{gamma_hat}
(\Gamma_p^\cl)_{\b x, \b y} \deq \frac{1}{\cal Z^\cl} \int  \mu_{(-\Delta/2+\kappa)^{-1}} (\dd \phi)\, \prod_{i = 1}^p \bar \phi(y_i) \prod_{i = 1}^p \phi(x_i) \,\ee^{-\frac{1}{2} \int_{\Lambda} \dd x\, \int_{\Lambda} \dd y\, |\phi(x)|^2 \,v(x-y)\, |\phi(y)|^2}\,,
\end{equation}
where $\b x, \b y \in \Lambda^p$. As observed in \cite{Symanzik_1968}, the relation between the classical correlation function of the interacting field theory $(\Gamma_p^\cl)_{\b x, \b y}$ from \eqref{gamma_hat} and the correlation function of the Symanzik loop ensemble $\gamma_p^\cl(\omega_1, \dots, \omega_p)$ from \eqref{gamma_p_intro} is given by
\begin{equation}  \label{symanzik_representation}
(\Gamma_p^\cl)_{\b x, \b y} =  \sum_{\pi \in S_p} \pBB{\prod_{i = 1}^p \int_0^\infty \dd T_i \, \ee^{-\kappa T_i} }\pBB{\prod_{i = 1}^p \int \bb W^{T_i}_{y_{\pi(i)}, x_i}(\dd \omega_i) }\, \gamma^\cl_p(\omega_1, \dots, \omega_p)\,.
\end{equation}
See Proposition \ref{Symanzik_representation_theorem} and Appendix \ref{sec:symanzik_proof} below.

\begin{figure}[!ht]
\begin{center}
\includegraphics{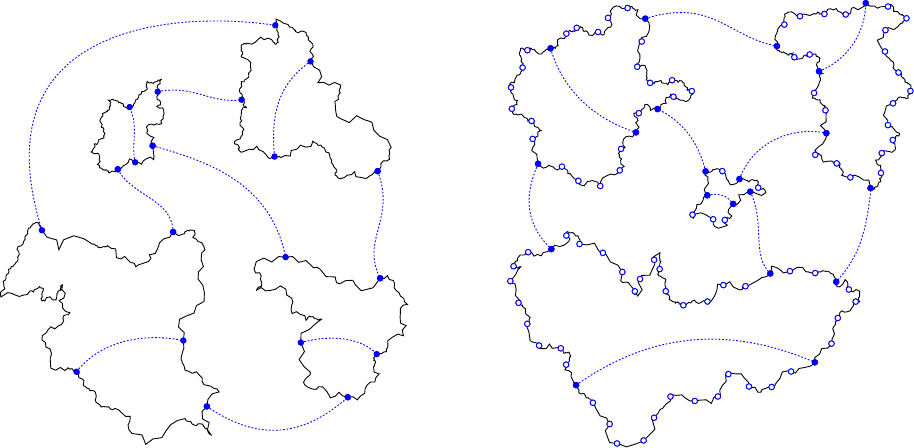}
\end{center}
\caption{An illustration of the Symanzik (left) and Ginibre (right) loop ensembles. The random loops $\omega$ are drawn in black. An interaction $\cal V(\omega, \tilde \omega)$ is drawn with a dotted blue line, joining the points $\omega(t)$ and $\tilde \omega(\tilde t)$ that appear in the argument of the interaction potential $v$. Note that each loop can interact with each other loop and with itself. In the Ginibre ensemble, the duration of the loops is a multiple of $\nu$, and the times at which the loops interact  differ by integer multiples of $\nu$, indicated using empty blue dots. In the Symanzik ensemble, all times are arbitrary real numbers.
\label{fig:loops}
}
\end{figure}

Next, we describe the \emph{Ginibre loop ensemble}. It depends on three parameters $\nu, \kappa, \lambda > 0$. It has the single-loop measure
\begin{equation} \label{L_nu}
\int \bb L^{\nu, \kappa}(\dd \omega) \deq
\nu \sum_{T \in \nu \N^*} \, \frac{\ee^{-\kappa T}}{T} \int \bb W^T(\dd \omega)\,.
\end{equation}
Note that \eqref{L_nu} is obtained from \eqref{L_cl} by replacing the Lebesgue integral $\int \dd T$ with its discrete (Riemann-sum) approximation $\nu \sum_{T \in \nu \N^*}$. In the Ginibre loop ensemble, the  two-loop interaction  is
\begin{equation}
\label{V_interaction_1}
\cal V^{\nu,\lambda}(\omega, \tilde \omega) \deq
\frac{\lambda}{\nu^2} \, \nu \sum_{s \in \nu \N} \ind{s < T(\omega)} \, \nu \sum_{\tilde s \in \nu \N} \ind{\tilde s < T(\tilde \omega)}  \, \frac{1}{\nu} \int_0^\nu \dd t \, v(\omega(s + t) - \tilde \omega(\tilde s + t))\,,
\end{equation}
and we denote the associated $n$-loop interaction potential by $V^{\nu, \lambda}(\omega_1, \dots, \omega_n)$ (see  \eqref{n_loop_interaction}).  Thus, the Ginibre loop ensemble can be regarded as a discretized version of the Symanzik loop ensemble, where the times are constrained to belong to the lattice $\nu \Z$. See Figure \ref{fig:loops} for an illustration of the Symanzik and Ginibre ensembles.

The Ginibre loop ensemble was introduced in \cite{Ginibre1,Ginibre2,Ginibre3,Ginibre} to describe the statistical mechanics of interacting Bose gases at positive temperatures.
A system of $n$ spinless bosons of mass $m>0$ confined to $\Lambda$ is governed by the Hamiltonian
\begin{equation}
\label{Hamiltonian_H_n_1}
\mathbb{H}_n 
\deq -\sum_{i = 1}^n \frac{\Delta_i}{2m} + \frac{\lambda}{2} \sum_{i,j = 1}^n v(x_i - x_j)\,,
\end{equation}
where $\Delta_i$ is the discrete Laplacian introduced in \eqref{discrete_laplacian} acting on the variable $x_i$, $\lambda \geq 0$ is a coupling constant, and $v \col \Lambda \rightarrow \R$ is a two-body interaction potential. The Hamiltonian \eqref{Hamiltonian_H_n_1} acts on 
the $n$-particle bosonic Hilbert space $\cal H_n \deq P^+_n L^2(\Lambda^n)$, where
\begin{equation*}
\label{P_n^+}
P^+_n f(x_1, \dots, x_n) \deq \frac{1}{n!} \sum_{\pi \in S_n} f(x_{\pi(1)}, \dots, x_{\pi(n)})
\end{equation*}
is the orthogonal projection onto the subspace of symmetric functions, and $L^2(\Lambda^n)$ is the $L^2$-space with respect to the counting measure 
on $\Lambda^n$.

We analyse the Bose gas in the \emph{grand-canonical ensemble} at positive temperature. Its equilibrium state is described by a sequence of density matrices  $(\rho_n)_{n \in \N}$ defined by 
\begin{equation*} 
\rho_n \deq \frac{1}{\Xi} \ee^{-\beta (\mathbb{H}_n-\mu n)} \quad \text{with} \quad  \Xi \deq \sum_{n \in \N} \tr \pb{\ee^{-\beta (\mathbb{H}_n-\mu n)}}\,.
\end{equation*}
Here  $\beta > 0$ is the inverse temperature, $\mu<0$ is the chemical potential, and $\Xi$ is the \textit{grand-canonical partition function}, which is a normalization constant chosen such that 
$\sum_{n} \text{Tr}( \rho_{n}) = 1$.

Without loss of generality, we set $\beta = 1$ as it can be absorbed into the other parameters, and we replace the parameters $m$ and $\mu$ with
\begin{equation}
\label{nu_kappa}
\nu \deq \frac{1}{m}>0\,,\qquad \kappa \deq -\mu m>0\,.
\end{equation}
Thus we find that the grand-canonical ensemble is characterized by the sequence $(\rho_n^{\nu, \kappa, \lambda})_{n \in \N}$, where
\begin{equation} \label{rho_intro}
\rho_n^{\nu, \kappa, \lambda} \deq \frac{1}{\Xi^{\nu, \kappa, \lambda}} \ee^{-(H_n^{\nu, \lambda} + \kappa \nu n)}\,, \qquad
\Xi^{\nu, \kappa, \lambda} \deq \sum_{n \in \N} \tr \pb{\ee^{-(H_n^{\nu, \lambda} + \kappa \nu n)}}\,,
\end{equation}
and
\begin{equation}
\label{Hamiltonian_H_n}
H_n^{\nu, \lambda} 
\deq - \frac{\nu}{2} \sum_{i = 1}^n \Delta_i + \frac{\lambda}{2} \sum_{i,j = 1}^n v(x_i - x_j)\,.
\end{equation}
The \emph{reduced $p$-particle density matrix} of the grand-canonical ensemble is defined as 
\begin{equation}
\label{gamma_p_definition}
\Gamma_{p}^{\nu, \kappa, \lambda} = \sum_{n \in \N} \frac{(p+n)!}{n!} \tr_{p+1, \dots, p+n} \pb{\rho_{p+n}^{\nu, \kappa, \lambda}}\,,
\end{equation}
where $\tr_{p+1, \dots, p+n}$ denotes the partial trace over the coordinates $x_{p+1}, \dots, x_{p+n}$. We denote by $(\Gamma_{p}^{\nu, \kappa, \lambda})_{\b x, \b y}$ the operator kernel of $\Gamma_{p}^{\nu, \kappa, \lambda}$. As observed in \cite{Ginibre1,Ginibre2,Ginibre3,Ginibre}, the relation between the reduced density matrix $(\Gamma_{p}^{\nu, \kappa, \lambda})_{\b x, \b y}$ from \eqref{gamma_p_definition} and the correlation function of the Ginibre loop ensemble $\gamma_p^{\nu, \kappa, \lambda}(\omega_1, \dots, \omega_p)$ from \eqref{gamma_p_intro} is given by
\begin{equation} \label{ginibre_representation}
(\Gamma_p^{\nu, \kappa, \lambda})_{\b x, \b y} =  \frac{1}{\nu^p} \sum_{\pi \in S_p} \pBB{\prod_{i = 1}^p \nu \sum_{T_i \in \nu \N^*} \ee^{-\kappa T_i}} \pBB{\prod_{i = 1}^p \int \bb W^{T_i}_{y_{\pi(i)}, x_i}(\dd \omega_i) }\, \gamma^{\nu, \kappa, \lambda}_p(\omega_1, \dots, \omega_p)\,.
\end{equation}
See Proposition \ref{Ginibre_loop_representation} and Appendix \ref{sec:ginibre_proof} below.

In this paper we analyse various limiting regimes of the Ginibre loop ensemble.

\begin{enumerate}
\item[(a)]
\emph{The mean-field (or classical field) limit:} $\nu \to 0$, $\lambda = \nu^2$, $\kappa$ fixed. Recalling \eqref{nu_kappa}, we see that this amounts to a high temperature and high density limit, or, equivalently, to a large mass and large chemical potential limit. Our main result is the convergence of the Ginibre loop ensemble to the Symanzik loop ensemble,
\begin{equation} \label{convergence_gin_sym}
\lim_{\nu \to 0} \gamma_p^{\nu, \kappa, \nu^2} = \gamma^\cl_p\,.
\end{equation}
At a formal level, this convergence is plausible after comparing the single-loop measures \eqref{L_cl}, \eqref{L_nu} and the two-loop interactions \eqref{V_symanzik}, \eqref{V_interaction_1}. As a corollary, using the representations \eqref{symanzik_representation}, \eqref{ginibre_representation}, we deduce the convergence of the rescaled reduced density matrices of the quantum Bose gas to the correlation functions of the classical field theory:
\begin{equation} \label{intro_mf_conv}
\lim_{\nu \to 0} \nu^p \,\Gamma_p^{\nu, \kappa, \nu^2} = \Gamma_p^\cl\,;
\end{equation}
see Theorem \ref{mean_field_convergence}.
\item[(b)]
\emph{The large-mass limit:} $\nu \to 0$, $\lambda = 1$, $\kappa = \kappa_0 / \nu$ for some fixed $\kappa_0$. Recalling \eqref{nu_kappa}, we see that this amounts to the limit of large $m$ for fixed $\beta$ and $\mu$. In this limit, we show that the Ginibre loop ensemble converges to an ensemble of interacting classical particles,
\begin{equation} \label{intro_lm_conv}
\lim_{\nu \to 0} \gamma_p^{\nu, \kappa_0/\nu, 1} = \gamma^\lm_p\,,
\end{equation}
where $\gamma^\lm_p$ is the correlation function of a process of interacting weighted particles which we describe in more detail below\footnote{In the continuum $\R^d$ instead of the lattice $\Z^d$, with the correct choice of $\kappa$ as a function of $\nu$, the limit $\nu \rightarrow 0$ describes a classical gas of point particles with two-body interactions given by the potential $v$. \label{footnote_continuum}}. We conclude convergence of the reduced density matrices
\begin{equation} \label{large_mass_convergence}
\lim_{\nu \to 0} \Gamma_p^{\nu, \kappa_0/\nu, 1} = \Gamma_p^\lm\,,
\end{equation}
where $\Gamma_p^\lm$ is a classical correlation function defined in terms of a process of interacting weighted particles. See Theorem \ref{Large_mass_lattice_4}.
\end{enumerate}

In addition, we extend both convergence results to the thermodynamic limit $L \to \infty$, under the assumption that the two-body potential $v$ is of short range and not too large. Indicating the $L$-dependence of all quantities with a superscript, we extend \eqref{intro_mf_conv} to
\begin{equation} \label{intro_mf_conv_infty}
\lim_{\nu \to 0} \lim_{L \to \infty} \nu^p \,\Gamma_p^{\nu, \kappa, \nu^2,L} = \lim_{L \to \infty} \Gamma_p^{\cl,L}
\end{equation}
where all limits exist; see Theorem \ref{Infinite_volume_theorem_1}. Similarly, we extend \eqref{large_mass_convergence} to
\begin{equation} \label{large_mass_convergence_infty}
\lim_{\nu \to 0} \lim_{L \to \infty} \Gamma_p^{\nu, \kappa_0/\nu, 1,L} = \lim_{L \to \infty}\Gamma_p^{\lm,L}\,,
\end{equation}
where all limits exist. See Theorem \ref{Infinite_volume_theorem_2}.

The convergence \eqref{intro_mf_conv} is not new and was previously established in \cite{Knowles_Thesis}, using different methods. The convergence of the loop ensembles \eqref{convergence_gin_sym}, however, and the other three results \eqref{large_mass_convergence}, \eqref{intro_mf_conv_infty}, and \eqref{large_mass_convergence_infty} appear to be new.

Finally, we  describe  the process of interacting weighted particles that emerges in the large-mass limit (b). It may be formulated as an ensemble of interacting stationary loops of integer time length. To describe this precisely, introduce the measure $\bb D_x^T$ as the atomic measure on $\Omega^T$ at the constant loop $\omega(t) = x$ for all $t \in [0,T]$. The loop ensemble in the large-mass limit has the single-loop measure
\begin{equation*}
\int \bb D^\lm(\dd \omega) \deq \sum_{k \in \N^*} \frac{\ee^{-\kappa_0 k}}{k} \int_\Lambda \dd x \int \bb D_x^k(\dd \omega)\,,
\end{equation*}
and the two-loop interaction is
\begin{equation*}
\cal V^\lm(\omega, \tilde \omega) \deq \sum_{0 \leq k < T(\omega)} \sum_{0 \leq \tilde k < T(\tilde \omega)} \int_0^1 \dd t \, v(\omega(k + t) - \tilde \omega(\tilde k + t))\,.
\end{equation*}
The right-hand side of \eqref{intro_lm_conv} is by definition the $p$-loop correlation function of this loop ensemble, and analogously the right-hand side of \eqref{large_mass_convergence} is given by
\begin{equation*}
(\Gamma_{p}^\lm)_{\b x, \b y} \deq  \sum_{\pi \in S_p}  \pBB{\prod_{i = 1}^p \sum_{k_i \in \N^*} \ee^{-\kappa_0 k_i} \, \delta(y_{\pi(i)} - x_i) \int \bb D^{k_i}_{x_i}(\dd \omega_i)}  \gamma_p^\lm(\omega_1, \dots, \omega_p)\,.
\end{equation*}
The interpretation is that in the large-mass limit, only loops of duration of order $\nu$ contribute, in which time the simple random walk cannot make even a single jump. Loops thus collapse to points. We can make this collapse more explicit by parametrizing a stationary loop $\omega$ with its location $x \in \Lambda$ and its duration $k \nu$, where $k \in \N^*$. The couple $(k,x)$ describes a weighted particle, where $k$ is the occupation number and $x$ the position. Using this parametrization, the single-loop measure and the two-loop interaction become
\begin{equation*}
\int \bb D^\lm(\dd (k,x)) = \sum_{k \in \N^*} \frac{\ee^{-\kappa_0 k}}{k} \int_\Lambda \dd x\,, \qquad \cal V^\lm((k,x), (\tilde k, \tilde x)) = k \, \tilde k \, v(x - \tilde x)\,.
\end{equation*}
It is of some interest to consider also interaction potentials with a hard core repulsion, i.e.\ $v(0) = + \infty$. In that case the interaction energy in \eqref{Hamiltonian_H_n} is always infinite, but we can renormalize it by omitting the diagonal terms $i = j$. Then the density matrices in \eqref{rho_intro} and the Ginibre loop ensembles are well defined. 
For systems with hard core two-body potentials, only loops of duration $\nu$ contribute in the large-mass limit, i.e.\ all occupation numbers $k$ are zero or one. 
In particular, one can see that loops of duration $k \nu$ with $k \geq 2$ are eliminated by a self-interaction, which is absent from loops of duration $\nu$.
The interpretation is that, due to hard core repulsion, multiple occupancies of a single site are excluded.

\subsection{Statement of the main results}
Let
\begin{equation}
\label{cal_Z_definition}
\cal Z^{\nu,\kappa, \lambda} \deq \frac{\Xi^{\nu,\kappa, \lambda}}{\Xi^{\nu,\kappa, 0}}
\end{equation}
be the relative grand-canonical partition function, where $\Xi^{\nu,\kappa, \lambda}$ was defined in \eqref{rho_intro}.

We first state our result on the mean-field regime on a finite lattice. 
We begin by stating precise assumptions on the interaction potential $v$.
\begin{assumption}
\label{interaction_potential_v}
We consider an interaction potential $v: \Lambda \rightarrow \R$ which is pointwise nonnegative and of positive type (i.e.\ its Fourier transform is pointwise nonnegative).
\end{assumption}

\begin{theorem}[Mean-field limit on a finite lattice]
\label{mean_field_convergence}
With $v$ as in Assumption \ref{interaction_potential_v}, the following limits hold.
\begin{itemize}
\item[(i)]$\lim_{\nu \rightarrow 0} \cal Z^{\nu,\kappa, \nu^2} =\cal Z^\cl$.
\item[(ii)] For all $p \in \N^*$ and $\b x, \b y \in \Lambda^p$, we have $\lim_{\nu \to 0} \nu^p\,(\Gamma_p^{\nu, \kappa, \nu^2})_{\b x, \b y} = (\Gamma_p^\cl)_{\b x, \b y}$.
\end{itemize}
\end{theorem}

Next, we state our result on the large-mass limit on a finite lattice.
In this regime, we modify Assumption \ref{interaction_potential_v} to account for interaction potentials that can have a hard core.
\begin{assumption}
\label{interaction_potential_v_2}
We consider an interaction potential $v: \Lambda \rightarrow [0,\infty]$ for which there exists $R \in \{0,1\}$ such that (i) $v(x) \in [0,\infty)$ for $|x| \geq R$. (ii) $v(x)=\infty$ for $|x|<R$.
\end{assumption}
We note that the analysis can be extended to consider arbitrary $R \geq 0$, but we consider $R \in \{0,1\}$ for simplicity.
As remarked in the overview, we take $\lambda=1$ and $\kappa=\kappa_0/\nu$ for some fixed $\kappa_0$. In light of Assumption \ref{interaction_potential_v_2}, we modify \eqref{Hamiltonian_H_n} and work with  
\begin{equation}
\label{H_n_mass_m}
H_n^{\nu,1}=
-\frac{\nu}{2}\,\sum_{i = 1}^n  \Delta_i+ \frac{1}{2} \sum_{i,j = 1}^n (1 - \ind{i = j} \ind{R = 1}) v(x_i - x_j)\,,
\end{equation}
thereby eliminating the infinite self-interaction in the presence of a hard core interaction potential.

Given $n \in \N^{*}, \b k \in (\N^{*})^n, \b x \in \Lambda^n$, we define
\begin{equation}
\label{V_classical_interaction}
V^\lm(\b k,\b x)\deq 
\begin{cases}
\frac{1}{2}\,\sum_{i,j=1}^{n} k_i k_j v(x_i-x_j)& \text{if } R=0
\\
\frac{1}{2}\,\sum_{i,j=1}^{n} v(x_i-x_j)\,\ind{i \neq j}\,\ind{\b k=\b 1_n}+\infty\,\ind{\b k \neq \b 1_n}
& \text{if } R=1\,.
\end{cases} 
\end{equation}
Here, we write
\begin{equation}
\label{b_1}
\b 1_n \deq (1,\ldots,1) \in (\N^*)^n\,.
\end{equation}
We then consider the (relative) classical partition function with infinite mass, which is defined as
\begin{equation}
\label{Z_infty}
\cal Z^\lm \deq \Biggl\{\sum_{n = 0}^\infty \frac{1}{n!} \sum_{\b k \in (\N^*)^n} \prod_{i = 1}^n \frac{1}{k_i} \int_{\Lambda^n} \dd \b x  \, \ee^{-\kappa_0 \abs{\b k}} \exp\bigl(-V^\lm(\b k,\b x)\bigr)\Biggr\}\Bigg/\exp\Biggl\{\sum_{k=1}^{\infty} \frac{\ee^{-\kappa_0 k}}{k} |\Lambda| \Biggr\}\,.
\end{equation}
Furthermore, for $p \in \N^*$ and $\b x, \b y \in \Lambda^p$, we consider 
\begin{equation}
\label{gamma_infty}
(\Gamma_{p}^\lm)_{\b x, \b y} \deq \sum_{\b k \in (\N^*)^p} \sum_{\pi \in S_p} \ee^{-\kappa_0  \abs{\b k}} \,\delta (\pi \b y - \b x)\,\frac{Z^\lm(\b k, \b x)}{Z^\lm}\,,
\end{equation}
where 
\begin{equation}
\label{Z_infty_lx}
Z^\lm(\b k,\b x) = \sum_{n = 0}^\infty \frac{1}{n!} \sum_{\b{\tilde k} \in (\N^*)^n} \prod_{i = 1}^n \frac{1}{\tilde{k}_i} \int_{\Lambda^n} \dd \tilde {\b x}  \, \ee^{-\kappa_0 \abs{\b{\tilde k}}} \exp\bigl(-V^\lm(\b k \b{\tilde k},\b x\b{\tilde{x}})\bigr)\,,\qquad Z^\lm = Z^\lm(\emptyset)\,.
\end{equation}
Here $S_p$ is the set of permutations of $\{1, \dots, p\}$, $\pi \b y = (y_{\pi(1)}, \dots, y_{\pi(p)})$, and we use the notation \eqref{xy_notation} below.

We prove the following result.
\begin{theorem}[Large-mass limit on a finite lattice]
\label{Large_mass_lattice_4}
With $v$ as in Assumption \ref{interaction_potential_v_2}, and with notations as in \eqref{Z_infty}--\eqref{gamma_infty}, the following holds.
\begin{itemize}
\item[(i)] $\lim_{\nu \rightarrow 0} \cal Z^{\nu,\kappa_0/\nu,1} =\cal Z^\lm$.
\item[(ii)] For all $p \in \N^*$ and $\b x, \b y \in \Lambda^p$, we have that
$\lim_{\nu \to 0} (\Gamma_p^{\nu, \kappa_0/\nu, 1})_{\b x, \b y} = (\Gamma_p^\lm)_{\b x, \b y}$.
\end{itemize}
\end{theorem}

Next, we extend the results of Theorems \ref{mean_field_convergence} and \ref{Large_mass_lattice_4} to the infinite lattice $\Z^d$. To that end, we explicitly include the side length $L$ of the cube $\Lambda \equiv \Lambda_L$ in our notation.
\begin{assumption}
\label{interaction_potential_v^{L}}
We consider $v: \Z^d \rightarrow \R$ which satisfies the following properties. (i) $v$ is pointwise nonnegative. (ii) $v$ is of positive type. (iii) $v \in \ell^1(\Z^d)$. (iv) $v$ is radially decreasing.
\end{assumption}
With $v$ given as in Assumption \ref{interaction_potential_v^{L}}, we work with $v^{L} : \Lambda_L \rightarrow \R$ given by
\begin{equation}
\label{v^{L}}
v^{L}(x) \deq \sum_{k \in (L \Z)^d} v(x+k)\,.
\end{equation}
Furthermore, we define the \emph{specific (relative) Gibbs potential} of the Bose gas by
\begin{equation}
\label{quantum_free_energy}
g^{\nu,\kappa,\lambda,L}  \deq \frac{1}{|\Lambda_L|}\,\log \cal Z^{\nu,\kappa,\lambda,L}
\end{equation}
and the \emph{classical specific (relative) Gibbs potential} by
\begin{equation}
\label{classical_free_energy}
g^{\cl,L} \deq \frac{1}{|\Lambda_L|}\,\log \cal Z^{\cl,L}\,.
\end{equation}
In order to study the convergence of the reduced density matrices, we need to define an appropriate norm.
Given $p \in \N^*$ and $L_0 \in \N^*$, we define
\begin{equation}
\label{Pi_{L,p}}
\Pi_{L_0,p} \deq P_{L_0}^{\otimes p} \,(\cdot)\,P_{L_0}^{\otimes p}\,,
\end{equation} 
where $P_{L_0}: \ell^2(\Z^d) \rightarrow \ell^2(\Lambda_{L_0})$ denotes the canonical orthogonal projection. With $\Pi_{L_0,p}$ given as in \eqref{Pi_{L,p}}, and for $A$ an operator on $\ell^2(\Z^d)^{\otimes p}$, we define
\begin{equation}
\label{L_0_p_norm}
\|A\|_{L_0,p} \deq \|\Pi_{L_0,p} \,A\|_{\ell^{\infty}_{\b x} \ell^1_{\b y}}\,.
\end{equation}

\begin{theorem}[Infinite volume mean-field limit]
\label{Infinite_volume_theorem_1}
Let $v$ be as in  Assumption \ref{interaction_potential_v^{L}} and let $\kappa>1$ be given.
If $\|v\|_{\ell^1(\Z^d)}$ is sufficiently small depending on $\kappa$ (or if $\kappa$ is large depending on $\|v\|_{\ell^1(\Z^d)}$), the following limits exist and satisfy the following relations.
\begin{itemize}
\item[(i)] We have
\begin{equation*}
\lim_{\nu \rightarrow 0} \lim_{L \rightarrow \infty} g^{\nu,\kappa,\nu^2,L} =\lim_{L \rightarrow \infty} g^{\cl,L}\,.
\end{equation*}
\item[(ii)] Fix $p \in \N^*$. Then for any $L_0 \in \N^*$ we have
\begin{equation*}
\lim_{\nu \rightarrow 0} \lim_{L \rightarrow \infty} \nu^p\,\Gamma^{\nu,\kappa,\nu^2,L}=
\lim_{L \rightarrow \infty} \Gamma^{\cl,L}_p
\end{equation*}
with respect to $\|\cdot\|_{L_0,p}$, uniformly in $L_0$.
\end{itemize}
\end{theorem}

When studying the large-mass limit in the infinite volume, we modify Assumption \ref{interaction_potential_v^{L}} as follows.
\begin{assumption}
\label{interaction_potential_v_m_infty}
We consider an interaction potential $v: \Z^d \rightarrow [0,\infty]$ for which there exists $R \in \{0,1\}$ such that  (i) $v\,\ind{|x| \geq R} \in \ell^1(\Z^d)$ is radially decreasing on $|x|  \geq R$ and (ii) $v(x)=\infty$ for $|x|<R$. 
\end{assumption}

We consider the Hamiltonian defined in \eqref{H_n_mass_m} with interaction potential $v^{L}: \Lambda_L \rightarrow [0,\infty]$, which is given by \eqref{v^{L}} with $v$ as in Assumption \ref{interaction_potential_v_m_infty}, and modify notations accordingly. 
\begin{theorem}[Infinite volume large-mass limit]
\label{Infinite_volume_theorem_2}
Let $v$ be as in Assumption \ref{interaction_potential_v_m_infty}. Let $\kappa_0>0$ be given.
If $\kappa_0$ is sufficiently large, the following limits exist and satisfy the following relations.
\begin{itemize}
\item[(i)] We have
\begin{equation*}
\lim_{\nu \rightarrow 0} \lim_{L \rightarrow \infty} g^{\nu,\kappa_0/\nu,1,L}=\lim_{L \rightarrow \infty} g^{\lm,L}\,,
\end{equation*}
where $g^{\lm,L}\deq\frac{1}{|\Lambda_L|}\,\log \cal Z^{\lm,L}$.
\item[(ii)] Fix $p \in \N^*$. Then for any $L_0 \in \N^*$ we have
\begin{equation*}
\lim_{\nu \rightarrow 0} \lim_{L \rightarrow \infty} \Gamma_{p}^{\nu,\kappa_0/\nu,1,L}=
\lim_{L \rightarrow \infty} \Gamma_p^{\lm,L}
\end{equation*}
with respect to $\|\cdot\|_{L_0,p}$, uniformly in $L_0$.
\end{itemize}
\end{theorem}

\begin{remark}
We note that the results stated in Theorems \ref{Infinite_volume_theorem_1} and \ref{Infinite_volume_theorem_2} extend to more general boundary conditions for \eqref{discrete_laplacian} by using very similar arguments. This can be seen by appropriately modifying the proofs of Propositions \ref{MF_cluster_expansion} and \ref{large_mass_cluster_expansion} below. We study periodic boundary conditions for concreteness.
\end{remark}

\begin{remark}
By using subadditivity arguments \cite{Ruelle} one can show convergence of the thermodynamic potentials (as $L \rightarrow \infty$) for all temperatures.
\end{remark}

\begin{remark}
When studying the mean-field limit, the assumption that $v$ is of positive type and pointwise nonnegative (see Assumption \ref{interaction_potential_v} and \ref{interaction_potential_v^{L}} above) is needed purely for mathematical reasons. More precisely, the assumption that $v$ is of positive type is needed to apply the Hubbard-Stratonovich formula \eqref{Hubbard_Stratonovich_formula}, which is the starting point of the proof of the Symanzik loop representation, stated in Proposition \ref{Symanzik_representation_theorem}. Furthermore, the proof of the Ginibre loop representation, stated in Proposition \ref{Ginibre_loop_representation}, relies on the pointwise nonnegativity of $v$.
\end{remark}

\subsection{Related results}

The methods used in this paper are inspired by representations of Bose gases and Euclidean field theories in terms of interacting random loops developed by Ginibre \cite{Ginibre1,Ginibre2,Ginibre3,Ginibre} and Symanzik \cite{Symanzik_1968}, respectively.
We note that the mean-field limit was studied for $d=1,2,3$ in the continuum in the work of Lewin, Nam, and Rougerie \cite{LNR_1,LNR_2,LNR_3,LNR_4}, in our previous work \cite{FKSS_2017,FKSS_2020_1,FKSS_2020_2,FKSS_2022}, in the work of the fourth author \cite{Sohinger_2019}, as well as in the work of Rout and the fourth author \cite{Rout_Sohinger_1}. When studying the limiting regime of Bose gases, as $\nu \rightarrow 0$, in dimensions $d =2,3$ in the continuum, it is necessary to introduce a renormalization in the form of Wick ordering.
This serves to control ultraviolet (short-distance) singularities. The latter do not occur in the study of lattice Bose gases (see also Appendix \ref{The heat kernel on the lattice}).
We note that the results in \cite{FKSS_2017} have been extended to gases with singular interaction potentials in \cite{Sohinger_2019}, and \textit{time-dependent} correlation functions in one dimension have been constructed and studied in \cite{FKSS_2019}.

Cluster expansions, which we use to extend our results to the infinite lattice, are ubiquitous in statistical mechanics. 
They were first applied to classical gases in \cite{Lebowitz_Penrose,Penrose_1,Penrose_2}.
For further information on cluster expansions, see \cite{Brydges_1984,GJS,Kotecky_Preiss_1986, Ueltschi_2004} and references given there.
Concerning the infinite-volume limit in the continuum, we note that 
the normalization of the classical Gibbs measure and its distributional properties in the limit $L \rightarrow \infty$ have been studied in \cite{Bourgain_4}. 

In the recent paper \cite{Salmhofer_2020}, a construction of regularized coherent-state functional integrals for ensembles of bosons on a lattice is given. As in \cite{FKSS_2020_1}, an important tool is the Hubbard-Stratonovich formula.

\section{The Symanzik and Ginibre representations} 
\label{The Symanzik and Ginibre representation} 

\subsection{Notation} Let us first introduce some notation that we will use throughout the paper.
We write $\N = \{0,1,2,\dots\}$ and $\N^* = \{1,2,3,\dots\}$.
We use the notation $A_{x,y}$ for the operator kernel of an operator $A$ with respect to the counting measure.
When working on a fixed spatial domain $\Lambda$, we write 
$\int \dd x$ to mean $\int_{\Lambda} \dd x \equiv \sum_{x \in \Lambda}$, if there is no possibility for confusion. We denote by $\mu_{\cal C}$ a Gaussian measure with covariance 
$\cal{C}$. We state all of the properties of $\mu_{\cal C}$ that we use throughout the paper in Appendix \ref{Remarks on Gaussian integrals}.

Given $a,b \geq 0$, we write $a \lesssim b$ if there exists $C>0$ such that $a \leq C b$. We sometimes also write this as $a=O(b)$. Furthermore, if $C$ depends on a set of parameters $p_1,\ldots,p_k$, we write this as $C=C_{p_1,\ldots,p_k}$ or $a \lesssim_{p_1,\ldots,p_k} b$, $a=O_{p_1,\ldots,p_k}(b)$.  $\ind{X}$ denotes the indicator function of a set $X$.

We use the notation 
\begin{equation}
\label{xy_notation}
\b x \b y = (x_1, \dots, x_p,y_1, \dots, y_n)
\end{equation}
for vectors $\b x = (x_1, \dots, x_p), \b y = (y_1, \dots, y_n)$. Moreover, the symmetric group $S_p$ acts on $p$-component vectors $\b x = (x_1, \dots, x_p)$ according to  $\pi \b x = (x_{\pi(1)}, \dots, x_{\pi(p)})$.

For $n \in \N^*$ and $\mathbf T \in (0,\infty)^n$, denote by $\Omega^{\b T} \deq \Omega^{T_1} \times \cdots \times \Omega^{T_n}$ and $\b \omega = (\omega_1, \dots, \omega_n) \in \Omega^{\b T}$ where $\omega_i \in \Omega^{T_i}$.
For $\b x, \b y \in \Lambda^n$, we define the product measure on $\Omega^{\b T}$
by $\bb W^{\b T}_{\b y, \b x}(\dd \b \omega) \deq \bb W^{T_1}_ {y_1,x_1}(\dd \omega_1) \cdots \bb W^{T_n}_{y_n,x_n}(\dd \omega_n)$.
Given  $\b T \in (0,\infty)^p$, we write $\abs{\b T} \deq \sum_{i = 1}^p T_i$.
 Analogously to \eqref{xy_notation}, given $\b \omega = (\omega_1, \dots, \omega_n) \in \Omega^{\b T}$ and $\tilde{\b \omega} = (\tilde{\omega}_1, \dots, \tilde{\omega}_m) \in \Omega^{\tilde{\b T}}$, we write
\begin{equation*}
\b \omega \tilde{\b \omega}=(\omega_1, \dots, \omega_n,\tilde{\omega}_1, \dots, \tilde{\omega}_m) \in \Omega^{\b T \tilde{\b T}}\,.
\end{equation*}

\subsection{The Symanzik representation}

In this section, we derive the Symanzik representation for the classical partition function \eqref{Z_hat} and for the classical $p$-particle correlation functions \eqref{gamma_hat} following \cite{Symanzik_1968}. The precise statement is given in Proposition \ref{Symanzik_representation_theorem} below. 
In Corollary \ref{correlation_inequality_theorem}, we use the methods from the proof of Proposition  \ref{Symanzik_representation_theorem} to give a proof of a correlation inequality for classical $p$-particle correlation functions. The proofs of both Proposition \ref{Symanzik_representation_theorem} and Corollary \ref{correlation_inequality_theorem} are given in Appendix \ref{sec:symanzik_proof}.

In this subsection, the interaction $V \equiv V^{\cl}$ is given by \eqref{n_loop_interaction} using \eqref{V_symanzik}.
For the single-loop measure, we will work with a suitable regularization of $\bb L^{\cl}$ given by \eqref{L_cl}. In particular, given $\epsilon>0$, we consider $\bb L^{\cl,\epsilon}$ given by
\begin{equation} 
\label{L_cl_epsilon}
\int \bb L^{\cl,\epsilon} (\dd \omega) \deq
\int_{\epsilon}^\infty \dd T \, \frac{\ee^{-\kappa T}}{T} \int \bb W^T(\dd \omega)\,.
\end{equation}
We define
\begin{equation} 
\label{cal_Z_epsilon}
\cal Z^{\cl,\epsilon} \deq \sum_{n=0}^{\infty} \frac{1}{n!} \int \bb L^{\cl,\epsilon} (\dd \omega_1) \cdots \int \bb L^{\cl,\epsilon} (\dd \omega_n)\,\exp(-V^{\cl}(\b \omega)) \, \exp(K^{\epsilon})\,,
\end{equation}
where
\begin{equation} 
\label{K_epsilon_definition}
K^{\epsilon}\deq- \int_{\epsilon}^{\infty} \frac{\dd T}{T} \,\ee^{-\kappa T} \int \bb W^{T}(\dd \omega)\,.
\end{equation}
Note that $\lim_{\epsilon \to 0} K^\epsilon = -\infty$. 
Furthermore, we let
\begin{equation} 
\label{gamma^epsilon_definition}
(\Gamma_p^{\cl,\epsilon})_{\b x, \b y} \deq  \sum_{\pi \in S_p} \int_{(0,\infty)^p} \dd \b T\,\ee^{-\kappa \abs{\b T}}\,\int \bb W^{\b T}_{\pi \b y,\b x}(\dd \b \omega)\,\frac{\cal Z^{\cl,\epsilon} (\b \omega)}{\cal Z^{\cl}}\,,
\end{equation}
where
\begin{equation}
\label{cal_Z_epsilon_omega}
\cal Z^{\cl,\epsilon} (\b \omega) \deq 
\sum_{n=0}^{\infty} \frac{1}{n!} \int \bb L^{\cl,\epsilon} (\dd \tilde \omega_1) \cdots \bb L^{\cl,\epsilon} (\dd \tilde \omega_n)\,\exp(-V^{\cl}(\b \omega \tilde{\b \omega})) \, \exp(K^{\epsilon})\,.
\end{equation}
Note that \eqref{cal_Z_epsilon_omega} reduces to \eqref{cal_Z_epsilon} when $\b \omega=\emptyset$.

\begin{proposition}[Symanzik  loop  representation]
\label{Symanzik_representation_theorem} Let $v$ be of positive type.
With notation as in \eqref{Z_hat}, \eqref{gamma_hat}, \eqref{cal_Z_epsilon}, and \eqref{gamma^epsilon_definition}, the following claims hold.
\begin{itemize}
\item[(i)] $\cal Z^{\cl} = \lim_{\epsilon \to 0} \cal Z^{\cl,\epsilon}$.
\item[(ii)] For all $p \in \N^*$ and $\b x, \b y \in \Lambda^p$, we have $(\Gamma_p^{\cl})_{\b x, \b y} = \lim_{\epsilon \to 0} (\Gamma_p^{\cl,\epsilon})_{\b x, \b y}$.
\end{itemize}
\end{proposition}

We now introduce a coupling constant $\lambda \geq 0$ in front of the interaction $v$. This changes the exponential weight in \eqref{Z_hat}.
Note that if $v$ is of positive type, then so is $\lambda v$.
The proof of Proposition \ref{Symanzik_representation_theorem} (see Appendix \ref{sec:symanzik_proof}) allows us to deduce a correlation inequality for the (unnormalized) classical $p$-point correlation function
\begin{equation}
\label{complex_correlation1}
\bigl(\widehat{\Gamma}_p^{\cl,\lambda}\bigr)_{\b x, \b y}\deq\int \mu_{(-\Delta/2+\kappa)^{-1}}(\dd \phi) \, \bar \phi(y_1)\,\cdots\,\bar \phi (y_p)\,\phi(x_1)\,\cdots\,\phi(x_p)\,
\ee^{-\frac{\lambda}{2}  \int \dd x\, \int \dd y\, |\phi(x)|^2 \,v(x-y)\, |\phi(y)|^2}\,.
\end{equation}

\begin{corollary}[Correlation inequality]
\label{correlation_inequality_theorem} Let $v$ be of positive type.
Given $p \in \N^*$, and $\b x, \b y \in \Lambda^p$, with notation as in \eqref{complex_correlation1}, we have
\begin{equation}
\label{correlation_inequality_1}
0 \leq \bigl(\widehat{\Gamma}_p^{\cl,\lambda}\bigr)_{\b x, \b y} \leq \bigl(\widehat{\Gamma}_p^{\cl,0}\bigr)_{\b x, \b y}\,.
\end{equation}
\end{corollary}

\subsection{The Ginibre representation}

In this section, we recall the \emph{Ginibre loop representation} of the reduced $p$-particle density matrices and of the partition function. These results appeared originally in the work of Ginibre  \cite{Ginibre1, Ginibre2, Ginibre3, Ginibre}.

\begin{proposition}[Ginibre loop representation]
\label{Ginibre_loop_representation} 
Let $v$ be pointwise nonnegative.
Recalling \eqref{gamma_p_intro}--\eqref{Z(omega)}, the operator kernel of $\Gamma_{p}^{\nu,\kappa,\lambda}$ defined in \eqref{gamma_p_definition} satisfies identity 
\eqref{ginibre_representation}. Here, the single loop measure $\bb L \equiv \bb L^{\nu,\lambda}$ is given by \eqref{L_nu} and the interaction $V \equiv V^{\nu,\lambda}$ is given by \eqref{n_loop_interaction} using \eqref{V_interaction_1}.
Furthermore, we can write \eqref{cal_Z_definition} as
\begin{equation}
\label{Z_nu}
\cal Z^{\nu,\kappa,\lambda}=  \Biggl\{\sum_{n = 0}^\infty 
\frac{1}{n!}\int \bb L^{\nu,\kappa}(\dd \omega_1) \cdots \bb L^{\nu,\kappa}(\dd \omega_n)
\,\exp(-V^{\nu,\lambda}({\b \omega}))\Biggr\} \Bigg/\exp\biggl\{\int \bb L^{\nu,\kappa}(\dd \omega) \biggr\}\,.
\end{equation}

\end{proposition}
For completeness, we give the proof of Proposition \ref{Ginibre_loop_representation} in Appendix \ref{sec:ginibre_proof}.

It is useful to rewrite the Ginibre representation of \eqref{gamma_p_definition} in terms of relative quantities.
\begin{remark}
\label{Ginibre_representation_relative}
Let us define
\begin{equation}
\label{Z_nu_omega}
\Xi^{\nu,\kappa,\lambda}(\b \omega) \deq  \Biggl\{\sum_{n = 0}^\infty 
\frac{1}{n!}\int \bb L^{\nu,\kappa}(\dd \tilde \omega_1) \cdots \bb L^{\nu,\kappa}(\dd \tilde \omega_n)
\,\exp(-V^{\nu,\lambda}(\b \omega \tilde{\b \omega}))\Biggr\}\,,\quad \Xi^{\nu,\kappa,\lambda} \deq \Xi^{\nu,\kappa,\lambda}(\emptyset)\,,
\end{equation}
and we let $\cal Z^{\nu,\kappa,\lambda}(\b \omega) \deq \frac{\Xi^{\nu,\kappa,\lambda}(\b \omega)}{\Xi^{\nu,\kappa,\lambda}}$. Then, Proposition \ref{Ginibre_loop_representation}  implies that for $p \in \N^*$ and $\b x, \b y\in \Lambda^p$, we have
\begin{equation}
\label{Ginibre_loop_representation_gamma_p_2}
(\Gamma_{p}^{\nu,\kappa,\lambda})_{\b x,\b y} = \sum_{\pi \in S_p} \sum_{\b T \in (\nu \N^*)^p} \ee^{-\kappa \abs{\b T}} \int \bb W^{\b T}_{\pi \b y,\b x}(\dd \b \omega) \, \frac{\cal{Z}^{\nu,\kappa,\lambda}(\b \omega)}{\cal{Z}^{\nu,\kappa,\lambda}}\,.
\end{equation}
\end{remark}

\section{The mean-field limit: 
Convergence of the Ginibre representation to the Symanzik representation}
\label{Mean_field_limit_finite_lattice}

In this section, we study the mean-field convergence on the finite lattice $\Lambda$ and we prove Theorem \ref{mean_field_convergence}.
Let us recall that in this regime, we consider $\nu \rightarrow 0$ with $\lambda=\nu^2$ and $\kappa$ fixed.
Throughout the section, we assume that $v$ satisfies Assumption \ref{interaction_potential_v}.
Recalling \eqref{V_symanzik} and \eqref{V_interaction_1}, by pointwise nonnegativity of $v$, we have that 
\begin{equation} 
\label{V_nonnegative}
\cal V^{\cl}(\omega, \tilde \omega) \geq 0\,,\quad \cal V^{\nu,\nu^2}(\omega, \tilde \omega) \geq 0\,.
\end{equation}
 Given $n \in \N^*$, and $\b T \in (0,\infty)^n$, in the sequel we write
\begin{equation}
\label{loop_measure_quantum}
\bb W^{\b T}(\dd \b \omega) \deq \int_{\Lambda^n} \dd \b x\,  \bb W_{\b x,\b x}^T (\dd \b \omega)\,.
\end{equation}

We denote by
\begin{equation} 
\label{heat_kernel_Lambda_L}
\psi^{t}(x) \deq (\ee^{t \Delta/2})_{x,0}
\end{equation}
the heat kernel on $\Lambda$.
Let us note that for $x,y \in \Lambda$ and $t>0$ we have
\begin{equation}
\label{heat_kernel_identity}
\int \bb W^{t}_{y,x}(\dd \omega)= \psi^{t}(y-x)\,.
\end{equation}
Further properties of the heat kernel on the lattice are given in Appendix \ref{The heat kernel on the lattice}.

\begin{proof}[Proof of Theorem \ref{mean_field_convergence}]
Let us first prove (i).
For $\b \omega \in \Omega^{\b T}$, we can write $V^{\nu,\nu^2}(\b \omega)= \frac{1}{2} \langle f, \nu v f \rangle$,
where
\begin{equation}
\label{V_interaction_3_rewritten_f}
f(x)=\sum_{i=1}^{n} \sum_{r_i \in \nu \N} \ind{r_i<T_i}\,\int_{0}^{\nu} \dd t\, \delta(x-\omega_i(t+r_i))\,.
\end{equation}
We rewrite $\cal Z^{\nu,\kappa,\nu^2}$ by starting from \eqref{Z_nu}.
By using the Hubbard-Stratonovich formula \eqref{Hubbard_Stratonovich_formula} 
for $f$ as in  \eqref{V_interaction_3_rewritten_f}, collecting terms in the exponential, and recalling \eqref{L_nu} we can write
\begin{equation}
\label{Z_beta_1A}
\cal Z^{\nu,\kappa,\nu^2}=\int \mu_{\nu v}(\dd \sigma)\,\exp\Biggl\{\int \bb L^{\nu,\kappa}(\dd \omega)\biggl[\ee^{\ii \sum_{r \in \nu \N} \ind{r <T(\omega)} \int_{0}^{\nu} \dd t \, \sigma(\omega(t+r))}-1\biggr]\Bigg\}\,.
\end{equation}
Given $\epsilon>0$, we let
\begin{equation}
\label{Z_beta_epsilon}
\cal Z^{\nu,\kappa,\nu^2,\epsilon} \deq\int \mu_{\nu v}(\dd \sigma)\,\exp\Biggl\{\nu \sum_{T \in \nu \N^* \cap [\epsilon, \infty)} \frac{\ee^{-\kappa T}}{T}\int \bb W^{T} (\dd \omega) \biggl[\ee^{\ii \sum_{r \in \nu \N} \ind{r < T} \int_{0}^{\nu} \dd t \, \sigma(\omega(t+r))}-1\biggr]\Bigg\}\,.
\end{equation}
We first show that there exists $\nu_0>0$ sufficiently small such that
\begin{equation}
\label{Z_beta_epsilon_convergence}
\lim_{\epsilon \rightarrow 0} \cal Z^{\nu,\kappa,\nu^2,\epsilon}=\cal Z^{\nu,\kappa,\nu^2}\,,\quad \mbox{uniformly in } \nu \in (0,\nu_0)\,.
\end{equation}
In order to prove \eqref{Z_beta_epsilon_convergence}, we use \eqref{Z_beta_1A}--\eqref{Z_beta_epsilon} to compute
\begin{multline}
\label{Z_beta_1}
\cal Z^{\nu,\kappa,\nu^2}-\cal Z^{\nu,\kappa,\nu^2,\epsilon}
\\
=\int \mu_{\nu v}(\dd \sigma)\,\exp\Biggl\{\nu \sum_{T \in \nu \N^* \cap [\epsilon, \infty)} \frac{\ee^{-\kappa T}}{T}\int \bb W^{T} (\dd \omega) \biggl[\ee^{\ii \sum_{r \in \nu \N} \ind{r <T} \int_{0}^{\nu} \dd t \, \sigma(\omega(t+r))}-1\biggr]\Bigg\}
\\
\times \Biggl(
\exp\Biggl\{\nu \sum_{T \in \nu \N^* \cap (0,\epsilon)} \frac{\ee^{-\kappa T}}{T}\int \bb W^{T} (\dd \omega) \biggl[\ee^{\ii \sum_{r \in \nu \N} \ind{r < T} \int_{0}^{\nu} \dd t \, \sigma(\omega(t+r))}-1\biggr]\Biggr\}-1\Biggr)\,.
\end{multline}
We estimate each factor of the integrand in \eqref{Z_beta_1} separately. 
For the first factor, we recall \eqref{W_path_measures} and \eqref{heat_kernel_identity} to note that $\int \bb W^{T} (\dd \omega) =\psi^{T} (0) |\Lambda|$. Furthermore, we use the elementary inequality $\abs{\ee^{\ii a}-1} \leq C \abs{a}$ for $a \in \R$ and consider Riemann sums with mesh size $\nu$ for the integral 
\begin{equation}
\label{Z_beta_integral}
\int_{\epsilon}^{\infty} \dd T\, \ee^{-\kappa T}\,\psi^T(0)=O_{\kappa}(1)\,,
\end{equation}
to deduce that 
\begin{equation}
\Biggl|\exp\Biggl\{\nu \sum_{T \in \nu \N^* \cap [\epsilon, \infty)} \frac{\ee^{-\kappa T}}{T}\int \bb W^{T} (\dd \omega) \biggl[\ee^{\ii \sum_{r \in \nu \N} \ind{r < T} \int_{0}^{\nu} \dd t \, \sigma(\omega(t+r))}-1\biggr]\Biggr\}\Biggr|
\leq \ee^{C_\kappa |\Lambda|  \|\sigma\|_{\infty} }\,.
\end{equation}
Note that for \eqref{Z_beta_integral} we used that $\psi^T(0) \leq 1$, which follows from Lemma \ref{estimates_heat_kernel} (i).
Similarly, for the second factor of the integrand in \eqref{Z_beta_1}, we use the elementary inequality $\abs{\ee^{\zeta}-1} \leq \abs{\zeta} \ee^{\abs{\zeta}}$ for $\zeta \in \C$
and consider Riemann sums with mesh size $\nu$ for the integral 
$\int_{0}^{\epsilon} \dd T\, \ee^{-\kappa T}\,\psi^T(0)\leq \epsilon$
to deduce that 
\begin{equation}
\label{Z_beta_3}
\Biggl|\exp\Biggl\{\nu \sum_{T \in \nu \N^* \cap (0,\epsilon)} \frac{\ee^{-\kappa T}}{T}\int \bb W^{T} (\dd \omega) \biggl[\ee^{\ii \sum_{r \in \nu \N} \ind{r <T} \int_{0}^{\nu} \dd t \, \sigma(\omega(t+r))}-1\biggr]\Biggr\}-1\Biggr| 
\leq C \epsilon |\Lambda|  \|\sigma\|_{\infty} \,\ee^{C \epsilon |\Lambda|  \|\sigma\|_{\infty} }\,.
\end{equation}
Combining \eqref{Z_beta_1}--\eqref{Z_beta_3}, it follows that
\begin{equation}
\label{Z_beta_4}
|\cal Z^{\nu,\kappa,\nu^2}-\cal Z^{\nu,\kappa,\nu^2,\epsilon}|\leq C \epsilon \, \int \dd \mu_{\nu v}(\sigma) \,\ee^{C_{\kappa} |\Lambda| \|\sigma\|_{\infty} }\,\,.
\end{equation}
Recalling \eqref{exp_Linfty_bound}, we note that \eqref{Z_beta_epsilon_convergence} follows from \eqref{Z_beta_4} if we prove that, for fixed $x \in \Lambda$, we have
\begin{equation}
\label{Z_beta_5}
\int \dd \mu_{\nu v}(\sigma) \,\ee^{C |\sigma(x)|}=O(1)\,, \quad \mbox{uniformly in }  \nu \in (0,\nu_0)\,,
\end{equation}
for $\nu_0>0$ sufficiently small. We deduce \eqref{Z_beta_5} by using \eqref{exp_Linfty_bound_application} (with $v$ replaced by $\nu v$). Namely, we note the elementary inequality $
\sqrt{\frac{(2i)!}{i!\,2^i}}\leq\frac{(2i)!}{i!\,2^i} \leq 2^i\,i!$ and thus obtain that the left-hand side of \eqref{Z_beta_5} is
$\leq
\sum_{i=0}^{\infty} \bigl(2C v(0)^{1/2}\nu^{1/2}\bigr)^{i} =O(1)$,
provided that $\nu \in (0,\nu_0)$ for $\nu_0>0$ sufficiently small.

By using \eqref{Hubbard_Stratonovich_formula} as in the proof of \eqref{Z_beta_1A}, we can rewrite \eqref{Z_beta_epsilon} as
\begin{equation}
\label{Z_hat_calculation3_Theorem_1.1_2}
\cal Z^{\nu,\kappa,\nu^2,\epsilon}=\sum_{n=0}^{\infty} \frac{\nu^n}{n!} \sum_{\b T \in  (\nu \N^* \cap [\epsilon, \infty))^n} \prod_{i=1}^{n} \frac{1}{T_i}\, \ee^{-\kappa \abs{\b T}} \int \bb W^{\b T} (\dd \b \omega) \exp(-V^{\nu,\nu^2} (\b \omega))\,\exp(K^{\epsilon}_{\nu})\,,
\end{equation}
where 
\begin{equation}
\label{K_{epsilon,nu}}
K^{\epsilon}_{\nu} \deq -\nu \sum_{T \in \nu \N^* \cap [\epsilon, \infty)} \frac{\ee^{-\kappa T}}{T} \psi^{T}(0)  |\Lambda| \leq 0\,.
\end{equation}
For the inequality above, we used Lemma \ref{estimates_heat_kernel} (i).

Recalling \eqref{cal_Z_epsilon}, we now show that for all $\epsilon>0$
\begin{equation}
\label{Z_beta_epsilon_convergence_2}
\lim_{\nu \rightarrow 0} \cal Z^{\nu,\kappa,\nu^2,\epsilon}=\cal Z^{\cl,\epsilon}\,.
\end{equation}
In order to prove \eqref{Z_beta_epsilon_convergence_2}, we show three auxiliary claims. Let us henceforth fix $\epsilon>0$.

\begin{itemize}
\item[(1)] For $u_1,u_2 \in \Lambda$, we have
\begin{multline}
\label{Z_beta_epsilon_proof1}
\nu^2 \sum_{T_1 \in \nu \N^* \cap [\epsilon, \infty)} \sum_{T_2 \in \nu \N^* \cap [\epsilon, \infty)} \frac{\ee^{-\kappa T_1} \ee^{-\kappa T_2}}{T_1 T_2} \int \bb{W}_{u_1, u_1}^{T_1} (\dd \omega_1)
\int \bb{W}_{u_2, u_2}^{T_2} (\dd \omega_2)\,
\\
\Bigl|\exp\bigl(-\cal V^{\nu,\nu^2}(\omega_1,\omega_2)\bigr)-\exp\bigl(-\cal V^{\cl}(\omega_1,\omega_2)\bigr)\Bigr| \leq C_{\kappa} \nu \|v\|_{\ell^{\infty}}\,.
\end{multline}

\item[(2)] For $u \in \Lambda$, we have
\begin{equation}
\label{Z_beta_epsilon_proof3}
\nu \sum_{T \in \nu \N^* \cap [\epsilon, \infty)} \frac{\ee^{-\kappa T}}{T}\, \int \bb W_{u, u}^{T}(\dd \omega) = O_{\epsilon,\kappa}(|\Lambda|)\,.
\end{equation}
\item[(3)] With $K^{\epsilon}_{\nu}$ as in \eqref{K_{epsilon,nu}} and $K^\epsilon$ as in \eqref{K_epsilon_definition}, we have
\begin{equation}
\label{Z_beta_epsilon_proof4}
\lim_{\nu \rightarrow 0} K^{\epsilon}_{\nu}=K^{\epsilon}\,.
\end{equation}
\end{itemize}

Let us assume \eqref{Z_beta_epsilon_proof1}--\eqref{Z_beta_epsilon_proof4} for now. We show how one can then deduce \eqref{Z_beta_epsilon_convergence_2}.
Let us define for $\nu>0$ the auxiliary quantity
\begin{equation}
\label{Z_tilde_beta_epsilon}
\wt{\cal Z}^{\nu,\kappa,\nu^2,\epsilon} \deq 
\sum_{n=0}^{\infty} \frac{\nu^n}{n!} \sum_{\b T \in (\nu \N^*) \cap [\epsilon, \infty)^n} \prod_{i=1}^{n} \frac{1}{T_i}\, \ee^{-\kappa \abs{\b T}} \int \bb W^{\b T} (\dd \b \omega) \exp(-V^{\cl} (\b \omega))\,\exp(K^{\epsilon}_{\nu})\,.
\end{equation}

Using  \eqref{V_nonnegative}, \eqref{K_{epsilon,nu}}, \eqref{Z_beta_epsilon_proof1}--\eqref{Z_beta_epsilon_proof3}, and applying a telescoping argument in comparing \eqref{Z_hat_calculation3_Theorem_1.1_2} and \eqref{Z_tilde_beta_epsilon}, we deduce that
\begin{equation}
\label{Z_beta_epsilon_convergence_2A}
\cal Z^{\nu,\kappa,\nu^2,\epsilon}-\wt{\cal Z}^{\nu,\kappa,\nu^2,\epsilon}= O_{\epsilon,v,\Lambda,\kappa} (\nu)\,.
\end{equation}
By using \eqref{cal_Z_epsilon}, \eqref{Z_beta_epsilon_proof4}, \eqref{Z_tilde_beta_epsilon}, and by considering Riemann sums with mesh size $\nu$ we deduce that
\begin{equation}
\label{Z_beta_epsilon_convergence_2B}
\lim_{\nu \rightarrow 0} \wt{\cal Z}^{\nu,\kappa,\nu^2,\epsilon}=\cal Z^{\cl,\epsilon}\,.
\end{equation}
We hence obtain \eqref{Z_beta_epsilon_convergence_2} from \eqref{Z_beta_epsilon_convergence_2A}--\eqref{Z_beta_epsilon_convergence_2B}.
Claim (i) then follows from Proposition \ref{Symanzik_representation_theorem} (i), \eqref{Z_beta_epsilon_convergence}, and \eqref{Z_beta_epsilon_convergence_2}.

Let us now show \eqref{Z_beta_epsilon_proof1}--\eqref{Z_beta_epsilon_proof4}.
We first show \eqref{Z_beta_epsilon_proof1}.
By using \eqref{V_nonnegative} and recalling \eqref{V_symanzik}, \eqref{V_interaction_1}, we have that the contribution to the left-hand side of \eqref{Z_beta_epsilon_proof1} for fixed $T_1, T_2$ is
\begin{align}
\notag
\leq 
\int \bb{W}_{u_1, u_1}^{T_1}  (\dd \omega_1)
\int \bb{W}_{u_2, u_2}^{T_2} &(\dd \omega_2) \,
\sum_{r \in \nu \N} \ind{r < T_1} \sum_{s \in \nu \N} \ind{s < T_2} \int_{0}^{\nu} \dd t_1 \int_{0}^{\nu} \dd t_2
\\
\label{Triangle_{l1,l2}_A}
&\bigl|v\bigl(\omega_1(t_1+r)-\omega_2(t_1+s)\bigr)-v\bigl(\omega_1(t_1+r)-\omega_2(t_2+s)\bigr)\bigr|\,.
\end{align}
The contribution to \eqref{Triangle_{l1,l2}_A} when $t_1 \leq t_2$ is
\begin{align}
\notag
\leq 
\sum_{r \in \nu \N} \ind{r < T_1} &\sum_{s \in \nu \N} \ind{s <T_2} \int_{0}^{\nu} \dd t_1 \int_{0}^{\nu} \dd t_2 \int \dd \zeta \int \dd \eta_1 \int \dd \eta_2\,\psi^{t_1+r}(\zeta-u_1)\,\psi^{T_1-t_1-r}(u_1-\zeta)\,
\\
\label{Triangle_{l1,l2}_B}
&\times \psi^{t_1+s}(\eta_1-u_2)\,\psi^{t_2-t_1}(\eta_2-\eta_1)\,
\psi^{T_2-t_2-s}(u_2-\eta_2)\, \bigl|v(\zeta-\eta_1)-v(\zeta-\eta_2)\bigr|\,,
\end{align}
which by Lemma \ref{estimates_heat_kernel} is\footnote{Note that \eqref{Triangle_{l1,l2}_B} vanishes if $\eta_1=\eta_2$, hence we estimate $\psi^{t_2-t_1}(\eta_2-\eta_1)$ by Lemma \ref{estimates_heat_kernel} (ii).}
\begin{equation}
\label{Triangle_{l1,l2}}
\lesssim \nu^3 \,\|v\|_{\ell^{\infty}} \,\sum_{r \in \nu \N} \ind{r < T_1} \sum_{s \in \nu \N} \ind{s < T_2} =\nu \,\|v\|_{\ell^{\infty}} \,T_1T_2\,.
\end{equation}
The contribution to \eqref{Triangle_{l1,l2}_A} when $t_1 > t_2$ is estimated analogously.
Using \eqref{Triangle_{l1,l2}}, it follows that the expression on the left-hand side of \eqref{Z_beta_epsilon_proof1} is
\begin{equation}
\label{Z_beta_epsilon_proof1_conclusion}
\lesssim \nu \,\|v\|_{\ell^{\infty}} \,\sum_{T_1 \in \nu \N^* \cap [\epsilon, \infty)} \sum_{T_2 \in \nu \N^* \cap [\epsilon, \infty)} \nu^2 \ee^{-\kappa T_1} \ee^{-\kappa T_2} \leq C_{\kappa}\nu \|v\|_{\ell^{\infty}}\,,
\end{equation}
as was claimed.
In order to obtain \eqref{Z_beta_epsilon_proof1_conclusion}, we considered a Riemann sum of mesh size $\nu$ for the integral
$\int_{\epsilon}^{\infty} \dd T_1 \int_{\epsilon}^{\infty} \dd T_2 \, \ee^{-\kappa T_1} \ee^{-\kappa T_2} = O_{\kappa}(1).$

We now show \eqref{Z_beta_epsilon_proof3}. By Lemma \ref{estimates_heat_kernel} (i), it follows that the expression on the left-hand side of \eqref{Z_beta_epsilon_proof3} is
\begin{equation}
\label{Z_beta_epsilon_proof3A}
\leq \nu \sum_{T \in \nu \N^* \cap [\epsilon, \infty)} \frac{\ee^{-\kappa T}}{T}|\Lambda|
=O_{\epsilon,\kappa}(|\Lambda|)\,,
\end{equation}
and we hence obtain \eqref{Z_beta_epsilon_proof3}.
In \eqref{Z_beta_epsilon_proof3A}, we considered a Riemann sum of mesh size $\nu$ for the integral
$\int_{\epsilon}^{\infty}\dd T\,\frac{\ee^{-\kappa T}}{T} = O_{\epsilon,\kappa}(1).$

Finally, in order to show \eqref{Z_beta_epsilon_proof4}, we rewrite \eqref{K_epsilon_definition} using Lemma \ref{FK_discrete} as
\begin{equation} \label{K_epsilon_rewritten}
K^{\epsilon}= -\int_{\epsilon}^{\infty} \frac{\dd T}{T}\,\ee^{-\kappa T} \psi^T(0)\,|\Lambda|\,.
\end{equation}
Using \eqref{K_{epsilon,nu}}, \eqref{K_epsilon_rewritten}, and considering Riemann sums of mesh size $\nu$ for the integral in \eqref{K_epsilon_rewritten}, we deduce \eqref{Z_beta_epsilon_proof4}. Claim (i) now follows.

The proof of (ii) is similar to that of (i). We just outline the main differences. We now start from \eqref{Ginibre_loop_representation_gamma_p_2} and hence work with paths that can also be open. 
For fixed $\b T \in (\nu {\N^*})^p, \tilde{\b T} \in (\nu {\N^*})^n, \b x, \b y \in \Lambda^p, (\b \omega,\tilde{\b \omega}) \in \Omega^{\b T}_{\b y, \b x} \times \Omega^{\tilde{\b T}}$, we can write $V^{\nu,\nu^2}(\b \omega \tilde{\b \omega})= \frac{1}{2} \langle f, \nu v f \rangle$,
where
\begin{equation}
\label{correlations_f_choice}
f(x)=\sum_{j=1}^{p} \sum_{r_j \in \nu \N} \ind{r_j<T_j}\,\int_{0}^{\nu} \dd t\, \delta(x-\omega_j(t+r_j))+\sum_{i=1}^{n} \sum_{s_i \in \nu \N} \ind{s_i<\tilde T_i}\,\int_{0}^{\nu} \dd t\, \delta(x-\tilde \omega_i(t+s_i))\,.
\end{equation}
We use \eqref{Ginibre_loop_representation_gamma_p_2} 
and apply \eqref{Hubbard_Stratonovich_formula} with $f$ as in \eqref{correlations_f_choice} to obtain that
\begin{equation}
\label{gamma_hat_identity}
(\Gamma^{\nu,\kappa,\nu^2}_p)_{\b x,\b y} =\frac{1}{\cal Z^{\nu,\kappa,\nu^2}}\,(\widehat{\Gamma}^{\nu,\kappa,\nu^2}_p)_{\b x,\b y}\,, 
\end{equation}
where $(\widehat{\Gamma}^{\nu,\kappa,\nu^2}_p)_{\b x,\b y}
$ is
\begin{align*}
& \deq \sum_{\b T \in (\nu \N^*)^p} \sum_{\pi \in S_p} \ee^{-\kappa \abs{\b T}} \int \bb W^{\b T}_{\pi \b y,\b x}(\dd \b \omega)\,\int \mu_{v} (\dd \sigma)\,
\exp\Biggl\{\ii \sum_{j=1}^{p} \sum_{r_j \in \nu \N} \ind{r_j < T_j} \int_{0}^{\nu} \dd t \, \sigma(\omega_j(t+r_j))\Biggr\}
\\
&\times \exp\Biggl\{\int \bb L^{\nu,\kappa}(\dd \tilde \omega) \biggl[\ee^{\ii \sum_{s\in \nu \N} \ind{s< T(\tilde \omega)} \int_{0}^{\nu} \dd t \, \sigma(\tilde \omega(t+s))}-1\biggr]\Biggr\}\,.
\end{align*}
By arguing analogously as for \eqref{Z_beta_epsilon_convergence} 
we have that there exists $\nu_0>0$ sufficiently small such that 
\begin{equation}
\label{gamma_beta_epsilon_convergence}
\lim_{\epsilon \rightarrow 0} \nu^p (\widehat{\Gamma}^{\nu,\kappa,\nu^2,\epsilon}_p)_{\b x,\b y}=\nu^p (\widehat{\Gamma}^{\nu,\kappa,\nu^2}_p)_{\b x,\b y}\,,\quad \mbox{uniformly in }  \nu \in (0,\nu_0)\,,
\end{equation}
where
\begin{multline}
\label{gamma_hat_calculation3_Theorem_1.1_2}
(\widehat{\Gamma}^{\nu,\kappa,\nu^2,\epsilon}_p)_{\b x,\b y} =\sum_{\b T \in (\nu \N^*)^p} \sum_{\pi \in S_p}  \ee^{-\kappa \abs{\b T}} \int  \bb W^{\b T}_{\pi \b y,\b x}(\dd \b \omega) \sum_{n=0}^{\infty} \frac{\nu^n}{n!} \sum_{\tilde{\b T} \in (\nu \N^*)^n \cap [\epsilon,\infty)^n} \prod_{i=1}^{n} \frac{1}{\tilde T_i} \ee^{-\kappa \abs{\tilde{\b T}}} 
\\
\times
\int \bb W^{\tilde{\b T}} (\dd \tilde{\b \omega}) \exp(-V^{\nu,\nu^2}(\b \omega \tilde{\b \omega}))\,\exp(K^{\epsilon}_{\nu})\,,
\end{multline}
for $K^{\epsilon}_{\nu}$ as given by \eqref{K_{epsilon,nu}}.
We now use \eqref{gamma_hat_calculation3_Theorem_1.1_2} and consider Riemann sums as in the proof of \eqref{Z_beta_epsilon_convergence_2} to deduce that for all $\epsilon>0$ we have
\begin{equation}
\label{correlations_convergence_4}
\lim_{\nu \rightarrow 0} \nu^p (\widehat{\Gamma}^{\nu,\kappa,\nu^2,\epsilon}_p)_{\b x,\b y}=\cal Z^{\cl}\,(\Gamma_p^{\cl,\epsilon})_{\b x, \b y}\,,
\end{equation}
where $(\Gamma_p^{\cl,\epsilon})_{\b x, \b y}$ is given by \eqref{gamma^epsilon_definition}. 
Claim (ii) now follows from \eqref{gamma_hat_identity}, \eqref{gamma_beta_epsilon_convergence}, \eqref{correlations_convergence_4}, Proposition \ref{Symanzik_representation_theorem} (ii), and from part (i) of the theorem.

\end{proof}

\section{The large-mass limit}
\label{The large-mass limit}

In this section, we analyse the large-mass limit on the finite lattice $\Lambda$ and provide the proof of Theorem \ref{Large_mass_lattice_4}. 
We recall that in this regime, we consider $\nu \rightarrow 0$ with $\lambda=1$ and $\kappa=\frac{\kappa_0}{\nu}$.
Throughout this section, we assume that $v$ satisfies Assumption \ref{interaction_potential_v_2}.
In light of \eqref{H_n_mass_m}, we need to slightly modify \eqref{n_loop_interaction} when studying the large-mass limit. Namely, recalling \eqref{V_interaction_1}, we work with 
\begin{equation}
\label{V_interaction_large_mass}
V^{\nu,1}(\b \omega) = \frac{1}{2} \mathop{\sum_{i,j=1}^{n}}_{i \neq j} \cal V^{\nu,1}(\omega_i, \omega_j)+ \frac{1}{2}\sum_{i=1}^{n}\wt{\cal V}^{\nu,1}(\omega_i)+ \frac{v(0)}{2\nu}  |\b T|\,\ind{R=0}\,,
\end{equation} 
where
\begin{equation}
\label{V_interaction_1_large_mass_B}
\wt{\cal V}^{\nu,1} (\omega) \deq  \frac{1}{\nu}\,\sum_{r,s \in \nu\N} \ind{r,s< T(\omega)}\,\ind{r \neq s} \int_0^{\nu} \dd t \, v\pb{\omega(t + r) - \omega(t + s)}\,.
\end{equation}
Note that \eqref{V_interaction_large_mass} differs from \eqref{n_loop_interaction} only in the presence of a hard core potential. With this modification, the result of Proposition \ref{Ginibre_loop_representation} holds. We use this without further comment below.

Before proceeding to the proof of Theorem \ref{Large_mass_lattice_4}, we first note several basic facts.
\begin{lemma}
\label{Large_mass_lattice_2}
Let $g_1: \Lambda \rightarrow \C$ and $g_2: \Lambda \times \Lambda \rightarrow \C$ be given. The following estimates hold.
\begin{itemize}
\item[(i)] For $x,x' \in \Lambda, T >0$, and $r,s,t \in [0, T]$ satisfying $r \leq s$ and $t+s \leq T$, we have
\begin{equation}
\label{Large_mass_lattice_2_Claim_i}
\int \bb{W}^{T}_{x',x} (\dd \omega) \,\bigl|g_1 \bigl(\omega(t+r)-\omega(t+s)\bigr)-g_1(0)\bigr| =  O_{\Lambda}\bigl(T\,\|g_1\|_{\ell^{\infty}}\bigr)\,.
\end{equation}
\item[(ii)]  For $x_1,x_2,x_1',x_2' \in \Lambda, T_1,T_2>0, t_1 \in [0, T_1], t_2 \in [0, T_2]$ we have
\begin{equation}
\label{Large_mass_lattice_2_Claim_ii}
\int \bb{W}^{T_1}_{x_1',x_1} (\dd \omega_1)\, \int \bb{W}^{T_2}_{x_2',x_2} (\dd \omega_2) \,\bigl|g_2 \bigl(\omega_1(t_1),\omega_2(t_2)\bigr)-g_2(x_1,x_2) \bigr| 
= O_{\Lambda}\bigl((T_1+T_2)\,\|g_2\|_{\ell^{\infty}}\bigr)\,.
\end{equation}
\end{itemize}
\end{lemma}
\begin{proof}
We first prove (i). Let us first consider the case when $x=x'$. The expression on the left-hand side of \eqref{Large_mass_lattice_2_Claim_i} is then given by
\begin{equation}
\label{Large_mass_lattice_2_i_proof_1A}
\int\, \dd \zeta \, \int \dd \eta\, 
\psi^{t+r}(\zeta-x)\,\psi^{s-r}(\eta-\zeta)\,\psi^{T-s-t}(x-\eta)\,\bigl|g_1(\zeta-\eta)-g_1(0)\bigr|\,.
\end{equation}
By Lemma \ref{estimates_heat_kernel} (i)--(ii), it follows that the expression in \eqref{Large_mass_lattice_2_i_proof_1A} is 
\begin{equation}
\label{Large_mass_lattice_2_i_proof_1B}
\leq
\int\, \dd \zeta \, \int \dd \eta\,\psi^{s-r}(\eta-\zeta)\,\bigl|g_1(\zeta-\eta)-g_1(0)\bigr|=O_{\Lambda}\bigl(T\,\|g_1\|_{\ell^{\infty}}\bigr)\,.
\end{equation}
Let us now consider the case when $x \neq x'$.
By using \eqref{heat_kernel_identity} and Lemma \ref{estimates_heat_kernel} (ii), it follows that the expression on the left-hand side of \eqref{Large_mass_lattice_2_Claim_i} is 
\begin{equation}
\label{Large_mass_lattice_2_i_proof_1C}
\leq 2 \|g_1\|_{\ell^{\infty}} \psi^{T}(x'-x) \lesssim T \|g_1\|_{\ell^{\infty}}\,.
\end{equation}
Claim (i) follows from \eqref{Large_mass_lattice_2_i_proof_1B}--\eqref{Large_mass_lattice_2_i_proof_1C}.

We now prove (ii).  Let us first consider the case when $x_1=x_1'$ and $x_2=x_2'$. Similarly as in the proof of (i), we need to estimate 
\begin{equation*}
\int\, \dd \zeta \int \, \dd \eta \, \psi^{t_1}(\zeta-x_1)\,\psi^{T_1-t_1}(x_1-\zeta)\,\psi^{t_2}(\eta-x_2)\,\psi^{T_2-t_2}(x_2-\eta)\, \bigl|g_2(\zeta,\eta)-g_2(x_1,x_2)\bigr|\,,
\end{equation*}
which by the triangle inequality and Lemma \ref{estimates_heat_kernel} (i) is 
\begin{equation*}
\leq \int \dd \eta \, \int \dd \zeta\,\psi^{t_1}(\zeta-x_1)\, \bigl|g_2(\zeta,\eta)-g_2(x_1,\eta)\bigr|
+  \int \dd \eta \, \int \dd \zeta\,\psi^{t_2}(\eta-x_2)\,\bigl|g_2(x_1,\eta)-g_2(x_1,x_2)\bigr|\,.
\end{equation*}
By Lemma \ref{estimates_heat_kernel} (ii), the above expression is 
\begin{equation}
\label{Large_mass_lattice_2_ii_proof_1A}
=O_{\Lambda}\bigl((T_1+T_2)\,\|g_2\|_{\ell^{\infty}}\bigr)\,.
\end{equation}
Let us now consider the case when $x_1 \neq x_1'$. By \eqref{heat_kernel_identity} and Lemma \ref{estimates_heat_kernel} (i)--(ii), we obtain that the expression on the left-hand side of \eqref{Large_mass_lattice_2_Claim_ii} is 
\begin{equation}
\label{Large_mass_lattice_2_ii_proof_1B}
\leq 2 \|g_2\|_{\ell^{\infty}} \,\psi^{T_1}(x_1'-x_1) \,\psi^{T_2}(x_2'-x_2) \lesssim T_1 \|g_2\|_{\ell^{\infty}} \,.
\end{equation}
Finally, when $x_2 \neq x_2'$, we obtain by analogous arguments that the left-hand side of \eqref{Large_mass_lattice_2_Claim_ii} is 
\begin{equation}
\label{Large_mass_lattice_2_ii_proof_1C}
\lesssim T_2 \|g_2\|_{\ell^{\infty}} \,.
\end{equation}
Claim (ii) follows from \eqref{Large_mass_lattice_2_ii_proof_1A}--\eqref{Large_mass_lattice_2_ii_proof_1C}.
\end{proof}
We define
\begin{equation}
\label{hat_U_full_interaction}
\wh{V}^{\nu,1}(\b \omega) \deq \frac{1}{2} \sum_{\substack {i,j=1 \\ i \neq j}}^{n} \wh{\cal V}^{\nu,1}(\omega_i, \omega_j)+\frac{1}{2}\sum_{i=1}^{n}\wh{\cal V}^{\nu,1}(\omega_i, \omega_i)\,\ind{R=0}\,,
\end{equation}
where we let
\begin{equation}
\label{hat_U_interaction}
\wh{\cal V}^{\nu,1}(\omega, \tilde \omega) \deq \frac{1}{\nu^2} \, T(\omega)\,T(\tilde \omega) v\bigl(x(\omega)-x(\tilde \omega)\bigr)\,.
\end{equation}
We note that $\wh{V}^{\nu,1}(\b \omega)$ depends only on the durations $\b T=(T(\omega_1), \ldots, T(\omega_n))$ and initial points $\b x=(x(\omega_1), \ldots, x(\omega_n))$ of the paths $\b \omega =(\omega_1, \ldots, \omega_n)$.
Let us note a consequence of Lemma \ref{Large_mass_lattice_2}.

\begin{lemma}
\label{Large_mass_lattice_3}
The following estimates hold.
\begin{itemize}
\item[(i)] Suppose that $R=0$ in Assumption \ref{interaction_potential_v_2}.
For $x,x' \in \Lambda$ and $T \in \nu \N^*$, we have
\begin{equation*}
\int \bb{W}^{T}_{x',x} (\dd \omega) \,\Bigl|\ee^{-\cal V^{\nu,1}(\omega,\omega)}-\ee^{-\wh{\cal V}^{\nu,1}(\omega,\omega)}\Bigr|=O_{\Lambda}\biggl(\frac{T^3}{\nu^2}\,\|v\|_{\ell^{\infty}}\biggr).
\end{equation*}
\item[(ii)] With $v$ as in Assumption \ref{interaction_potential_v_2}, we define the function $\tilde{v} : \Lambda \rightarrow \R$ by 
\begin{equation}
\label{tilde_v_definition}
\tilde{v}(x) \deq v(x) \, \ind{|x| \geq R}\,.
\end{equation}
For $x_1,x_2,x_1',x_2' \in \Lambda$ and $T_1,T_2 \in \nu \N^*$, we have
\begin{align}
\notag
\int \bb{W}^{T_1}_{x_1',x_1} (\dd \omega_1)\,\int \bb{W}^{T_2}_{x_2',x_2} (\dd \omega_2)\,&\Bigl|\ee^{-\cal V^{\nu,1}(\omega_1,\omega_2)}-\ee^{-\wh{\cal V}^{\nu,1}(\omega_1,\omega_2)}\Bigr|
\\
\label{Large_mass_lattice_3_ii}
&=O_{\Lambda}\biggl(\frac{ T_1 T_2 (T_1+T_2)}{\nu^2}\,\|\tilde{v}\|_{\ell^{\infty}}\biggr)+O(T_1^2+T_2^2)\,.
\end{align}
\end{itemize}
\end{lemma}

\begin{proof}
In order to prove (i), we note that for $\omega \in \Omega^{T}_{x',x}$, we have by  \eqref{V_interaction_large_mass}, \eqref{hat_U_interaction}, and Assumption \ref{interaction_potential_v_2} 
\begin{equation}
\label{Large_mass_lattice_3_i_proof_1}
\Bigl|\ee^{-\cal V^{\nu,1}(\omega,\omega)}-\ee^{-\wh{\cal V}^{\nu,1}(\omega,\omega)}\Bigr|
\leq 
\frac{1}{2\nu}\sum_{r \in \nu \N} \ind{r<T} \sum_{s \in \nu \N} \ind{s<T} \int_{0}^{\nu}\dd t\,\bigl|v \bigl(\omega(t+r)-\omega(t+s)\bigr)-v(0)\bigr|\,.  
\end{equation}
Using \eqref{Large_mass_lattice_3_i_proof_1}, Fubini's theorem, and Lemma \ref{Large_mass_lattice_2} (i) with $g_1=v$, we first integrate in $\omega$ for fixed $t \in [0,\nu],r,s \in \nu \N$ with $r,s<T$. Then, we integrate in $t$ and sum in $r,s$ to deduce claim (i).

We now prove (ii). We consider the cases $R=0$ and $R=1$ separately.

\paragraph{\textbf{Case 1:} $R=0$}
Let $\omega_1 \in \Omega^{T_1}_{x_1',x_1}$ and $\omega_2 \in \Omega^{T_2}_{x_2',x_2}$ be given. As in \eqref{Large_mass_lattice_3_i_proof_1}, we get that
\begin{multline}
\label{Large_mass_lattice_3_ii_proof_1}
\Bigl|\ee^{-\cal V^{\nu,1}(\omega_1,\omega_2)}-\ee^{-\wh{\cal V}^{\nu,1}(\omega_1,\omega_2)}\Bigr|
\\
\leq
\frac{1}{2\nu}\sum_{r \in \nu \N} \ind{r<T_1} \sum_{s \in \nu \N} \ind{s<T_2} \int_{0}^{\nu}\dd t\,\bigl|v \bigl(\omega_1(t+r)-\omega_2(t+s)\bigr)-v(x_1-x_2)\bigr|\,.  
\end{multline}
We use Fubini's theorem and first integrate in $\omega_1$ and $\omega_2$ in \eqref{Large_mass_lattice_3_ii_proof_1}. 
In doing so, for fixed $t \in [0,\nu]$, and $r,s \in \nu\N$, with $r<T_1,s<T_2$, we use Lemma \ref{Large_mass_lattice_2} (ii) with $t_1=t+r,t_2=t+s ,g_2(\zeta,\eta)=v(\zeta-\eta)$. We then integrate in $t$ and sum in $r,s$ 
and recall that by \eqref{tilde_v_definition} we have $\tilde{v}=v$. Claim (ii) for $R=0$ now follows.

\paragraph{\textbf{Case 2:} $R=1$}
If $x_1=x_2$, then by the right continuity of $\omega_1$ and $\omega_2$, there exists $\epsilon \in (0,\nu)$ such that $\omega_1=\omega_2$ on $[0,\epsilon)$. Hence, recalling \eqref{V_interaction_1} and \eqref{hat_U_interaction}, the expression on the left-hand side of \eqref{Large_mass_lattice_3_ii} is zero by Assumption \ref{interaction_potential_v_2} (ii).

We henceforth assume that $x_1 \neq x_2$.
Let us separately consider the integral over two regions in $(\omega_1,\omega_2) \in \Omega^{T_1}_{x_1',x_1} \times \Omega^{T_2}_{x_2',x_2}$, which are defined depending on whether $\omega_1$ and $\omega_2$ intersect.
\begin{itemize}
\item[(i)] $\omega_1$ and $\omega_2$ do not intersect. In other words, we consider the region
\begin{equation}
\label{R_1_definition}
\cal R_1 \deq \Bigl\{(\omega_1,\omega_2) \in \Omega^{T_1}_{x_1',x_1} \times \Omega^{T_2}_{x_2',x_2}\,, \; \forall t_1 \in [0,T_1] \; \forall t_2 \in [0,T_2]\,,\;\omega_1(t_1) \neq \omega_2(t_2)\Bigr\}\,.
\end{equation}
Using \eqref{tilde_v_definition} and \eqref{R_1_definition}, we see that in the contribution 
from $\cal{R}_1$ to the left-hand side of \eqref{Large_mass_lattice_3_ii}, we can replace $v$ by $\tilde{v}$ in \eqref{V_interaction_1} and \eqref{hat_U_interaction}. Therefore, we can argue analogously as in Case 1 and obtain the same upper bound.

\item[(ii)] $\omega_1$ and $\omega_2$ intersect. We consider the region
\begin{equation*}
\cal R_2 \deq \Bigl\{(\omega_1,\omega_2) \in \Omega^{T_1}_{x_1',x_1} \times \Omega^{T_2}_{x_2',x_2}\,,\; \exists t_1 \in [0,T_1]\; \exists t_2 \in [0,T_2]\,,\; \omega_1(t_1) = \omega_2(t_2)\Bigr\}\,.
\end{equation*}
Since $x_1 \neq x_2$, we have $\cal R_2  \subset \cal R_2^{(1)} \cup \cal R_2^{(2)}$, where
\begin{equation*}
\cal R_2^{(j)} \deq \Bigl\{(\omega_1,\omega_2) \in \Omega^{T_1}_{x_1',x_1} \times \Omega^{T_2}_{x_2',x_2}\,,\; \exists t_j \in [0,T_j]\,,\;  \omega_j(t_j) \neq x_j \Bigr\}\,.
\end{equation*}
By using $|\ee^{-\cal V^{\nu,1}(\omega_1,\omega_2)}-\ee^{-\wh{\cal V}^{\nu,1}(\omega_1,\omega_2)}|\leq 1$ and recalling \eqref{heat_kernel_identity}, we get that the contribution from $\cal{R}_2^{(1)}$ to the left-hand side of \eqref{Large_mass_lattice_3_ii} is
\begin{equation}
\label{cal_L}
\biggl[\int_{\Lambda \setminus \{x_1\}}\dd y\, \int_{0}^{T_1}\dd t_1\, \psi^{t_1}(y-x_1)\,\psi^{T_1-t_1}(x_1'-y)\biggr]\,\psi^{T_2}(x_2'-x_2)=O(T_1^2)\,.
\end{equation}
In the last step above, we used Lemma \ref{estimates_heat_kernel}.
Similarly, the contribution from $\cal{R}_2^{(2)}$ to the left-hand side of \eqref{Large_mass_lattice_3_ii} is $O(T_2^2)$ and the claim follows.
\end{itemize}
\end{proof}

We now have the necessary ingredients to prove Theorem \ref{Large_mass_lattice_4}.
\begin{proof}[Proof of Theorem \ref{Large_mass_lattice_4}]
We first prove (i). 

Let us note that it suffices to consider unnormalized quantities. In particular, recalling  \eqref{Z_infty_lx} and \eqref{Z_nu_omega}, it suffices to show that
\begin{equation}
\label{Large_mass_lattice_4_equivalent}
\lim_{\nu \rightarrow 0}  \Xi^{\nu,\kappa_0/\nu,1}=Z^{\lm}\,.
\end{equation}
Namely, by \eqref{L_nu}, \eqref{heat_kernel_identity}, Lemma \ref{estimates_heat_kernel} (i)--(ii), and the dominated convergence theorem, we have that 
\begin{equation}
\label{Large_mass_lattice_4_corollary}
\lim_{\nu \rightarrow 0} \int \bb L^{\nu,\kappa_0/\nu}(\dd \omega) = \sum_{k=1}^{\infty} \frac{\ee^{-\kappa_0 k}}{k} |\Lambda|\,.
\end{equation}
Therefore, \eqref{Large_mass_lattice_4_equivalent} implies claim (i) by \eqref{Z_infty}, \eqref{Z_nu}, and \eqref{Large_mass_lattice_4_corollary}.

We note that \eqref{Large_mass_lattice_4_equivalent} follows if we show that for all $n \in \N^*,   \b k \in (\N^*)^n$, we have
\begin{equation}
\label{Large_mass_lattice_1_DCT}
\lim_{\nu \rightarrow 0} \int \bb W^{\nu \b k}(\dd \b \omega) \, \exp(-V^{\nu,1}(\b \omega))= \int_{\Lambda^n} \dd \b x  \, \exp\bigl(-V^{\lm}(\b k,\b x)\bigr)\,.
\end{equation}
Namely, since $\int  \bb{W}^{k \nu}(\dd \omega) \leq |\Lambda|$, we have
\begin{equation}
\label{Large_mass_lattice_1_proof2}
\sum_{k=1}^{\infty} \frac{\ee^{-\kappa_0 k}}{k}\int \bb{W}^{k \nu}(\dd \omega)=O_{\kappa_0}(|\Lambda|)\,.
\end{equation}
Using \eqref{Large_mass_lattice_1_proof2}, $V^{\nu,1}(\b \omega) \geq 0$, and the dominated convergence theorem, we deduce that the claim of the theorem indeed follows from \eqref{Large_mass_lattice_1_DCT}.
We now show \eqref{Large_mass_lattice_1_DCT}. In doing so, we consider the two cases $R=0$ and $R=1$ separately.

\paragraph{\textbf{Case 1:} $R=0$}

Recalling \eqref{V_interaction_large_mass}, \eqref{hat_U_full_interaction}, and using Lemma \ref{Large_mass_lattice_3} together with a telescoping argument, it follows that
\begin{equation}
\label{Large_mass_lattice_2_DCT}
\int \bb W^{\nu \b k}(\dd \b \omega) \, \exp(-V^{\nu,1}(\b \omega))=
\int \bb W^{\nu \b k}(\dd \b \omega) \, \exp(-\wh{V}^{\nu,1}(\b \omega))+O_{\Lambda,v,\abs{\b k},n}(\nu)\,.
\end{equation}
We use \eqref{V_classical_interaction}, \eqref{heat_kernel_identity}, and \eqref{hat_U_full_interaction}--\eqref{hat_U_interaction} to write
\begin{equation}
\label{Large_mass_lattice_3_DCT}
\int \bb W^{\nu \b k}(\dd \b \omega) \, \exp(-\wh{V}^{\nu,1}(\b \omega))=\prod_{i = 1}^n \psi^{k_i \nu}(0)\,\int_{\Lambda^n} \dd \b x \, \exp\bigl(-V^\lm(\b k, \b x)\bigr)\,.
\end{equation}
Note that by Lemma \ref{estimates_heat_kernel} (ii), we have that for all $k \in \N^*$
\begin{equation}
\label{Large_mass_lattice_4_DCT}
\psi^{k \nu}(0)=1+O(k\nu)\,.
\end{equation}
Combining \eqref{Large_mass_lattice_2_DCT}--\eqref{Large_mass_lattice_4_DCT}, we deduce 
\eqref{Large_mass_lattice_1_DCT}.

\paragraph{\textbf{Case 2:} $R=1$}
We recall \eqref{V_classical_interaction} and deduce that \eqref{Large_mass_lattice_1_DCT} is the consequence of the following two claims.
\begin{itemize}
\item[(1)] For all $n \in \N^*$ and $\b k \in (\N^*)^{n} \setminus \{{\b 1}_n\}$, we have
\begin{equation}
\label{HC_claim_i}
\lim_{\nu \rightarrow 0} \int \bb W^{\nu \b k}(\dd \b \omega) \, \exp\bigl(-V^{\nu,1}(\b \omega)\bigr)=0\,.
\end{equation}
\item[(2)] For all $n \in \N^*$, we have
\begin{equation}
\label{HC_claim_ii}
\lim_{\nu \rightarrow 0} \int \bb W^{\nu \b 1}(\dd \b \omega) \, \exp\bigl(-V^{\nu,1}(\b \omega)\bigr)=\int_{\Lambda^n} \dd \b x  \, \exp\bigl(-V^\lm(\b 1_n,\b x)\bigr)\,.
\end{equation}
Here we recall \eqref{b_1}.
\end{itemize}

We first prove claim (1). We recall 
\eqref{V_interaction_large_mass}--\eqref{V_interaction_1_large_mass_B}, and use the nonnegativity of $v$, to deduce that  \eqref{HC_claim_i} follows if we show that for all $k \geq 2$ and $x \in \Lambda$, we have
\begin{equation}
\label{HC_claim_i_B}
\lim_{\nu \rightarrow 0} \int \bb W^{k \nu}_{x,x}(\dd \omega) \, \exp\bigl(-\wt{\cal V}^{\nu,1} (\omega) \bigr)=0\,.
\end{equation}
Namely, we know that there exists a component of $\b k$ which is at least $2$. Let's assume without loss of generality that $k_1 \geq 2$.
When integrating in $\omega_1$, we use \eqref{HC_claim_i_B}. When integrating in $\omega_j, 2 \leq j \leq n$, we use the nonnegativity of $v$, \eqref{heat_kernel_identity} and Lemma \ref{estimates_heat_kernel} (i). Therefore, claim (i) indeed follows from \eqref{HC_claim_i_B}.

Let us now prove \eqref{HC_claim_i_B}. We define
$\cal L^{T}_{x} \deq \Bigl\{\omega \in \Omega^T_{x,x}\,, \exists t \in [0,T]\,, \omega(t) \neq x\Bigr\}$ and $\cal{S}^{T}_{x} \deq \Omega^T_{x,x} \setminus \cal L^{T}_{x}$.
With this notation, we have
\begin{equation}
\label{HC_claim_i_C}
\int \bb W^{k \nu}_{x,x}(\dd \omega) \, \exp\bigl(-\wt{\cal V}^{\nu,1} (\omega) \bigr) \leq 
\int \bb W^{k \nu}_{x,x}(\dd \omega)\,\ind{{\cal L}^{k \nu}_{x}}(\omega)+
\int \bb W^{k \nu}_{x,x}(\dd \omega)\, \exp\bigl(-\wt{\cal V}^{\nu,1} (\omega) \bigr) \,\ind{{\cal S}^{k \nu}_{x}}(\omega)\,.
\end{equation}
By arguing analogously as for \eqref{cal_L}, it follows that the first term on the right-hand side of \eqref{HC_claim_i_C} is $O(k^2 \nu^2)$. Furthermore, by \eqref{V_interaction_1_large_mass_B} and the assumptions on $v$, the second term is equal to zero (here, we are crucially using the fact that $k \geq 2$). We therefore obtain \eqref{HC_claim_i_B} and claim (i) follows.

Let us now prove claim (2). We apply Lemma \ref{Large_mass_lattice_3} (ii) with $T=\nu$ and use a telescoping argument to deduce that 
\begin{equation}
\label{HC_claim_ii_B}
\lim_{\nu \rightarrow 0}\biggl[\int \bb W^{\nu \b 1}(\dd \b \omega) \, \exp\bigl(-V^{\nu,1}(\b \omega)\bigr)-\int \bb W^{\nu \b 1}(\dd \b \omega) \, \exp\bigl(-V^\lm(\b 1_n, \b x)\bigr)\biggr]=0\,.
\end{equation}
Here, we recall \eqref{V_interaction_large_mass}--\eqref{V_interaction_1_large_mass_B}, \eqref{V_classical_interaction}, \eqref{hat_U_interaction},
and keep in mind that the left-hand side of \eqref{HC_claim_ii} does not contain any self-interactions. 
Claim (ii) now follows from \eqref{HC_claim_ii_B} by iteratively applying \eqref{heat_kernel_identity} and Lemma \ref{estimates_heat_kernel} (ii) in the second term and thus obtaining the expression on the right-hand side of \eqref{HC_claim_ii} in the limit. 
We deduce claim (i).

Let us now show claim (ii). The proof is similar to that of (i), so we will just outline the main differences.
Let $p \in \N^*$ and $\b x, \b y \in \Lambda^p$ be given.
By \eqref{gamma_infty}, \eqref{Z_nu_omega}--\eqref{Ginibre_loop_representation_gamma_p_2}, and \eqref{Large_mass_lattice_4_equivalent}, the claim follows if we show that 
\begin{equation}
\label{large_mass_correlations_proof1}
\lim_{\nu \rightarrow 0}  \sum_{\b T \in (\nu \N^*)^p} \sum_{\pi \in S_p} \ee^{-\kappa_0\abs{\b T}/\nu} \int \bb W^{\b T}_{\pi \b y,\b x}(\dd \b \omega) \, \Xi^{\nu,\kappa_0/\nu,1}(\b \omega) = \sum_{\b k \in (\N^*)^p} \sum_{\pi \in S_p} \ee^{-\kappa_0 \abs{\b k}} \,\delta (\pi \b y - \b x)\,Z^{\lm}(\b k, \b x)\,.
\end{equation}
Analogously as for \eqref{Large_mass_lattice_1_DCT}, it suffices to show that given $\b k \in (\N^*)^p, \pi \in S_p, n \in \N^*, \b{\tilde k} \in (\N^*)^n$ we have 
\begin{equation}
\label{correlation_functions_1_DCT}
\lim_{\nu \rightarrow 0} \int \bb W^{\nu \b k}_{\pi \b y,\b x}(\dd \b \omega)  \int \bb W^{\nu \b{\tilde k}}(\dd \tilde{\b \omega}) \, \exp(-V^{\nu,1}(\b \omega \tilde{\b \omega}))= \delta (\pi \b y - \b x)\,\int_{\Lambda^n} \dd \tilde{\b x}  \, \exp\bigl(-V^\lm(\b k \b{\tilde k},\b x \tilde{\b x})\bigr)\,.
\end{equation}
When $R=0$, the proof of \eqref{correlation_functions_1_DCT} proceeds as  that of \eqref{Large_mass_lattice_1_DCT}. The only difference is that we now apply Lemma \ref{Large_mass_lattice_3} for both open and closed paths. The $p$ open paths give rise to delta functions by Lemma \ref{estimates_heat_kernel} (ii). We omit the details. 

We henceforth show \eqref{correlation_functions_1_DCT} when $R=1$. Arguing as for \eqref{HC_claim_i}--\eqref{HC_claim_ii}, it suffices to show the following two claims.
\begin{itemize}
\item[(a)] For all $n \in \N^*$, $(\b k, \b{\tilde k}) \in (\N^*)^{p} \times (\N^*)^{n} \setminus \{{\b 1}_{p+n}\}$, and $\pi \in S_p$, we have
\begin{equation}
\label{HC_claim_i_correlations}
\lim_{\nu \rightarrow 0} \int \bb W^{\nu \b k}_{\pi \b y,\b x}(\dd \b \omega) \int \bb W^{\nu \b{\tilde k}}(\dd \tilde{\b \omega}) \, \exp\bigl(-V^{\nu,1}(\b \omega \tilde{\b \omega})\bigr)=0\,.
\end{equation}
\item[(b)] For all $n \in \N^*$, we have
\begin{equation}
\label{HC_claim_ii_correlations}
\lim_{\nu \rightarrow 0} \int \bb W^{\nu {\b 1}_{p}}_{\pi \b y,\b x}(\dd \b \omega)  \int \bb W^{\nu {\b 1}_{n}}(\dd \tilde{\b \omega}) \, \exp\bigl(-V^{\nu,1}(\b \omega \tilde{\b \omega})\bigr)
=\delta (\pi \b y-\b x)\int_{\Lambda^n} \dd \tilde{\b x}  \, \exp\bigl(-V^\lm(\b x \tilde{\b x})\bigr)\,.
\end{equation}
\end{itemize}

We first prove claim (a). Note that if $\b{\tilde k} \neq \b 1_n$, then \eqref{HC_claim_i_correlations} follows from \eqref{HC_claim_i_B} by using $V^{\nu,1}(\b \omega \tilde{\b \omega}) \geq V^{\nu,1}(\b \omega)$, and by recalling \eqref{heat_kernel_identity} and Lemma \ref{estimates_heat_kernel} (i). If $\b k \neq \b 1_p$, then we consider two subcases. 
If $\pi \b y \neq \b x$, then \eqref{HC_claim_i_correlations} follows from \eqref{heat_kernel_identity} and Lemma \ref{estimates_heat_kernel} (i)--(ii), since $V^{\nu,1}(\b \omega \tilde{\b \omega}) \geq 0$.  
If $\pi \b y=\b x$, then \eqref{HC_claim_i_correlations} again follows from \eqref{HC_claim_i_B}, since $V^{\nu,1}(\b \omega \tilde{\b \omega})$ contains self-interactions \eqref{V_interaction_1_large_mass_B}.

Let us now prove claim (b). If $\pi \b y \neq \b x$, then the limit is zero by arguing as in the proof of (a). If $\pi \b y=\b x$, then \eqref{HC_claim_ii_correlations} follows by using Lemma \ref{Large_mass_lattice_3} (ii) analogously as in the proof of \eqref{HC_claim_ii}. Note that now we are integrating only over the endpoints of the closed paths.  We hence obtain (b) and claim (ii) follows. 
\end{proof}

\section{The infinite-volume limit}
\label{The infinite-volume limit}

In this section, we study the infinite-volume limit. In Section \ref{Infinite_volume_limit_mean_field}, we study the mean-field regime and prove Theorem \ref{Infinite_volume_theorem_1}. In Section \ref{Infinite_volume_limit_large_mass}, we study the large-mass regime and prove Theorem \ref{Infinite_volume_theorem_2}.

\subsection{Infinite-volume limit of the specific relative Gibbs potential and reduced density matrices I: the mean-field regime}
\label{Infinite_volume_limit_mean_field}
In this subsection, we work in the mean-field regime. 
In the sequel, we vary the size $L \in \N^*$ of the box $\Lambda_L$. Throughout, we assume that the interaction potential on $\Lambda_L$ is given by \eqref{v^{L}} for $v$ as in Assumption \ref{interaction_potential_v^{L}}.
As was mentioned in the introduction, we keep track of the $L$-dependence of all the quantities by adding a superscript $L$.
More precisely, we write $\Omega^{L,T}$ and $\Omega^{L,T}_{y,x}$ for the appropriate space of c\`adl\`ag paths on $\Lambda_L$.
Analogously, we write $\bb W_{y,x}^{L,T}(\dd \omega), \bb W^{L,T}(\dd \omega)$ for \eqref{W_path_measures} respectively.
Finally, we write $\psi^{t} \equiv \psi^{L,t}$ for the heat kernel \eqref{heat_kernel_Lambda_L} on $\Lambda_L$.

Let us rewrite the reduced $p$-particle density matrix \eqref{gamma_p_definition} as a power series representation amenable to a cluster expansion. 
Our starting point is the Ginibre representation given in Proposition \ref{Ginibre_loop_representation} above. 
Before stating the precise formula, we introduce some notation.
We consider $\cal V^{\nu,\lambda,L}(\omega, \tilde \omega)$ given by \eqref{V_interaction_1} with interaction as in \eqref{v^{L}} and define positive measures on $\bigcup_{T > 0} \Omega^{L,T}$ through
\begin{align}
\label{mu_measure_1}
&{\mu}^{L}(\dd \omega) \deq \nu \sum_{T\in \nu \N^*} \frac{\ee^{-\kappa T}}{T} \,  \bb W^{L,T}(\dd \omega)
\,\ee^{-\cal{V}^{\nu,\lambda,L}(\omega,\omega)/2}\,,
\\
\label{mu_measure_2}
&\hat \mu^{L}_{y,x}(\dd \omega) \deq \sum_{T \in \nu \N^*} \ee^{-\kappa T} \, \bb W^{L,T}_{y,x}(\dd \omega)\,\ee^{-\cal{V}^{\nu,\lambda,L}(\omega,\omega)/2}\,.
\end{align}
Given $p \in \N$ and $\omega_1,\ldots,\omega_p \in \Omega$ we define
\begin{equation}
\label{Ursell_function_representation}
X^L(\omega_1, \dots, \omega_p) \deq \sum_{n \geq p \vee 1} \frac{n!}{(n - p)!} \int {\mu}^{L}(\dd \omega_{p+1}) \cdots {\mu}^{L}(\dd \omega_n)\, \varphi^L(\omega_1, \dots, \omega_n)\,,\quad  X^{L} \deq X^{L}(\emptyset)\,,
\end{equation}
with the \emph{Ursell function} given by
\begin{equation}
\label{Ursell_function}
\varphi^L(\omega_1, \dots, \omega_n)
\deq
\frac{1}{n!} \sum_{\cal G \in {\fra G}^c_n} \prod_{\{i,j\} \in \cal G} \zeta^L(\omega_i, \omega_j)\,, \qquad \zeta^L(\omega, \tilde \omega) \deq \exp\bigl(-\cal{V}^{\nu,\lambda,L}(\omega,\tilde \omega)\bigr) - 1\,.
\end{equation}
Here ${\fra G}^c_n$ denotes the set of all connected graphs on $[n]=\{1,2,\ldots,n\}$.
In this section, we work in the mean-field limit and therefore set $\lambda=\nu^2$ in the definitions above \footnote{In order to simplify notation in the sequel, we write $\mu^L, \hat \mu^L_{y,x}, X^L$, etc. instead of $\mu^{\nu,\kappa,\lambda,L}, \hat \mu^{\nu,\kappa,\lambda,L}_{y,x}, X^{\nu,\lambda,L}$, etc. We bear in mind that all of the above quantities depend on $\nu,\kappa,\lambda$ as well. In this section we fix $\kappa$ and take $\lambda=\nu^2$.}.

Throughout the sequel, we use the identity
\begin{equation}
\label{v^{L}_ell^1}
\|v^{L}\|_{\ell^1(\Lambda_L)} = \|v\|_{\ell^1(\Z^d)}\,,
\end{equation}
which follows directly from \eqref{v^{L}}. We henceforth write $\|v\|_{\ell^1} \equiv \|v\|_{\ell^1(\Z^d)}$.
Throughout this section, we assume
\begin{equation}
\label{kappa>1}
\kappa>1\,.
\end{equation}

\begin{proposition}[The cluster expansion]
\label{gamma_p_cluster_expansion}
With notation as in \eqref{Ursell_function_representation} and assuming that $\|v\|_{\ell^1}$ is sufficiently small depending on $\kappa$ (or if $\kappa$ is sufficiently large depending on $\|v\|_{\ell^1}$), we have the following identities.
\begin{itemize}
\item[(i)] $\log \cal Z^{\nu,\kappa,\nu^2,L}=X^{L}-X^{L,0}$, where $X^{L,0}$ is given by \eqref{Ursell_function_representation} with interaction potential set to zero (i.e.\ $\varphi^L=1$).
\item[(ii)]
Let $p \in \N^*$ and $\b x, \b y \in \Lambda^p$ be given. We have
\begin{equation}
\label{gamma_p_cluster_expansion_claim}
(\Gamma^{\nu,\kappa,\nu^2,L}_{p})_{\b x,\b y} = \sum_{\pi \in S_p} \int \hat \mu^{L}_{y_{\pi(1)},x_1}(\dd \omega_1) \cdots \hat \mu^{L}_{y_{\pi(p)},x_p}(\dd \omega_p) \sum_{\Pi \in \fra P_p} \prod_{\xi \in \Pi} X^{L}((\omega_i)_{i \in \xi})\,,
\end{equation}
where $\fra P_p$ is the set of partitions of $\{1, \dots, p\}$.
\end{itemize}
\end{proposition}

\begin{proof}
The identity in (i) follows from \cite[Theorem 1]{Ueltschi_2004}.
The application of the latter result is precisely justified in Remark \ref{cluster_expansion_remark} below.
 In order to prove (ii), we define 
\begin{equation}
\label{Z_omega_tilde}
\tilde X^{L}(\omega_1, \dots, \omega_p) \deq \sum_{n = p}^\infty \frac{1}{(n - p)!} \int \mu^{L}(\dd \omega_{p+1}) \cdots \mu^{L}(\dd \omega_n) \prod_{1 \leq i<j \leq n} \exp\bigl(-\cal{V}^{\nu,\nu^2,L}(\omega_i, \omega_j)\bigr)
\end{equation}
and $\tilde X^{L} \deq \tilde X^{L}(\emptyset)$.
By Proposition \ref{Ginibre_loop_representation}, \eqref{mu_measure_1}--\eqref{mu_measure_2}, and \eqref{Z_omega_tilde}, we have
\begin{equation}
\label{gamma_p_integral}
(\Gamma^{\nu,\kappa,\nu^2,L}_{p})_{\b x,\b y} = \sum_{\pi \in S_p} \int \hat \mu^{L}_{y_{\pi(1)},x_1}(\dd \omega_1) \cdots \hat \mu^{L}_{y_{\pi(p)},x_p}(\dd \omega_p) \frac{\tilde X^{L}(\omega_1, \dots, \omega_p)}{\tilde X^{L}}\,.
\end{equation}
By \cite[Theorem 2]{Ueltschi_2004}, we have
\begin{equation}
\label{gamma_p_integral_2}
\frac{\tilde X^{L}(\omega_1, \dots, \omega_p)}{\tilde X^{L}} = \sum_{\Pi \in \fra P_p} \prod_{\xi \in \Pi} X^{L}((\omega_i)_{i \in \xi})
\end{equation}
and the claim follows from \eqref{gamma_p_integral}--\eqref{gamma_p_integral_2}.
\end{proof}

The cluster expansion from Proposition \ref{gamma_p_cluster_expansion} allows us to prove the bounds on the specific relative Gibbs potential and the reduced density matrices, which are uniform in $\nu$, for $\nu$ small enough, and in $L$. For the remainder of the section, we assume that
\begin{equation} 
\label{nu_small}
\nu \in (0, 1/\kappa]\,.
\end{equation}
Given $p \in \N^*$, we consider the norm on operators acting on $\ell^2(\Lambda_L)^{\otimes p}$ given by
$\|A\|_{\ell^{\infty}_{\b x} \ell^1_{\b y}} \equiv \sup_{\b x \in \Lambda_L^p}\, \int_{\Lambda_L^p} \dd \b y\, |A_{\b x,\b y}|$.
By Schur's test, we note that for self-adjoint operators $A$ we have
$\|A\| \leq \|A\|_{\ell^{\infty}_{\b x} \ell^1_{\b y}}$, where $\|\cdot\|$ denotes the operator norm on $\ell^2(\Lambda_L)^{\otimes p}$.

\begin{proposition} [Bounds on the specific relative Gibbs potential and reduced density matrices in a finite volume]
\label{operator_norm_bound}
For $\|v\|_{\ell^1}$ sufficiently small depending on $\kappa$ (or if $\kappa$ is sufficiently large depending on $\|v\|_{\ell^1}$), we have the following bounds for all $L \in \N^*$.
\begin{itemize}
\item[(i)] The specific relative Gibbs potential \eqref{quantum_free_energy} satisfies
\begin{equation}
\label{free_energy_bound_1}
g^{\nu,\kappa,\nu^2,L} =O_{\kappa,\|v\|_{\ell^1}}(1)\,.
\end{equation}
\item[(ii)] For $p \in \N^*$, the $p$-particle reduced density matrix $\Gamma^{\nu,\kappa,\nu^2,L}_p$  satisfies 
\begin{equation} 
\label{operator_norm_bound_1}
\nu^p \,\bigl\|\Gamma^{\nu,\kappa,\nu^2,L}_p\bigr\|_{\ell^{\infty}_{\b x} \ell^1_{\b y}}=O_{\kappa,p,\|v\|_{\ell^1}}(1)\,.
\end{equation}
\end{itemize}
\end{proposition}
From Proposition \ref{operator_norm_bound}, we can deduce the existence of 
reduced density matrices in the infinite volume.

\begin{proposition}[Specific relative Gibbs potential and reduced density matrices in the infinite volume]
\label{operator_norm_bound_corollary}
With assumptions as in Proposition \ref{operator_norm_bound}, let $p \in \N^*$ and given. 
We can take $L=\infty$ in \eqref{gamma_p_cluster_expansion_claim} and obtain an operator $\Gamma^{\nu,\kappa,\nu^2,\infty}_p$ on $\ell^2(\Z^{d})^{\otimes p}$ which satisfies
\begin{equation} 
\label{operator_norm_bound_2}
\nu^p \,\bigl\|\Gamma^{\nu,\kappa,\nu^2,\infty}_p\bigr\|_{\ell^{\infty}_{\b x} \ell^1_{\b y}}=O_{\kappa,p,\|v\|_{\ell^1}}(1)\,. 
\end{equation}
In \eqref{operator_norm_bound_2}, we are taking the $\|\cdot\|_{\ell^{\infty}_{\b x} \ell^1_{\b y}}$ norm on operators acting on $\ell^2(\Z^{d})^{\otimes p}$.
\end{proposition}
We present the details of the proof of Proposition \ref{operator_norm_bound_corollary} in Appendix \ref{Corollary_5.3_proof}.

We can use Proposition \ref{operator_norm_bound_corollary} to show that specific relative Gibbs potential and reduced density matrices converge to corresponding infinite-volume limits uniformly in $\nu$.

\begin{proposition}[Convergence to the infinite-volume limit]
\label{MF_cluster_expansion}
With assumptions as in Proposition \ref{operator_norm_bound}, the following claims hold.
\begin{itemize}
\item[(i)] The quantity 
\begin{equation}
\label{operator_norm_bound_corollary_i}
\lim_{L \rightarrow \infty} g^{\nu,\kappa,\nu^2,L}  = : g^{\nu,\kappa,\nu^2,\infty}
\end{equation}
exists. Moreover, the convergence in \eqref{operator_norm_bound_corollary_i} holds uniformly in $\nu \in (0,1/\kappa]$.
\item[(ii)] Let $p \in \N^*$ and $L_0 \in \N^*$ be given. 
Let $\Gamma^{\nu,\kappa,\nu^2,\infty}_p$ be as in Proposition \ref{operator_norm_bound_corollary} (ii). Then, we have 
\begin{equation}
\label{gamma_infty_convergence}
\Gamma^{\nu,\kappa,\nu^2,\infty}_p=\lim_{L \rightarrow \infty} \Gamma^{\nu,\kappa,\nu^2,L}_p\,.
\end{equation}
The convergence in \eqref{gamma_infty_convergence}  holds in $\|\cdot\|_{L_0,p}$  given by \eqref{L_0_p_norm} and is uniform in $\nu \in (0,1/\kappa]$ and $L_0 \in \N^*$.
\end{itemize}
\end{proposition}

\begin{remark}
\label{infinite_volume_remark}
We note that the result in Proposition \ref{MF_cluster_expansion} (i) is not new. Furthermore, it  holds under more general assumptions that do not require the interaction to be small, see \cite{Ruelle}. Below, we give a short proof under our assumptions using cluster expansions for completeness and for expository purposes. This proof allows us to deduce more specific properties of the thermodynamic limit, such as analyticity.
\end{remark}

Before proceeding to the proofs of Proposition \ref{operator_norm_bound}, Proposition \ref{operator_norm_bound_corollary}, and Proposition \ref{MF_cluster_expansion}, we 
note several auxiliary results.
The first one is a useful estimate on the Ursell function \eqref{Ursell_function}.

\begin{lemma}[Tree bound]
\label{Ursell_function_estimate}
For $n \in \N$ and $\omega_1, \ldots, \omega_n \in \Omega$ we have 
\begin{equation*}
|\varphi^L(\omega_1, \dots, \omega_n)| \leq \frac{1}{n!} \sum_{\cal T \in {\fra T}_n} \prod_{\{i,j\} \in \cal T}|\zeta^L(\omega_i,\omega_j)|\,.
\end{equation*}
Here ${\fra T}_n$ denotes the set of all trees on $[n]=\{1,2,\ldots,n\}$.
\end{lemma}
We prove Lemma \ref{Ursell_function_estimate} in Appendix \ref{Appendix C.1}. The proof is based on \emph{Kruskal's algorithm} \cite{Kruskal_1956}, which we also recall in Appendix \ref{Appendix C.1} for completeness.
The second auxiliary result is the basis of an algorithm that allows us to iteratively integrate all of the paths $\omega_j$ in the representation of Proposition \ref{gamma_p_cluster_expansion} (recalling \eqref{mu_measure_1}--\eqref{Ursell_function}). We now state the integration lemma.

\begin{lemma}[Integrating out a vertex] 
\label{integration_lemma}
Let $\omega \in \Omega^{L,T(\omega)}$ with $T(\omega) \in \nu \N^*$,  $q \in \N$, and $x \in \Lambda_L$ be given.
Then, the following estimates hold.
\begin{itemize}
\item[(i)] $\int \mu^{L} (\dd \tilde \omega)\,T(\tilde \omega)^{q} \,|\zeta^L(\omega,\tilde \omega)| \lesssim \frac{T(\omega)}{\kappa^{q+1}}\,q!\,\|v\|_{\ell^1}\,.$
\item[(ii)] $\nu \int_{\Lambda_L} \dd y\,\int \hat{\mu}^{L}_{y,x} (\dd\tilde \omega) \, T(\tilde \omega)^{q}\,|\zeta^L(\omega,\tilde \omega)| \lesssim \frac{T(\omega)}{\kappa^{q+2}}\,(q+1)!\,\|v\|_{\ell^1}.$ 
\item[(iii)] $\int \mu^{L} (\dd \tilde \omega) \,T(\tilde \omega)^{q} \lesssim  \frac{(q-1)!}{\kappa^{q}}\,|\Lambda_L|$ if $q\in \N^*$.
\item[(iv)] $\nu \int_{\Lambda_L} \dd y\,\int \hat{\mu}^{L}_{y,x}(\dd \tilde \omega) \, T(\tilde \omega)^{q} \lesssim \frac{q!}{\kappa^{q+1}}.$
\end{itemize}
\end{lemma}

\begin{remark}
\label{cluster_expansion_remark}
The results in Lemma \ref{integration_lemma} (i) and (iii) combined with \eqref{kappa>1} allow us to justify the application of \cite[Theorem 1]{Ueltschi_2004} in the proof of Proposition \ref{gamma_p_cluster_expansion} (i) above.
Namely, Lemma \ref{integration_lemma} (i) implies that 
\begin{equation*}
\int \mu^{L} (\dd \tilde \omega)\,\ee^{T(\tilde \omega)}\, |\zeta^L(\omega,\tilde \omega)| =\sum_{q=0}^{\infty} \int \mu^L(\dd \tilde \omega)\, \frac{T(\tilde \omega)^q}{q!}\, |\zeta^L(\omega,\tilde \omega)| \lesssim \sum_{q=0}^{\infty} \frac{T(\omega)}{\kappa^{q+1}} \|v\|_{\ell^1} \lesssim_{\kappa} T(\omega) \,\|v\|_{\ell^1} \leq T(\omega)
\end{equation*}
if $\|v\|_{\ell^1}$ is sufficiently small, depending on $\kappa$, or vice versa.
Similarly, from Lemma \ref{integration_lemma} (iii), we have
\begin{equation*}
\int \mu^L(\dd \tilde \omega)\, \ee^{T(\tilde \omega)} =\int \mu^L(\dd \tilde \omega)\, \sum_{q=0}^{\infty}\frac{T(\tilde \omega)^q}{q!} \lesssim \sum_{q=0}^{\infty}  \frac{1}{\kappa^q} |\Lambda_L| \lesssim_{\kappa} |\Lambda_L|\,.
\end{equation*}
Finally, we note that from \eqref{Ursell_function} and the nonnegativity of $v$, we have 
\begin{equation*}
|1+\zeta^L(\omega,\tilde \omega)| \leq 1\,.
\end{equation*}
In particular, in the terminology and notation of \cite{Ueltschi_2004}, we can take $a(A) \equiv T(\omega)$.
\end{remark}

In the proof of Lemma \ref{integration_lemma}, we use the following lemma concerning Riemann sums, whose proof is given in Appendix \ref{Appendix C.2}.

\begin{lemma}
\label{Riemann_sum_lemma}
Given $q \in \N$ and assuming \eqref{nu_small}, we have
\begin{equation}
\label{integration_lemma_i_3a_B}
\nu \sum_{T \in \nu \N^*} \ee^{-\kappa T}\,T^{q} \lesssim \frac{q!}{\kappa^{q+1}}\,.
\end{equation}
\end{lemma}

\begin{proof}[Proof of Lemma \ref{integration_lemma}]
Throughout the proof, we use the observation that 
\begin{equation}
\label{Ursell_function_bound}
|\zeta^L(\omega,\tilde \omega) | \leq \cal{V}^{\nu,\nu^2,L}(\omega,\tilde \omega)\,,
\end{equation}
which follows from \eqref{Ursell_function} and \eqref{V_nonnegative}.
Let us first prove (i). By using \eqref{V_interaction_1}, \eqref{V_nonnegative}, \eqref{mu_measure_1}, and \eqref{Ursell_function_bound}, we need to estimate
\begin{align} 
\notag
&\int \mu^{L} (\dd \tilde \omega)\, T(\tilde \omega)^{q} \, \cal{V}^{\nu,\nu^2,L}(\omega,\tilde \omega) 
\\
\label{integration_lemma_i_1}
&
\leq\nu^2 \sum_{\tilde T \in \nu\N^*} \ee^{-\kappa \tilde T}\,\tilde{T}^{q-1}\,   \sum_{r \in \nu \N} \ind{r<T(\omega)}\sum_{s\in \nu \N}\ind{s<\tilde T}\,
\int_0^{\nu} \dd t\, \int_{\Lambda_L}  \dd x\,\int \bb W^{L,\tilde T}_{x,x}(\dd \tilde \omega)\,v^{L}\pb{\omega(t + r) - \tilde \omega(t+s)}\,.
\end{align}
We note that, for fixed $\tilde T \in \nu \N^*, y \in \Lambda_L, t \in [0,\nu], s\in \nu \N$ with $s< \tilde T$ we have
\begin{equation} \label{integration_lemma_i_2_a}
\int_{\Lambda_L}  \dd x\, \int \bb W^{L,\tilde T}_{x,x}(\dd \tilde \omega)\, v^{L}\pb{y- \tilde \omega(t+s)}
=\int_{\Lambda_L} \dd x\, \int_{\Lambda_L}  \dd z\, \int \bb W^{L,t+s}_{z,x} (\dd \tilde \omega_1)  \int \
\bb W^{L,\tilde T-(t+s)}_{x,z} (\dd \tilde \omega_2)\,v^{L}(y-z)\,,
\end{equation}
which is 
\begin{align}
\notag
&=\int_{\Lambda_L}  \dd z\, \biggl[\int_{\Lambda_L}  \dd x\, \int \bb W^{L,\tilde T-(t+s)}_{x,z}(\dd \tilde \omega_2)  \int \bb W^{L,t+s}_{z,x}(\dd \tilde \omega_1) \biggr] \, v^{L}(y-z) 
\\
\label{integration_lemma_i_2}
&= \int_{\Lambda_L}  \dd z\,v^{L}(y-z)\, \int \bb W^{L,\tilde T}_{z,z} (\dd \tilde \omega)= \int_{\Lambda_L}  \dd z\, v^{L}(y-z) \,\psi^{\tilde T}(0) \leq \|v^{L}\|_{\ell^1(\Lambda_L)}=\|v\|_{\ell^1}\,.
\end{align}
In order to deduce \eqref{integration_lemma_i_2}, we used Fubini's theorem, the time-reversibility of the random walk, \eqref{heat_kernel_identity}, Lemma \ref{estimates_heat_kernel} (i),  the assumption that $v$ is even, and \eqref{v^{L}_ell^1}. 

Using \eqref{integration_lemma_i_2} we get that the expression in \eqref{integration_lemma_i_1} is
\begin{equation}
\label{integration_lemma_i_3a}
\leq \nu^3 \sum_{\tilde T \in \nu \N^*} \ee^{-\kappa \tilde T}\,\tilde T^{q-1}\sum_{r \in \nu \N} \ind{r<T(\omega)}\sum_{s \in \nu \N} \ind{s< \tilde T}\, \|v\|_{\ell^1}=\nu \sum_{\tilde T \in \nu \N^*} \ee^{-\kappa \tilde T}\,\tilde T^{q}\,T(\omega)\,\|v\|_{\ell^1}\,.
\end{equation}
We hence deduce (i) from \eqref{integration_lemma_i_3a} and Lemma \ref{Riemann_sum_lemma}.

We now prove (ii).  Similarly as in \eqref{integration_lemma_i_1}, we have
\begin{align}
\notag
&\nu \int_{\Lambda_L} \dd y\,\int \hat{\mu}^{L}_{y,x} (\dd\tilde \omega) \, T(\tilde \omega)^{q}\,\cal{V}^{\nu,\nu^2,L}(\omega,\tilde \omega)
\\
\label{integration_lemma_ii_1}
&\leq \nu^2 \int_{\Lambda_L}  \dd y\, \sum_{\tilde T \in \nu \N^*} \ee^{-\kappa \tilde T}\,\tilde T^{q} \sum_{r \in \nu \N} \ind{r<T(\omega)} \sum_{s\in \nu \N} \ind{s<\tilde T} \int \bb W^{L,\tilde T}_{y,x}(\dd \tilde \omega) \int_0^{\nu} \dd t\,v^{L}\pb{\omega(t + r) - \tilde \omega(t+s)}\,.
\end{align}
Let us consider fixed $\tilde T \in \nu \N^*$, $r, s\in \nu \N$ with $r<T(\omega), s< \tilde T$, and $t \in [0,\nu]$. We then write
\begin{align}
\label{integration_lemma_ii_1_b}
&\int_{\Lambda_L}  \dd y\,\int \bb W^{L,\tilde T}_{y,x}(\dd \tilde \omega) \,v^{L}\pb{\omega(t + r) - \tilde \omega(t+s)}
\\
\notag
&=\int_{\Lambda_L}  \dd y\,\int_{\Lambda_L}  \dd z\, \int \bb W^{L,t+ s}_{z,x}(\dd \tilde \omega_1) \int \bb W^{L,\tilde T-(t+s)}_{y,z}(\dd \tilde\omega_2)\,v^{L}\pb{\omega(t + r)-z}\,,
\end{align}
which we rewrite by Fubini's theorem as 
\begin{equation}
\label{integration_lemma_ii_2}
\int_{\Lambda_L}  \dd z\, \int \bb W^{L,t+s}_{z,x}(\dd \tilde \omega_1)\,\biggl(\int_{\Lambda_L}  \dd y \,\int \bb W^{L,\tilde T-(t+s)}_{y,z}(\dd \tilde \omega_2)\biggr)\,v^{L}\pb{\omega(t + r)-z}
\leq \|v\|_{\ell^1}\,.
\end{equation}
Above, we used \eqref{heat_kernel_identity}, Lemma \ref{estimates_heat_kernel} (i), (iii), and \eqref{v^{L}_ell^1}. 
Using \eqref{integration_lemma_ii_2}  and Lemma \ref{Riemann_sum_lemma}, we get that the expression in \eqref{integration_lemma_ii_1} is
\begin{equation} \label{integration_lemma_ii_1_A}
\leq \nu \sum_{\tilde T \in \nu \N^*} \ee^{-\kappa \tilde T}\,\tilde T^{q+1}\,T(\omega)\,\|v\|_{\ell^1} \lesssim \frac{T(\omega)}{\kappa^{q+2}}\,(q+1)!\,\|v\|_{\ell^1}\,.
\end{equation}
This proves (ii).

We now prove (iii). Let $q \in \N^*$ be given. By using \eqref{mu_measure_1}, \eqref{V_nonnegative}, and recalling \eqref{heat_kernel_identity}, we have
\begin{equation}
\label{integration_lemma_iii_1_a}
\int \mu^{L} (\dd \tilde \omega) \, T(\tilde \omega)^{q} \leq  \nu\, \sum_{\tilde T \in \nu \N^*} \ee^{-\kappa \tilde T}\,\tilde T^{q-1}  \int_{\Lambda_L}  \dd x \,\int \bb W^{L,\tilde T}_{x,x}(\dd \tilde \omega) =\Biggl(
\nu\, \sum_{\tilde T \in \nu \N^*} \ee^{-\kappa \tilde T}\,\tilde T^{q-1} \,\psi^{L,\tilde T}(0)\Biggr)\,|\Lambda_L| \,,
\end{equation}
which by Lemma \ref{estimates_heat_kernel} (i) and Lemma \ref{Riemann_sum_lemma} is 
\begin{equation} 
\label{integration_lemma_iii_1}
\lesssim\Biggl(
\nu\, \sum_{\tilde T \in \nu \N^*} \ee^{-\kappa \tilde T}\,\tilde T^{q-1}\Biggr)\,|\Lambda_L|  \lesssim \frac{(q-1)!}{\kappa^{q}}\,|\Lambda_L|\,.
\end{equation}
We hence deduce (iii).

Finally, we prove (iv). By using \eqref{mu_measure_2}, \eqref{V_nonnegative}, followed by \eqref{heat_kernel_identity}, Lemma \ref{estimates_heat_kernel} (iii), and Lemma \ref{Riemann_sum_lemma}, we have
\begin{equation} 
\label{integration_lemma_iv_1}
\nu \int_{\Lambda_L}  \dd y\,\int \hat{\mu}^{L}_{y,x}(\dd \tilde \omega)\,T(\tilde \omega)^{q} \leq \nu\, \sum_{\tilde T \in \nu \N^*} \ee^{-\kappa \tilde T}\,\tilde T^{q}  \int_{\Lambda_L}  \dd y \int \bb W^{L,\tilde T}_{y,x}(\dd \tilde \omega)=
\nu\, \sum_{\tilde T \in \nu \N^*} \ee^{-\kappa \tilde T}\,\tilde T^{q} \lesssim \frac{q!}{\kappa^{q+1}}\,.  
\end{equation}
We hence deduce (iv).
\end{proof}

We now introduce some terminology and notation. Given $n \in \N^*$ and $(\delta_1,\ldots,\delta_n) \in \N^n$, we define
\begin{equation} 
\label{tree_delta}
{\fra{T}}_n^{\delta_1,\ldots,\delta_n} \deq \bigl\{\cal T \in {\fra T}_n\,, \mathrm{deg}(i)=\delta_i\,, i=1\,,\ldots,n \bigr\}\,.
\end{equation}
In \eqref{tree_delta}, $\mathrm{deg}(i)$ denotes the degree of $i$ in $\cal T$, i.e.\ the number of vertices in $[n] \setminus \{i\}$ with which $i$ is connected by edges of $\cal T$. 
We note that ${\fra T}_1^{0}$ consists of a single element, i.e.\ the tree with one vertex and no edges and that ${\fra T}_1^{\delta_1}=\emptyset$ if $\delta_1 \neq 0$. For $n \geq 2$,
${\fra T}_n^{\delta_1,\ldots,\delta_n}$ is nonempty only if $(\delta_1,\ldots,\delta_n) \in (\N^*)^n$ and $\sum_{i=1}^{n} \delta_i=2(n-1)$.

Let us now note how we can use Lemma \ref{integration_lemma} (i)--(ii) to integrate all of the paths $\omega_j$ which correspond to vertices of a tree that are not the root (which we henceforth designate to be $1$).

\begin{lemma}[Integration algorithm] 
\label{tree_integration}
Let $n \in \N^*$, $(\delta_1, \ldots, \delta_n) \in (\N^*)^n$ with $\sum_{i=1}^{n} \delta_i=2(n-1)$ and $\cal T \in  {\fra T}_n^{\delta_1,\ldots,\delta_n}$ be given. Furthermore, let $\cal O, \cal C \subset \{2,3,\ldots,n\}$ such that $\cal O \sqcup \cal C=\{2,3,\ldots,n\}$ and $(x_i)_{i \in \cal O} \in \Lambda^{\cal O}$ be given. For $i \in \{2,3,\ldots,n\}$ we let 
\begin{equation*} 
\hat \delta_i \deq
\begin{cases}
\delta_i & \text{if } i \in \cal C
\\
\delta_i+1 & \text{if } i \in \cal O
\end{cases}
\qquad
\Theta_i (\dd \omega) \deq
\begin{cases}
\mu^{L} (\dd \omega) & \text{if } i \in \cal C
\\
\nu \,\int_{\Lambda_L} \dd y \,\hat{\mu}^{L}_{y,x_i} (\dd \omega) & \text{if } i \in \cal O\,.
\end{cases}
\end{equation*}
With this notation, we have that for $\omega_1 \in \Omega^{T_1}$
\begin{equation}  
\label{tree_integration_bound}
\int \Theta_2(\dd \omega_2)\,\Theta_3(\dd \omega_3) \cdots \Theta_n(\dd \omega_n)\,\prod_{(i,j) \in \cal T} |\zeta^L(\omega_i,\omega_j)| \leq C^{n-1} \|v\|_{\ell^1}^{n-1} \prod_{i=2}^{n} \Bigl(\kappa^{-\hat \delta_i}\,(\hat \delta_i-1)!\Bigr)\,T(\omega_1)^{\delta_1}\,. 
\end{equation}
\end{lemma}
We note that, in the statement above, $\cal O, \cal C$, denotes the set of vertices $\neq 1$ which correspond to open paths and  closed paths respectively.

\begin{proof}
We prove the claim by induction on $n$.
\paragraph{Base} The claim trivially holds when $n=1$. 

\paragraph{Step} Suppose that $n \geq 2$ and that \eqref{tree_integration_bound} holds for all trees on at most $n-1$ vertices. 

Let $k \deq \delta_1$. Then $\omega_1$ is connected to $\omega_{i_1},\omega_{i_2}, \ldots, \omega_{i_k}$ for $1 < i_1 <i_2 < \cdots <i_k \leq n$.
By deleting the edges $(1,i_1), \ldots, (1,i_k)$ from $\cal T$, we get a forest
$\cal T_1 \sqcup \cal T_2 \sqcup \cdots \sqcup \cal T_k$,
where $i_{\ell} \in \mathrm{V}(\cal T_\ell)$ for $\ell=1,\ldots,k$. Here, $\mathrm{V}(\cdot)$ denotes the vertex set of a graph.

For $\ell=1,\ldots,k$, let $n_{\ell} \deq |\mathrm{V}(\cal T_{\ell})|$. By the inductive assumption, we have 
\begin{equation}
\label{tree_integration_induction}
\int \prod_{i \in \mathrm{V}(\cal T_{\ell}) \setminus \{i_\ell\}} \Theta_i(\dd \omega_i) \, \prod_{\{i,j\} \in \cal T_\ell} |\zeta^L(\omega_i,\omega_j)|
\leq C^{n_\ell-1}\,\|v\|_{\ell^1}^{n_\ell-1}\,\prod_{i \in \mathrm{V}(\cal T_{\ell}) \setminus \{i_\ell\}} \Bigl(\kappa^{-\hat \delta_i}\,(\hat \delta_i-1)!\Bigr)\,T(\omega_{i_\ell})^{\delta_{i_\ell}-1}\,,
\end{equation}
for all $\ell=1,\ldots,k$.

We use \eqref{tree_integration_induction} to deduce that the expression on the left-hand side of \eqref{tree_integration_bound} is
\begin{equation}
\label{tree_integration_induction_2}
\leq C^{n-k-1}\,\|v\|_{\ell^1}^{n-k-1}\,\prod_{i \in \{2,\ldots,n\} \setminus \{i_1,\ldots,i_k\}} \Bigl(\kappa^{-\hat \delta_i}\,(\hat \delta_i-1)!\Bigr)\,\prod_{\ell=1}^{k} \Biggl(\int \Theta_{i_\ell}(\dd \omega_{i_\ell})\,T(\omega_{i_\ell})^{\delta_{i_\ell}-1}\,|\zeta^L(\omega_1,\omega_{i_\ell})|\Biggr)\,.
\end{equation}
We now apply Lemma \ref{integration_lemma} in each of the $k$ factors in \eqref{tree_integration_induction_2}. More precisely, we apply Lemma \ref{integration_lemma} (i) if $i_\ell \in \cal C$ and Lemma \ref{integration_lemma}  (ii) if $i_\ell \in \cal O$ and deduce \eqref{tree_integration_bound}.
\end{proof}

We record a well-known result about the cardinality of ${\fra T}_n^{\delta_1,\ldots,\delta_n}$, which can be obtained e.g.\ from \cite[Theorem 5.3.4]{Stanley_EC2}.
\begin{lemma} 
\label{number_of_trees}
Let $n \geq 2$ and $(\delta_1,\ldots,\delta_n) \in (\N^*)^n$ such that $\sum_{i=1}^{n} \delta_i=2(n-1)$ be given. We then have 
$\bigl|{\fra T}_n^{\delta_1,\ldots,\delta_n}\bigr|=\frac{(n-2)!}{(\delta_1-1)! \cdots (\delta_n-1)!}$.
\end{lemma}
We now have the necessary tools to prove Proposition \ref{operator_norm_bound}.

\begin{proof}[Proof of Proposition \ref{operator_norm_bound}]
Throughout the proof, we work with $v$ such that $\|v\|_{\ell^1}$ is sufficiently small depending on $\kappa$ in a way to be precisely determined later.
We first prove (i). 
By using Proposition \ref{gamma_p_cluster_expansion} (i), \eqref{Ursell_function_representation}, \eqref{Ursell_function}, and Lemma \ref{Ursell_function_estimate}, we deduce that 
\begin{equation}
\label{hat_Z_1}
\Bigl|\log \cal Z^{\nu,\kappa,\nu^2,L}\Bigr|  \leq \sum_{n=2}^{\infty} \frac{1}{n!} \sum_{\cal T \in {\fra T}_n}\int \mu^{L}(\dd \omega_{1}) \cdots \mu^{L}(\dd \omega_n) \prod_{\{i,j\} \in \cal T} |\zeta^L(\omega_i,\omega_j)|\,.
\end{equation}
By recalling \eqref{tree_delta}, we rewrite the right-hand side of \eqref{hat_Z_1} as 
\begin{equation}
\label{hat_Z_2}
\sum_{n=2}^{\infty} \frac{1}{n!} \sum_{\substack{(\delta_1,\ldots,\delta_n) \in \N^n \\ \delta_1+\cdots+\delta_n=2(n-1)}} \sum_{\cal T \in {\fra T}_n^{\delta_1,\ldots,\delta_n}}
\int \mu^{L}(\dd \omega_1) \cdots \mu^{L}(\dd \omega_n) \prod_{\{i,j\} \in \cal T} 
|\zeta^L(\omega_i,\omega_j)|\,.
\end{equation}
By using  \eqref{hat_Z_1}--\eqref{hat_Z_2} as well as Lemma \ref{tree_integration} with 
$\cal C=\{2,\ldots,n\}$ in each term of \eqref{hat_Z_2}, we deduce that 
\begin{equation} 
\label{hat_Z_3}
 \Bigl|\log \cal Z^{\nu,\kappa,\nu^2,L}\Bigr| 
\lesssim \sum_{n=2}^{\infty} \frac{C^{n-1}}{n!}\,(n-2)!\,\|v\|_{\ell^1}^{n-1} \,\frac{1}{\kappa^{2n-2}}\,|\Lambda_L|=\sum_{n=2}^{\infty} \frac{1}{n(n-1)}\,\biggl(\frac{C}{\kappa^2}\,\|v\|_{\ell^1}\biggr)^{n-1}\,|\Lambda_L|\,.
\end{equation}
Here, we used the observation that for $(\delta_1,\ldots,\delta_n)$ as above, we have
\begin{equation}
\label{hat_Z_3_proof}
\prod_{i=2}^{n} \Bigl(\kappa^{-\delta_i}\,(\delta_i-1)!\Bigr)\, \int \mu^{L} (\dd \omega_1)\,T_1^{\delta_1} \lesssim \frac{1}{\kappa^{2n-2}}\, \prod_{i=1}^{n} (\delta_i-1)!\,|\Lambda_L|\,,
\end{equation}
which follows from Lemma \ref{integration_lemma} (iii). We deduce \eqref{hat_Z_3} from \eqref{hat_Z_3_proof}
by noting that for $n \geq 2$, there are $\binom{2n-3}{n-1} \leq C^{n-1}$ possible choices of $(\delta_1,\ldots,\delta_n) \in {(\N^*)^{n}}$ such that $\delta_1+\cdots+\delta_n=2(n-1)$ and by using Lemma \ref{number_of_trees}.  
Claim (i) now follows.
 
We now prove (ii). By Proposition \ref{gamma_p_cluster_expansion} (ii), it suffices to show that for all $\b x \in \Lambda^p$, we have
\begin{equation} 
\label{Schur_test_2}
\nu^p \,\int_{\Lambda_L^p} \dd \b y \, 
\int \hat \mu^{L}_{y_1,x_1}(\dd \omega_1) \cdots \hat \mu^{L}_{y_p,x_p}(\dd \omega_p) \sum_{\Pi \in \fra P_p} \prod_{\xi \in \Pi} \bigl|X^{L}((\omega_i)_{i \in \xi})\bigr|= O_{\kappa,p,\|v\|_{\ell^1}}(1)\,.
\end{equation}
Let us first estimate the contribution to the left-hand side of \eqref{Schur_test_2} coming from the trivial partition $\Pi=[p]$. To this end, we define
\begin{equation*} 
(\widetilde{\Gamma}^{\nu,\kappa,\nu^2,L}_p)_{\b x, \b y} \deq 
\int \hat \mu^{L}_{y_1,x_1}(\dd \omega_1) \cdots \hat \mu^{L}_{y_p,x_p}(\dd \omega_p) \,\bigl|X^{L}(\omega_1,\ldots,\omega_p)\bigr|
\end{equation*}
for $\b x, \b y \in \Lambda^p$.
Recalling \eqref{Ursell_function_representation} and arguing analogously as in \eqref{hat_Z_1}--\eqref{hat_Z_2}, we deduce that 
\begin{align}
\notag
(\widetilde{\Gamma}^{\nu,\kappa,\nu^2,L}_p)_{\b x, \b y} \leq  \sum_{n \geq p} \frac{1}{(n-p)!} \sum_{\substack{(\delta_1,\ldots,\delta_n) \in \N^n \\ \delta_1+\cdots+\delta_n=2(n-1)}}  &\sum_{\cal T \in {\fra T}_n^{\delta_1,\ldots,\delta_n}}
\int  \hat \mu^{L}_{y_1,x_1}(\dd \omega_1) \cdots \hat \mu^{L}_{y_p,x_p}(\dd \omega_p) 
\\
\label{hat_gamma}
&
\times \mu^{L}(\dd \omega_{p+1}) \cdots \mu^{L}(\dd \omega_n) \prod_{\{i,j\} \in \cal T} |\zeta^L(\omega_i,\omega_j)|\,.
\end{align}
We first integrate \eqref{hat_gamma} with respect to $y_2,\ldots,y_p$.
By using Lemma \ref{tree_integration} with $\cal O=\{2,3,\ldots,p\}, \cal C=\{p+1,\ldots,n\}$ in each term of \eqref{hat_gamma}, we deduce that 
\begin{multline} 
\label{hat_gamma_3}
\nu^p\,\int_{\Lambda_L^{p-1}}\dd y_2 \,\cdots \dd y_p\, (\widetilde{\Gamma}^{\nu,\kappa,\nu^2,L}_p)_{\b x, \b y} \leq  \ind{p=1} \int \hat{\mu}^{L}_{y_1,x_1} (\dd \omega_1)
\\
+ \sum_{n \geq p \vee 2} \frac{1}{(n-p)!}  \sum_{\substack{(\delta_1,\ldots,\delta_n) \in \N^n \\ \delta_1+\cdots+\delta_n=2(n-1)}} \sum_{\cal T \in {\fra T}_n^{\delta_1,\ldots,\delta_n}}  C^{n-1} \|v\|_{\ell^1}^{n-1} \prod_{i=2}^{n} \Bigl(\kappa^{-\hat \delta_i}\,(\hat \delta_i-1)!\Bigr)\, \int \hat{\mu}^{L}_{y_1,x_1} (\dd \omega_1)\,T_1^{\delta_1}\,.
\end{multline}
We now use \eqref{hat_gamma_3}, Lemma \ref{integration_lemma} (iv), the fact that $\delta_1,\ldots,\delta_n \leq n-1$, and argue analogously as for \eqref{hat_Z_3} to deduce that
\begin{equation} 
\label{hat_gamma_4}
\nu^p\,\int_{\Lambda_L^p} \dd \b y\, (\widetilde{\Gamma}^{\nu,\kappa,\nu^2,L}_p)_{\b x, \b y} 
\leq \frac{C}{\kappa} + \sum_{n \geq p \vee 2} \biggl(\frac{(n-1)^2}{\kappa}\biggr)^{p}\,\biggl(\frac{C}{\kappa^2}\,\|v\|_{\ell^1}\biggr)^{n-1}\,.
\end{equation}
We use \eqref{hat_gamma_4} to estimate the contribution to the left-hand side of \eqref{Schur_test_2} coming from $\Pi=[p]$.

By arguing analogously as for \eqref{hat_gamma_4} we deduce that for a general nonempty set $\xi \subset [p]$
\begin{equation} 
\label{hat_gamma_5}
\nu^{|\xi|}\,\int_{\Lambda_L^{|\xi|}} \prod_{i \in \xi} \dd y_i\, \prod_{i \in \xi} \int \hat{\mu}^{L}_{y_i,x_i}(\dd \omega_i)\,\bigl|X^{L}((\omega_i)_{i \in \xi})\bigr|
\leq \frac{C}{\kappa} + \sum_{n \geq |\xi| \vee 2} \biggl(\frac{(n-1)^2}{\kappa}\biggr)^{|\xi|}\,\biggl(\frac{C}{\kappa^2}\,\|v\|_{\ell^1}\biggr)^{n-1}\,.
\end{equation}
In the proof of \eqref{hat_gamma_5}, it is important that the last vertex over which we integrate corresponds to an open path (which in the proof of \eqref{hat_gamma_4} was $\omega_1$). This is possible to do by construction. 
We deduce \eqref{Schur_test_2} from \eqref{hat_gamma_5} and claim (ii) follows.
\end{proof}

Before proving Proposition \ref{MF_cluster_expansion}, we note two technical lemmas that we apply in the proof. We first introduce some notation.
Throughout the sequel, we denote by $|\cdot|_L$ the (periodic) Euclidean norm on $\Lambda_L$.
Furthermore, we fix $c>0$ small and let
\begin{equation}
\label{D^{L,c}}
\cal D^{L}_{c} \deq \Bigl\{\omega \in \bigcup_{T > 0} \Omega^{L,T}\,,|\omega(s)-\omega(t)|_L \geq cL \;\; \mbox{for some} \;\; s, t \in [0,T(\omega)]\Bigr\}\,.
\end{equation}
All of the estimates below depend on $c$, but we do not keep explicit track of this dependence.

The first lemma is a modification of Lemma \ref{integration_lemma} telling us that we get small contributions if we integrate over long paths. Before stating it, let us first introduce some notation.

\begin{definition} 
\label{o_L_definition}
Let $(A_L)_{\in \N^*}$ be a sequence of nonnegative numbers. For a quantity $B \geq 0$ and a  given set of parameters $\alpha_1,\ldots,\alpha_m$, we say that 
\begin{equation*}
A_L \lesssim_{\alpha_{1},\ldots,\alpha_{m}} o_{L}(B) 
\end{equation*}
when the following holds.
\begin{itemize}
\item[(i)] There exists $C \equiv C(\alpha_{1},\ldots,\alpha_{m})>0$ which does not depend on $L$ such that $A_L \leq C B$ for all $L \in \N^*$.
\item[(ii)] We have  $\lim_{L \rightarrow \infty} A_{L}=0$.
\end{itemize}
If $m=0$, we just write $A_L=o_L(B)$.
\end{definition}

\begin{lemma}
\label{integration_lemma_M}
Let $\omega \in \Omega^{L,T(\omega)}$ with $T(\omega) \in \nu \N^*$ and $q \in \N$ be given.
Then, the following estimates hold.
\begin{itemize}
\item[(i)] $\int \mu^{L} (\dd \tilde \omega)\,T(\tilde \omega)^{q} \,|\zeta^{L}(\omega,\tilde \omega)|\,\ind{\cal D^{L}_{c}}(\tilde \omega) \lesssim_{\kappa,q,d} o_L\bigl(T(\omega)\,\|v\|_{\ell^1}\bigr)\,.$ 
\item[(ii)] Let $L_0 \in \N^*$ with 
\begin{equation}
\label{L_0_choice}
L_0 \leq L/4
\end{equation}
be given. Then, for $x \in \Lambda_{L_0}$, we have 
\begin{equation*}
\nu\,\int_{\Lambda_{L_0}} \dd y\,\int \hat{\mu}^{L}_{y,x} (\dd\tilde \omega) \,  T(\tilde \omega)^{q}\,|\zeta^{L}(\omega,\tilde \omega)|\,\ind{\cal D^{L}_{c}}(\tilde \omega) \lesssim_{\kappa,q,d} o_L\Bigl(T(\omega)\,[\|v\|_{\ell^{\infty}}+\|v\|_{\ell^1}]\Bigr)\,.
\end{equation*}
\item[(iii)] $\frac{1}{|\Lambda_L|}\,\int {\mu}^{L}(\dd \tilde{\omega})\,T(\tilde{\omega})^{q}\,\ind{\mathcal{D}_{c}^{L}}(\tilde{\omega}) \lesssim_{\kappa,q,d} o_{L} (1)\,.$
\item[(iv)] For $L_0$ as in \eqref{L_0_choice} and $x \in \Lambda_{L_0}$, we have
\begin{equation}
\label{integration_lemma_M_iv}
\nu\,\int_{\Lambda_{L_0}} \dd y\,\int \hat{\mu}^{L}_{y,x}(\dd \tilde \omega) \,T(\tilde \omega)^{q} \,\ind{\cal D^{L}_{c}}(\tilde \omega)\lesssim_{\kappa,q,d}\,o_L(1)\,.
\end{equation}
\end{itemize}
\end{lemma}

The second lemma states that the contribution from two interacting paths which are far away is small. In order to state this precisely, we need to introduce some notation.
With $c>0$ as earlier, we let $\cal V^{\nu,\nu^2,L}_{c}(\omega,\tilde \omega)$ denote the quantity given as in \eqref{V_interaction_1} with interaction potential given by
\begin{equation}
\label{v^{L}_{c}}
{v^{L}_{c}}(x) \deq v^{L}(x) \,\ind{|x|_L \geq cL}\,.
\end{equation}
Note that, by \eqref{v^{L}} and Assumption \ref{interaction_potential_v^{L}} (iii), we have that
\begin{equation}
\label{v^{L}_{c}_estimate}
\lim_{L \rightarrow \infty} \|{v^{L}_{c}}\|_{\ell^1(\Lambda_L)}=0\,.
\end{equation}
In analogy with \eqref{Ursell_function}, we define
\begin{equation}
\label{zeta^{L,c}}
\zeta^{L}_{c}(\omega,\tilde \omega) \deq \exp\bigl(-\cal V^{\nu,\nu^2,L}_{c}(\omega,\tilde \omega)\bigr)-1\,.
\end{equation}
With this notation, we have the following lemma.
\begin{lemma}
\label{integration_lemma_M_2}
With assumptions as in Lemma \ref{integration_lemma_M}, the following estimates hold.
\begin{itemize}
\item[(i)] $\int \mu^{L} (\dd \tilde \omega)\, T(\tilde \omega)^{q} \,|\zeta^{L}_{c}(\omega,\tilde \omega)| \lesssim_{\kappa,q} T(\omega)\,\|v^{L}_{c}\|_{\ell^1(\Lambda_L)}\,.$
\item[(ii)] 
$\nu\,\int_{\Lambda_{L_0}} \dd y\,\int \hat{\mu}^{L}_{y,x} (\dd \tilde \omega) \, T(\tilde \omega)^{q}\,|\zeta^{L}_{c}(\omega,\tilde \omega)|  \lesssim_{\kappa,q} T(\omega)\,\|v^{L}_{c}\|_{\ell^1(\Lambda_L)}$
for all $x \in \Lambda_{L_0}$. 
\end{itemize}
\end{lemma}
We prove Lemmas \ref{integration_lemma_M} and \ref{integration_lemma_M_2} in Appendix \ref{Appendix C.3}. We now have all the necessary tools to prove Proposition \ref{MF_cluster_expansion}.
\begin{proof}[Proof of Proposition \ref{MF_cluster_expansion}]

We first show claim (ii) and then explain how the proof can be modified to obtain claim (i). By Proposition \ref{operator_norm_bound} (ii) and Proposition \ref{operator_norm_bound_corollary} (ii), we can consider fixed $n \in \N^*
$ in \eqref{Ursell_function_representation} and fixed $\pi \in S_p$ in \eqref{gamma_p_cluster_expansion_claim}. Without loss of generality, we can assume that $\pi$ is the identity and reduce to proving that 
\begin{multline}
\label{(ii)_star}
\lim_{L \rightarrow \infty} \nu^p \,\biggl\|\int \hat \mu^{L}_{y_1,x_1}(\dd \omega_1) \cdots \hat \mu^{L}_{y_p,x_p}(\dd \omega_p)\,\mu^{L}(\dd \omega_{p+1}) \cdots \mu^{L}(\dd \omega_n)\,\varphi^{L}(\omega_1,\ldots,\omega_n)
\\
-\int \hat \mu^{\infty}_{y_1,x_1}(\dd \omega_1) \cdots \hat \mu^{\infty}_{y_p,x_p}(\dd \omega_p)\,\mu^{\infty}(\dd \omega_{p+1}) \cdots \mu^{\infty}(\dd \omega_n)\,\varphi^{\infty}(\omega_1,\ldots,\omega_n)
 \biggr\|_{L_0,p} = 0\,,
\end{multline}
where we recall \eqref{L_0_p_norm} and also consider \eqref{mu_measure_1}--\eqref{mu_measure_2} and \eqref{Ursell_function} on the infinite lattice $\Z^d$, with analogous definitions. Here $v^{\infty} \equiv v$.
In \eqref{(ii)_star} and in the remainder of the proof, all of the convergence claims are interpreted as being uniform in $\nu \in (0,1/\kappa]$.

Let us now prove \eqref{(ii)_star}. Throughout the sequel, we assume that $L \in \N^*$ satisfies \eqref{L_0_choice} above.
Given such an $L$, we take $L_1 \in \N^*$ to be the smallest even integer
\begin{equation}
\label{L_1_choice}
L_1 \geq L/3\,,
\end{equation}
and define
\begin{equation}
\label{A^{L}_B^{L}}
\cal A^{L} \deq \biggl\{\omega \in \bigcup_{T > 0} \Omega^{\infty,T}\,,\omega(t) \in \Lambda_{L_1}\, \forall t \in [0,T(\omega)]\biggr\}\,,\quad \cal B^{L} \deq  \bigcup_{T > 0} \Omega^{\infty,T} \setminus \cal A^{L}\,.
\end{equation}
We first show that for all $k \in \{1,\ldots,n\}$, we have
\begin{equation}
\label{(ii)_star_proof1}
\lim_{L \rightarrow \infty} \nu^p\,\biggl\|\int \hat \mu^{L}_{y_1,x_1}(\dd \omega_1) \cdots \hat \mu^{L}_{y_p,x_p}(\dd \omega_p) \,\mu^{L}(\dd \omega_{p+1}) \cdots \mu^{L}(\dd \omega_n)\,\ind{\cal B^{L}}(\omega_k)\,\varphi^{L}(\omega_1,\ldots,\omega_n)\biggr\|_{L_0,p}=0\,.
\end{equation}
By Lemma \ref{Ursell_function_estimate}, \eqref{(ii)_star_proof1} follows if we show that
\begin{equation}
\label{(ii)_star_proof2}
\lim_{L \rightarrow \infty} \nu^p\,\biggl\|\int \hat \mu^{L}_{y_1,x_1}(\dd \omega_1) \cdots \hat \mu^{L}_{y_p,x_p}(\dd \omega_p)\,\mu^{L}(\dd \omega_{p+1}) \cdots  \mu^{L}(\dd \omega_n)\,\ind{\cal B^{L}}(\omega_k)\,\prod_{\{i,j\} \in \cal T} |\zeta^{L}(\omega_i,\omega_j)|\biggr\|_{L_0,p}=0\,,
\end{equation}
for a fixed $\cal T \in {\fra T}_n$.
We note that \eqref{(ii)_star_proof2} follows by applying the triangle inequality together with Lemmas \ref{integration_lemma_M} and \ref{integration_lemma_M_2}.
More precisely, we note that in \eqref{(ii)_star_proof2}, the nonzero contribution comes from $x_1 \in \Lambda_{L_0}$. For such $x_1$, we recall that by \eqref{L_0_choice} and \eqref{A^{L}_B^{L}}, there exists a point $z$ on $\omega_k$ with the property that $|z-x_1|_{L} \gtrsim L$.
In particular, by the triangle inequality,
 it follows that, for the paths over which we are integrating in \eqref{(ii)_star_proof2}, at least one of the following cases occur with suitable $c \gtrsim \frac{1}{n}$.
\begin{itemize}
\item[(1)] There exists $i \in \{1,\ldots,n\}$ such that $\omega_i \in \cal{D}^{L,c}$. Here, we recall \eqref{D^{L,c}}.
\item[(2)] There exists $\{i,j\} \in \cal T$ such that $\zeta^{L}(\omega_i,\omega_j)=\zeta^{L}_{c}(\omega_i,\omega_j)$. In other words, the two-particle interaction is given by \eqref{v^{L}_{c}}.
\end{itemize}
We now prove \eqref{(ii)_star_proof2} by arguing analogously as in the proof of Proposition \ref{operator_norm_bound} (ii). We just modify the proof to keep track of cases (1) and (2) above. If (1) occurs, we apply Lemma \ref{integration_lemma_M} when integrating $\omega_i$. If (2) occurs, we apply Lemma \ref{integration_lemma_M_2} when integrating the path $\omega_i$ or $\omega_j$ (determined by the algorithm from Lemma \ref{(ii)_star_proof2}). In this case, we also recall \eqref{v^{L}_{c}_estimate} to note that the estimates that we get from Lemma \ref{integration_lemma_M_2} tend to zero as $L \rightarrow \infty$. We hence deduce \eqref{(ii)_star_proof1}.
 
By arguing analogously as for \eqref{(ii)_star_proof1}, we get that for all $k \in \{1,\ldots,n\}$,
\begin{equation}
\label{(ii)_star_proof3}
\lim_{L \rightarrow \infty} \nu^p\,\biggl\|\int \hat \mu^{\infty}_{y_1,x_1}(\dd \omega_1) \cdots \hat \mu^{\infty}_{y_p,x_p}(\dd \omega_p)\,\mu^{\infty}(\dd \omega_{p+1}) \cdots \mu^{\infty}(\dd \omega_n)\,\ind{\cal B^{L}}(\omega_k)\,\varphi^{\infty}(\omega_1,\ldots,\omega_n)\biggr\|_{L_0,p}=0\,.
\end{equation}
Combining \eqref{A^{L}_B^{L}}, \eqref{(ii)_star_proof1}, \eqref{(ii)_star_proof3}, and recalling \eqref{L_0_p_norm}, we deduce that \eqref{(ii)_star} follows if we prove that 
\begin{multline}
\label{(ii)_star_proof4}
\lim_{L \rightarrow \infty} \nu^p\,\biggl\|\int \hat \mu^{L}_{y_1,x_1}(\dd \omega_1) \cdots \hat \mu^{L}_{y_p,x_p}(\dd \omega_p)\,\mu^{L}(\dd \omega_{p+1}) \cdots \mu^{L}(\dd \omega_n)\,\prod_{k=1}^{n}\ind{\cal A^{L}}(\omega_k)\,\varphi^{L}(\omega_1,\ldots,\omega_n)
\\
-\int \hat \mu^{\infty}_{y_1,x_1}(\dd \omega_1) \cdots \hat \mu^{\infty}_{y_p,x_p}(\dd \omega_p)\,\mu^{\infty}(\dd \omega_{p+1}) \cdots \mu^{\infty}(\dd \omega_n)\,\prod_{k=1}^{n}\ind{\cal A^{L}}(\omega_k)\,\varphi^{\infty}(\omega_1,\ldots,\omega_n)
\biggr\|_{\ell^{\infty}_{\b x}\ell^1_{\b y}} = 0\,.
\end{multline}

We now show \eqref{(ii)_star_proof4}. Let us first introduce some notation. We first consider the following modifications of the measures \eqref{mu_measure_1}--\eqref{mu_measure_2}.

\begin{align}
\label{mu_measure_1_*}
&\mu^{L,*}(\dd \omega) \deq \nu \sum_{T\in \nu \N^*} \frac{\ee^{-\kappa T}}{T} \,  \bb W^{L,T}(\dd \omega)
\,\ee^{-\cal{V}^{\nu,\nu^2,\infty}(\omega,\omega)/2}\,,
\\
\label{mu_measure_2_*}
&\hat \mu^{L,*}_{y,x}(\dd \omega) \deq \sum_{T \in \nu \N^*} \ee^{-\kappa T} \, \bb W^{L,T}_{y,x}(\dd \omega)\,\ee^{-\cal{V}^{\nu,\nu^2,\infty}(\omega,\omega)/2}\,.
\end{align}
Namely, we replace the self-interaction term $\ee^{-\cal{V}^{\nu,\nu^2,L}(\omega,\omega)/2}$ in \eqref{mu_measure_1}--\eqref{mu_measure_2} with $\ee^{-\cal{V}^{\nu,\nu^2,\infty}(\omega,\omega)/2}$.
Furthermore, we consider the following modifications, where we omit the self-interaction term.

\begin{align}
\label{mu_measure_1_0}
&\mu^{L,0}(\dd \omega) \deq \nu \sum_{T\in \nu \N^*} \frac{\ee^{-\kappa T}}{T} \,  \bb W^{L,T}(\dd \omega)\,,
\\
\label{mu_measure_2_0}
&\hat \mu^{L,0}_{y,x}(\dd \omega) \deq \sum_{T \in \nu \N^*} \ee^{-\kappa T} \, \bb W^{L,T}_{y,x}(\dd \omega)\,.
\end{align}
From \eqref{mu_measure_1}--\eqref{mu_measure_2_*}, \eqref{mu_measure_1_*}--\eqref{mu_measure_2}, \eqref{mu_measure_1_0}--\eqref{mu_measure_2_0}, we have 
\begin{equation}
\label{mu_measure_bound}
\mu^{L}(\dd \omega)\,, \,\mu^{L,*}(\dd \omega) \leq \mu^{L,0}(\dd \omega)\,, \qquad 
\hat \mu^{L}_{y,x}(\dd \omega)\,,\,\hat \mu^{L,*}_{y,x}(\dd \omega) \leq \hat \mu^{L,0}_{y,x}(\dd \omega)\,.
\end{equation}

By \eqref{mu_measure_1_*}--\eqref{mu_measure_2_*} and \eqref{A^{L}_B^{L}}, we have that for all integrable $F$ 
\begin{multline}
\label{(ii)_star_proof5}
\int \hat \mu^{L,*}_{y,x}(\dd \omega) \,\ind{\cal A^{L}}(\omega)\,F(\omega)=\int \hat \mu^{\infty}_{y,x}(\dd \omega) \,\ind{\cal A^{L}}(\omega)\,F(\omega)\,, 
\\
\int \mu^{L,*}(\dd \omega) \,\ind{\cal A^{L}}(\omega)\,F(\omega)=\int \mu^{\infty}(\dd \omega) \,\ind{\cal A^{L}}(\omega)\,F(\omega)\,.
\end{multline}
In other words, the presence of the indicator functions $\ind{\cal A^{L}}(\omega)$ erases the boundary effects of $\Lambda_L$.

Claim \eqref{(ii)_star_proof4} follows from 

\begin{multline}
\label{(ii)_star_proof4_circle_1}
\lim_{L \rightarrow \infty} \nu^p\,\biggl\|\int \hat \mu^{L}_{y_1,x_1}(\dd \omega_1) \cdots \hat \mu^{L}_{y_p,x_p}(\dd \omega_p)\,\mu^{L}(\dd \omega_{p+1}) \cdots \mu^{L}(\dd \omega_n)\,\prod_{k=1}^{n}\ind{\cal A^{L}}(\omega_k)\,\varphi^{L}(\omega_1,\ldots,\omega_n)
\\
-\int \hat \mu^{L,*}_{y_1,x_1}(\dd \omega_1) \cdots \hat \mu^{L,*}_{y_p,x_p}(\dd \omega_p)\,\mu^{L,*}(\dd \omega_{p+1}) \cdots \mu^{L,*}(\dd \omega_n)\,\prod_{k=1}^{n}\ind{\cal A^{L}}(\omega_k)\,\varphi^{L}(\omega_1,\ldots,\omega_n)
\biggr\|_{\ell^{\infty}_{\b x}\ell^1_{\b y}} = 0
\end{multline}
and 
\begin{multline}
\label{(ii)_star_proof4_circle_2}
\lim_{L \rightarrow \infty} \nu^p\,\biggl\|\int \hat \mu^{L,*}_{y_1,x_1}(\dd \omega_1) \cdots \hat \mu^{L,*}_{y_p,x_p}(\dd \omega_p)\,\mu^{L,*}(\dd \omega_{p+1}) \cdots \mu^{L,*}(\dd \omega_n)\,\prod_{k=1}^{n}\ind{\cal A^{L}}(\omega_k)\,\varphi^{L}(\omega_1,\ldots,\omega_n)
\\
-\int \hat \mu^{\infty}_{y_1,x_1}(\dd \omega_1) \cdots \hat \mu^{\infty}_{y_p,x_p}(\dd \omega_p)\,\mu^{\infty}(\dd \omega_{p+1}) \cdots \mu^{\infty}(\dd \omega_n)\,\prod_{k=1}^{n}\ind{\cal A^{L}}(\omega_k)\,\varphi^{\infty}(\omega_1,\ldots,\omega_n)
\biggr\|_{\ell^{\infty}_{\b x}\ell^1_{\b y}} = 0\,.
\end{multline}

Let us first prove \eqref{(ii)_star_proof4_circle_1}.
By recalling \eqref{mu_measure_1}--\eqref{mu_measure_2}, \eqref{mu_measure_1_*}--\eqref{mu_measure_2_*}, \eqref{mu_measure_bound}, and by using a telescoping argument, as well as the mean-value theorem, we reduce  \eqref{(ii)_star_proof4_circle_1} to showing that for any $\ell \in \{1,\ldots, n\}$ we have

\begin{multline}
\label{(ii)_star_proof4_circle_1_A}
\lim_{L \rightarrow \infty} \nu^p\,\biggl\|\int \hat \mu^{L,0}_{y_1,x_1}(\dd \omega_1) \cdots \hat \mu^{L,0}_{y_p,x_p}(\dd \omega_p)\,\mu^{L,0}(\dd \omega_{p+1}) \cdots \mu^{L,0}(\dd \omega_n)\,\prod_{k=1}^{n}\ind{\cal A^{L}}(\omega_k)\,|\varphi^{L}(\omega_1,\ldots,\omega_n)|\,
\\
\times
\bigl|\mathcal{V}^{\nu,\nu^2,L}(\omega_{\ell},\omega_{\ell})-\mathcal{V}^{\nu,\nu^2,\infty}(\omega_{\ell},\omega_{\ell})\bigr|
\biggr\|_{\ell^{\infty}_{\b x}\ell^1_{\b y}} = 0\,.
\end{multline}
By Lemma \ref{Ursell_function_estimate}, \eqref{(ii)_star_proof4_circle_1_A} follows if we show that for all $\cal T \in {\fra T}_n$, we have 
\begin{multline}
\label{(ii)_star_proof4_circle_1_B}
\lim_{L \rightarrow \infty} \nu^p\,\biggl\|\int \hat \mu^{L,0}_{y_1,x_1}(\dd \omega_1) \cdots \hat \mu^{L,0}_{y_p,x_p}(\dd \omega_p)\,\mu^{L,0}(\dd \omega_{p+1}) \cdots \mu^{L,0}(\dd \omega_n)\,\prod_{k=1}^{n}\ind{\cal A^{L}}(\omega_k)\,\prod_{\{i,j\} \in \cal T}|\zeta^L(\omega_i,\omega_j)|\,
\\
\times
\bigl|\mathcal{V}^{\nu,\nu^2,L}(\omega_{\ell},\omega_{\ell})-\mathcal{V}^{\nu,\nu^2,\infty}(\omega_{\ell},\omega_{\ell})\bigr|
\biggr\|_{\ell^{\infty}_{\b x}\ell^1_{\b y}} = 0\,.
\end{multline}
Since $|\zeta^L(\omega_i,\omega_j)| \leq 1$, \eqref{(ii)_star_proof4_circle_1_B} follows if we show
\begin{multline}
\label{(ii)_star_proof4_circle_1_C}
\lim_{L \rightarrow \infty} \nu^p\,\biggl\|\int \hat \mu^{L,0}_{y_1,x_1}(\dd \omega_1) \cdots \hat \mu^{L,0}_{y_p,x_p}(\dd \omega_p)\,\mu^{L,0}(\dd \omega_{p+1}) \cdots \mu^{L,0}(\dd \omega_n)\,\prod_{k=1}^{n}\ind{\cal A^{L}}(\omega_k)\,\mathop{\prod_{\{i,j\} \in \cal T}}_{i,j \neq \ell} |\zeta^L(\omega_i,\omega_j)|\,
\\
\times
\bigl|\mathcal{V}^{\nu,\nu^2,L}(\omega_{\ell},\omega_{\ell})-\mathcal{V}^{\nu,\nu^2,\infty}(\omega_{\ell},\omega_{\ell})\bigr|
\biggr\|_{\ell^{\infty}_{\b x}\ell^1_{\b y}} = 0\,.
\end{multline}
In order to obtain \eqref{(ii)_star_proof4_circle_1_C}, we first integrate in $\omega_{\ell}$ using the estimate
\begin{equation}
\label{integral_over_omega_ell_closed}
\int \mu^{L,0} (\dd \omega_{\ell})\,\bigl|\mathcal{V}^{\nu,\nu^2,L}(\omega_{\ell},\omega_{\ell})-\mathcal{V}^{\nu,\nu^2,\infty}(\omega_{\ell},\omega_{\ell})\bigr|\,\ind{\cal{A}^L} (\omega_l) \lesssim \frac{1}{\kappa^2}\,\|v-v^L\|_{\ell^1(\Lambda_L)}\,\,\, \text{if } p+1 \leq \ell \leq n
\end{equation}
and 
\begin{equation}
\label{integral_over_omega_ell_open}
\nu\,\int_{\Lambda_L} \dd y\,\int \hat{\mu}^{L,0}_{y,x}(\dd \omega_{\ell})\,\bigl|\mathcal{V}^{\nu,\nu^2,L}(\omega_{\ell},\omega_{\ell})-\mathcal{V}^{\nu,\nu^2,\infty}(\omega_{\ell},\omega_{\ell})\bigr|\,\ind{\cal{A}^L} (\omega_{\ell}) \lesssim \frac{1}{\kappa^3}\,\|v-v^L\|_{\ell^1(\Lambda_L)}\,\,\, \text{if } 1 \leq \ell \leq p\,.
\end{equation}
The estimates \eqref{integral_over_omega_ell_closed}--\eqref{integral_over_omega_ell_open}
follow by using the same arguments as in the proof of Lemma \ref{integration_lemma} (i) and (ii) respectively. The only difference is that we are now integrating out a self-interaction and that the interaction potential is $v-v^L$. Moreover, we use that $\omega_{\ell}(\cdot)-\omega_{\ell}(\cdot)$ always lies within $\Lambda_L$ when the $\omega_L$ lies within $\Lambda_{L_1}$.

Having integrated out $\omega_{\ell}$, we decompose $\mathcal{T} \setminus \{\ell\}$ into disjoint trees and we integrate over $\omega_k$ for $k \neq \ell$ using the integration algorihm from Lemma \ref{tree_integration} on each of these trees. This gives us an $\cal{O}(1)$ contribution. We hence deduce \eqref{(ii)_star_proof4_circle_1_C} since
\eqref{v^{L}} and the nonnegativity of $v$ imply
\begin{equation}
\label{v-v^L_ell^1}
\|v-v^L\|_{\ell^1(\Lambda_L)}=\sum_{k \in (L \Z)^d \setminus \{0\}} \int_{\Lambda_L} v(x+k)=\|v\|_{\ell^1(\Z^d \setminus \Lambda_L)} \rightarrow 0  \quad \text{as } L \rightarrow \infty\,.
\end{equation}
For the last step, we used the integrability of $v$ given by Assumption \ref{interaction_potential_v^{L}} (iii). We hence obtain \eqref{(ii)_star_proof4_circle_1}.

We now prove \eqref{(ii)_star_proof4_circle_2}.
By using \eqref{(ii)_star_proof5}, it suffices to show
\begin{multline}
\label{(ii)_star_proof4_circle_2_A}
\lim_{L \rightarrow \infty} \nu^p\,\biggl\|\int \hat \mu^{\infty}_{y_1,x_1}(\dd \omega_1) \cdots \hat \mu^{\infty}_{y_p,x_p}(\dd \omega_p)\,\mu^{\infty}(\dd \omega_{p+1}) \cdots \mu^{\infty}(\dd \omega_n)\,\prod_{k=1}^{n}\ind{\cal A^{L}}(\omega_k)
\\
\times \Bigl[\varphi^{L}(\omega_1,\ldots,\omega_n)-\varphi^{\infty}(\omega_1,\ldots,\omega_n)\Bigr]
\biggr\|_{\ell^{\infty}_{\b x}\ell^1_{\b y}} = 0\,.
\end{multline}
By recalling \eqref{Ursell_function}, it suffices to show that for a fixed $\cal G \in {\fra G}^c_n$, we have 
\begin{multline}
\label{(ii)_star_proof4_circle_2_B}
\lim_{L \rightarrow \infty} \nu^p\,\biggl\|\int \hat \mu^{\infty}_{y_1,x_1}(\dd \omega_1) \cdots \hat \mu^{\infty}_{y_p,x_p}(\dd \omega_p)\,\mu^{\infty}(\dd \omega_{p+1}) \cdots \mu^{\infty}(\dd \omega_n)\,\prod_{k=1}^{n}\ind{\cal A^{L}}(\omega_k)
\\
\times \Bigl[\prod_{\{i,j\} \in \cal G}\zeta^L(\omega_i,\omega_j)-\prod_{\{i,j\} \in \cal G}\zeta^{\infty}(\omega_i,\omega_j)
\Bigr]
\biggr\|_{\ell^{\infty}_{\b x}\ell^1_{\b y}} = 0\,.
\end{multline}
Let $<$ denote an arbitrary total order on the edges of $\cal G$. We write an edge $e \in \mathcal{G}$ as $\{i_e,j_e\}$ with $i_e<j_e$. We telescope and write
\begin{multline}
\label{Ursell_function_difference}
\Biggl|\prod_{\{i,j\} \in \cal G} \zeta^L(\omega_i, \omega_j)-\prod_{\{i,j\} \in \cal G} \zeta^{\infty}(\omega_i, \omega_j)\Biggr| 
\\
\leq \sum_{e \in \cal G} \mathop{\prod_{\{i,j\} \in \mathcal{G}}}_{\{i,j\} < e} \bigl|\zeta^L(\omega_i,\omega_j) \bigr|
\,\bigl|(\zeta^L-\zeta^{\infty})(\omega_{i_e},\omega_{j_e})\bigr|\, \mathop{\prod_{\{i,j\} \in \mathcal{G}}}_{e < \{i,j\}} \bigl|\zeta^{\infty}(\omega_i,\omega_j)\bigr|\,.
\end{multline}
Given $e \in \cal G$, let $\cal T_e \in {\fra T}_n$ be an arbitrarily chosen spanning tree for $\cal G$ with the property that $e \in \cal T_e$. 
Since 
\begin{equation*}
|\zeta^L(\omega, \tilde \omega)| \leq 1\,,\qquad |\zeta^{\infty}(\omega,\tilde \omega)| \leq 1\,,
\end{equation*}
we obtain
\begin{equation}
\label{Ursell_function_difference_2}
\eqref{Ursell_function_difference} \leq 
 \sum_{e \in \cal G} \mathop{\prod_{\{i,j\} \in \cal T_e}}_{\{i,j\} < e} \bigl|\zeta^L(\omega_i,\omega_j) \bigr|
\,\bigl|(\zeta^L-\zeta^{\infty})(\omega_{i_e},\omega_{j_e})\bigr|\, \mathop{\prod_{\{i,j\} \in \cal T_e}}_{e < \{i,j\}} \bigl|\zeta^{\infty}(\omega_i,\omega_j)\bigr|\,.
\end{equation}
Since $v \geq 0$, we have that
\begin{multline}
\label{zeta_difference}
\bigl|\zeta^L(\omega,\tilde{\omega})-\zeta^{\infty}(\omega,\tilde{\omega})\bigr| \leq \bigl|V^{\nu,\nu^2,L}(\omega,\tilde{\omega})-V^{\nu,\nu^2}(\omega,\tilde{\omega})\bigr|
\\
\leq \nu \sum_{s \in \nu \N} \ind{s < T(\omega)} \, \sum_{\tilde s \in \nu \N} \ind{\tilde s < T(\tilde \omega)}  \, \int_0^\nu \dd t \, |v-v^L|(\omega(s + t) - \tilde \omega(\tilde s + t))\,.
\end{multline}
If $\omega,\tilde{\omega} \in \mathcal{A}_L$, we recall the choice of $L_1$ from \eqref{L_1_choice} and take $L_1 \leq L/2$. In particular, 
we obtain from \eqref{zeta_difference}, we have 
\begin{multline}
\label{zeta_difference_2}
\bigl|\zeta^L(\omega,\tilde{\omega})-\zeta^{\infty}(\omega,\tilde{\omega})\bigr| \,\ind{\cal{A}^L} (\omega) \,\ind{\cal{A}^L} (\tilde \omega)\leq 
\\
\leq \nu \sum_{s \in \nu \N} \ind{s < T(\omega)} \, \sum_{\tilde s \in \nu \N} \ind{\tilde s < T(\tilde \omega)}  \, \int_0^\nu \dd t \, \Bigl|(v^L-v)\big|_{\Lambda_L}\Bigr|\,(\omega(s + t) - \tilde \omega(\tilde s + t))\,.
\end{multline}
Namely, $\tilde{\omega}(\cdot)-\omega(\cdot)$ always lies within $\Lambda_L$ when the $\tilde{\omega}(\cdot)$ and $\omega(\cdot)$ lie within $\Lambda_{L_1}$.
Similarly
\begin{multline}
\label{zeta_restriction}
\Bigl(\bigl|\zeta^L(\omega,\tilde{\omega})|+|\zeta^{\infty}(\omega,\tilde{\omega})\bigr|\Bigr)\,\ind{\cal{A}^L} (\omega) \,\ind{\cal{A}^L} (\tilde \omega)\leq 
\\
\leq \nu \sum_{s \in \nu \N} \ind{s < T(\omega)} \, \sum_{\tilde s \in \nu \N} \ind{\tilde s < T(\tilde \omega)}  \, \int_0^\nu \dd t \,\Bigl(\big|v^L|_{\Lambda_L}\big|+\big|v|_{\Lambda_L}\big|\Bigr)\,(\omega(s + t) - \tilde \omega(\tilde s + t))\,.
\end{multline}

We recall \eqref{A^{L}_B^{L}}, and use \eqref{Ursell_function_difference}--\eqref{Ursell_function_difference_2}, \eqref{zeta_difference_2}--\eqref{zeta_restriction}, combined with the integration algorithm from Lemma \ref{tree_integration} to deduce that

\begin{multline}
\label{(ii)_star_proof4_circle_2_C}
\nu^p\,\biggl\|\int \hat \mu^{\infty}_{y_1,x_1}(\dd \omega_1) \cdots \hat \mu^{\infty}_{y_p,x_p}(\dd \omega_p)\,\mu^{\infty}(\dd \omega_{p+1}) \cdots \mu^{\infty}(\dd \omega_n)\,\prod_{k=1}^{n}\ind{\cal A^{L}}(\omega_k)
\\
\times \Bigl[\prod_{\{i,j\} \in \cal G}\zeta^L(\omega_i,\omega_j)-\prod_{\{i,j\} \in \cal G}\zeta^{\infty}(\omega_i,\omega_j)
\Bigr]
\biggr\|_{\ell^{\infty}_{\b x}\ell^1_{\b y}} \lesssim_{n,\kappa} \bigl(\|v^L\|_{\ell^1(\Lambda_L)}+\|v\|_{\ell^1(\Lambda_L)}\bigr)^{n-1}\,\|v-v^L\|_{\ell^1(\Lambda_L)}
\\
\lesssim \|v\|_{\ell^1(\Z^d)}^{n-1}\,\|v-v^L\|_{\ell^1(\Lambda_L)} \rightarrow 0
\quad \text{as } L \rightarrow \infty\,.
\end{multline}
In the second estimate, we recalled \eqref{v^{L}_ell^1}  and \eqref{v-v^L_ell^1}. We hence obtain \eqref{(ii)_star_proof4_circle_2}.

Combining \eqref{(ii)_star_proof4_circle_1} and \eqref{(ii)_star_proof4_circle_2}, we deduce that \eqref{(ii)_star_proof4} holds. By construction, the convergence is uniform in $\nu$ and $L_0$. Claim (ii) now follows.

We now show (i). The arguments are similar to that of (ii). We now outline the main differences. By the cluster expansion in Proposition \ref{gamma_p_cluster_expansion} (i), we can write 
\begin{multline}
\label{f_v_i_v_1}
g^{\nu,\kappa,\nu^2,L}=\frac{1}{|\Lambda_L|} \log \mathcal{Z}^{\nu,\kappa,\nu^2,L}=
\frac{1}{|\Lambda_L|} \int_{\Lambda_L} \dd {x}_{1}\,\sum_{n=2}^{\infty} \frac{1}{n!}\, \nu \sum_{T_1 \in \nu \N^*} \frac{\ee^{-\kappa T_1}}{T_1}\, \int {\bb W}^{L,T_1}_{x_1,x_1}(\dd \omega_1)\, 
\\
\times \ee^{-\mathcal{V}^{\nu,\nu^2,L}({\omega}_{1},{\omega}_{1})/2}\, {\mu}^{L}(\dd \omega_2) \cdots {\mu}^{L}(\dd \omega_n)\, {\varphi}^{L}({\omega}_{1},\ldots,\omega_n)\,.
\end{multline}
We analyse \eqref{f_v_i_v_1} using several steps.

The first step is to reduce to integrating in points $x_1$ which are away from the boundary of $\Lambda_L$. To this end, let $\varepsilon>0$ be small which we henceforth fix. We let 
\begin{equation}
\label{L_epsilon}
L_{\varepsilon} \deq \lfloor (1-\varepsilon)L \rfloor\,.
\end{equation}
Using \eqref{L_epsilon}, let us rewrite \eqref{f_v_i_v_1} using
\begin{equation}
\label{x_1_average}
\frac{1}{|\Lambda_L|} \int_{\Lambda_L} \dd {x}_{1}=
\frac{1}{|\Lambda_L|} \int_{
\Lambda_{L_{\varepsilon}}} \dd {x}_{1}+\frac{1}{|\Lambda_L|} \int_{\Lambda_L \setminus \Lambda_{L_{\varepsilon}}} \dd {x}_{1}\,.
\end{equation}
An analogous argument as the proof of Lemma \eqref{integration_lemma} (iii) shows that 
\begin{equation}
\label{f_v_i_v_2}
\int_{\Lambda_L \setminus \Lambda_{L_{\varepsilon}}} \nu \sum_{T \in \nu \N^*} \frac{z^T}{T} \int {\bb W}^{L,T}_{x,x}(\dd 
\omega) \, T^q \lesssim \ \frac{(q-1)!}{\kappa^q}\ \left| \Lambda_L \setminus \Lambda_{L_{\varepsilon}} \right| \lesssim 
 \nu^q\, \frac{(q-1)!}{\kappa^q}\ |\Lambda_L|\,\varepsilon^d\,.
\end{equation}
for all $q \in \N$. 

We now apply the integration algorithm from Lemma \ref{tree_integration} as in the proof of Proposition  \ref{operator_norm_bound} (i) in such a way that we integrate out the root of the tree using \eqref{f_v_i_v_2} instead of Lemma \ref{integration_lemma}  (iii). 
We hence obtain \begin{equation}
\label{f_v_i_v_3}
g^{\nu,\kappa,\nu^2,L}=I^{\nu,\kappa,\nu^2,L}+ \mathcal{O}_{d,\kappa,v} (\varepsilon)\,,
\end{equation}
uniformly in $\nu$,
where
\begin{multline}
\label{f_v_i_v_4}
I^{\nu,\kappa,\nu^2,L} \deq \
\frac{1}{|\Lambda_L|} \int_{\Lambda_{L_{\varepsilon}}}  \dd {x}_{1}\,\sum_{n=2}^{\infty} \frac{1}{n!}\, \nu \sum_{T_1 \in \nu \N^*} \frac{\ee^{-\kappa T_1}}{T_1}\, \int {\bb W}^{L,T_1}_{x_1,x_1}(\dd \omega_1)\, 
\\
\times \ee^{-\mathcal{V}^{\nu,\nu^2,L}({\omega}_{1},{\omega}_{1})/2}\, {\mu}^{L}(\dd \omega_2) \cdots {\mu}^{L}(\dd \omega_n)\, {\varphi}^{L}({\omega}_{1},\ldots,\omega_n)\,. 
\end{multline}

The second step of the analysis is to reduce to integrating over loops $\omega_j$ which are close to $x_1$.
In order to set this up, given $n \in \N^*$, we define the set 
\begin{multline}
\label{f_v_i_v_5}
\mathcal{C}_{n}^{L,\varepsilon} \deq \Biggl\{({\omega}_{1}, \ldots,{\omega}_{n}) \in (\Omega^L)^n,\, \,\, |\omega_k(t_{k})-\omega_{\ell}(t_{\ell})|=|\omega_k(t_{k})-\omega_{\ell}(t_{\ell})|_{L} \leq \frac{L\varepsilon}{2}\,
\\
\forall 1 \leq k,\ell \leq n\,,\,\, \forall t_{k} \in [0,T(\omega_k)]\,,\,\, \forall t_{\ell} \in [0,T(\omega_{\ell})]\Biggr\}\,.
\end{multline}
Using \eqref{f_v_i_v_4}--\eqref{f_v_i_v_5}, and arguing analogously as for 
\eqref{(ii)_star_proof1}--\eqref{(ii)_star_proof2}, we have that
\begin{equation}
\label{f_v_i_v_6}
I^{\nu,\kappa,\nu^2,L}=II^{\nu,\kappa,\nu^2,L}+ {A}_{1}(\nu,L,\varepsilon)\,,\quad |{A}_{1}(\nu,L,\varepsilon)| \lesssim_{d,\kappa,v} o_{L}^{\varepsilon}(1)\,,
\end{equation}
where
\begin{multline}
\label{f_v_i_v_7}
II^{\nu,\kappa,\nu^2,L} \deq \
\frac{1}{|\Lambda_L|} \int_{\Lambda_{L_{\varepsilon}}} \dd {x}_{1}\,\sum_{n=2}^{\infty} \frac{1}{n!}\, \nu \sum_{{T}_{1} \in \nu \N^{*}} \frac{\ee^{-\kappa T_{1}}}{T_{1}}\, \int {\bb W}^{L,T_1}_{x_1,x_1}(\dd {\omega}_{1})\, 
\\
\times \ee^{-\mathcal{V}^{\nu,\nu^2,L}({\omega}_{1},{\omega}_{1})/2}\, {\mu}^{L}(\dd \omega_2) \cdots {\mu}^{L}(\dd {\omega}_{n})\, {\varphi}^{L}({\omega}_{1},\ldots,{\omega}_{n})\,\mathbf{1}_{\mathcal{C}_{n}^{L,\varepsilon}}({\omega}_{1},\ldots,{\omega}_{n})\,.
\end{multline}
In \eqref{f_v_i_v_7} as well as in the sequel, we denote by $o_{L}^{\varepsilon}(1)$ a quantity which tends to zero as $L \rightarrow \infty$ depending on $\varepsilon$ (which in our argument is fixed).

In the third step, we replace the interaction potential $v^L$ by $v^{\infty} \equiv v$.
We argue analogously as for \eqref{(ii)_star_proof4_circle_1}--\eqref{(ii)_star_proof4_circle_2} and recall \eqref{mu_measure_1_*}, \eqref{integral_over_omega_ell_closed}--\eqref{integral_over_omega_ell_open}, \eqref{(ii)_star_proof4_circle_2_C} to deduce that
\begin{equation}
\label{f_v_i_v_8}
II^{\nu,\kappa,\nu^2,L} =III^{\nu,\kappa,\nu^2,L} +\mathcal{O}_{d,\kappa,v}\left(\|v^L-v\|_{L^1(\Lambda_L)}\right)\,,
\end{equation}
where
\begin{multline}
\label{f_v_i_v_9}
III^{\nu,\kappa,\nu^2,L} \deq \
\frac{1}{|\Lambda_L|} \int_{\Lambda_{L_{\varepsilon}}} \dd {x}_{1}\,\sum_{n=2}^{\infty} \frac{1}{n!}\, \nu \sum_{T_1 \in \nu \N^*} \frac{\ee^{-\kappa T_1}}{T_1}\, \int {\bb W}^{L,T_1}_{x_1,x_1}(\dd {\omega}_{1})\, 
\\
\times \ee^{-\mathcal{V}^{\nu,\nu^2,\infty}({\omega}_{1},\omega_{1})/2}\, \mu^{L,*}(\dd \omega_2) \cdots \mu^{L,*}(\dd {\omega}_{n})\, \varphi^{\infty}({\omega}_{1},\ldots,{\omega}_{n})\,\prod_{i=1}^{n} \mathbf{1}_{\mathcal{C}^{L,\varepsilon}}(\omega_i)\,.
\end{multline}
For \eqref{f_v_i_v_8}--\eqref{f_v_i_v_9}, we used (as for \eqref{(ii)_star_proof4_circle_1}--\eqref{(ii)_star_proof4_circle_2} in the proof of part (ii)) the observation that 
\begin{equation*}
\omega_{k}(\cdot)-\omega_{\ell}(\cdot) \in \Lambda_L \qquad \forall 1 \leq k, \ell \leq n\,,
\end{equation*} 
whenever $({\omega}_{1},\ldots,{\omega}_{n}) \in \mathcal{C}_{n}^{L,\varepsilon}$.

By \eqref{f_v_i_v_5} and the assumption that ${\omega}_{1}(0)=x_1 \in \Lambda_{L_{\varepsilon}}$, we have that
\begin{equation}
\label{f_v_i_v_10}
\omega_{k}: [0,T(\omega_{k})] \rightarrow \Bigl(1-\frac{\varepsilon}{2}\Bigr)  \,\Lambda_L \cap \Z^d \quad \forall k \in \N^*
\end{equation}
in \eqref{f_v_i_v_9}. By \eqref{f_v_i_v_10}, we can erase the boundary effects in \eqref{f_v_i_v_9}. In particular, we have that
\begin{multline}
\label{f_v_i_v_11}
III^{\nu,\kappa,\nu^2,L} 
=
\frac{1}{|\Lambda_L|} \sum_{n=2}^{\infty} \frac{1}{n!}\, \int_{\Lambda_{L_{\varepsilon}}} \dd {x}_{1}\, \int_{\Lambda_L} \dd x_2\, \cdots \int_{\Lambda_L} \dd x_n
\\
\prod_{j=1}^{n} \left[\nu \sum_{T_j \in \nu \N^*} \frac{\ee^{-\kappa T_j}}{T_j}\, \int {\bb W}^{\infty,T_j}_{x_j,x_j}(\dd \omega_{j})\, 
\ee^{-\mathcal{V}^{\nu,\nu^2,\infty}({\omega}_{j},\omega_{j})/2}\,  \right]
\,\varphi^{\infty}({\omega}_{1},\ldots,{\omega}_{n})\,
\prod_{i=1}^{n} \mathbf{1}_{\mathcal{C}^{L,\varepsilon}}(\omega_i)\,.
\end{multline}

In the fourth step, we relax the assumption that the loops $\omega_j$ are close to $x_1$.
To this end, we observe that
\begin{multline}
\label{f_v_i_v_12_C}
\frac{1}{|\Lambda_L|} \sum_{n=1}^{\infty} \frac{1}{n!}\, \int_{\Lambda_{L_{\varepsilon}}} \dd {x}_{1}\, \int_{\Z^d} \dd x_2\, \cdots \int_{\Z^d} \dd x_n
\\
\prod_{j=1}^{n} \left[\nu \sum_{T_j \in \nu \N^*} \frac{\ee^{-\kappa T_j}}{T_j}\, \int {\bb W}^{\infty,T_j}_{x_j,x_j}(\dd \omega_{j})\, 
\ee^{-\mathcal{V}^{\nu,\nu^2,\infty}({\omega}_{j},\omega_{j})/2}\right]
\,\left|\varphi^{\infty}({\omega}_{1},\ldots,{\omega}_{n})\right|\,
\left(1-\prod_{i=1}^{n} \mathbf{1}_{\mathcal{C}^{L,\varepsilon}}(\omega_i)\right)
\\
\lesssim_{d,\kappa,v} o_{L}^{\varepsilon}(1)\,.
\end{multline}

In order to show \eqref{f_v_i_v_12_C}, let first fix $n \geq 1$. We apply the tree bound from Lemma \ref{Ursell_function_estimate} to estimate $\left|\varphi^{\infty}({\omega}_{1},\ldots,{\omega}_{n})\right|$ in \eqref{f_v_i_v_12_C}. When estimating the contribution from the trees, we always choose the root to be the vertex corresponding to the label $1$, i.e.\ ${\omega}_{1}$.
When integrating with respect to ${\omega}_{1}$ using the integration algorithm from Lemma \ref{tree_integration}, we use one of the following estimates.
\begin{equation}
\label{root_1_omega_1}
\int_{\Lambda_{L_{\varepsilon}}} \dd {x}_{1}\,\nu \sum_{T_1 \in \nu \N^*} \frac{\ee^{-\kappa T_1}}{T_1}\, \int {\bb W}^{\infty,T_1}_{x_1,x_1}(\dd {\omega}_{1})\, 
\ee^{-\mathcal{V}^{\nu,\infty}({\omega}_{1})}\,T_1^q \lesssim_{\zeta,q,d} |\Lambda_{L_{\varepsilon}}|\,.
\end{equation}
\begin{equation}
\label{root_2_omega_1}
\frac{1}{|\Lambda_{L_{\varepsilon}}|}\,\int_{\Lambda_{L_{\varepsilon}}} \dd {x}_{1}\,\nu \sum_{T_1 \in \nu \N^*} \frac{\ee^{-\kappa T_1}}{T_1}\, \int {\bb W}^{\infty,T_1}_{x_1,x_1}(\dd {\omega}_{1})\, 
\ee^{-\mathcal{V}^{\nu,\infty}({\omega}_{1})}\,\mathbf{1}_{\mathcal{D}^{\infty,L}_{c\varepsilon}}({\omega}_{1})\,T_1^q  \lesssim_{\zeta,q,d} o^{\varepsilon}_{L}\left(1\right)\,.
\end{equation}
In \eqref{root_2_omega_1}, we write
\begin{equation*}
\label{D^{L,tilde_L}_c}
\mathcal{D}^{\infty,L}_c:=\Bigl\{\omega \in \bigcup_{T>0} \Omega^{\infty,T}\,,\,|\omega(t)-\omega(s)| \geq cL \,\,\text{for some} \,\,s,t \in [0,T(\omega)]\Bigr\}\,.
\end{equation*}
We obtain the estimate \eqref{root_1_omega_1} by arguing as for Lemma \ref{integration_lemma} (iii). The estimate \eqref{root_2_omega_1} follows by arguing as for Lemma \ref{integration_lemma_M} (iii).

Recalling \eqref{f_v_i_v_5}, if $({\omega}_{1},\ldots,{\omega}_{n}) \notin \mathcal{C}_{n}^{L,\varepsilon}$, then there exist $1 \leq k, \ell \leq n$ and  $t_{k} \in [0,T(\omega_{k})], t _{\ell} \in [0,T(\omega_{\ell})]$ such that 
\begin{equation}
\label{distance_paths_L_delta_2}
|\omega_k(t_{k})-\omega_{\ell}(t_{\ell})| > L \varepsilon/2\,.
\end{equation}
Using \eqref{distance_paths_L_delta_2}, followed by \eqref{root_1_omega_1}--\eqref{root_2_omega_1}, and suitable modifications of Lemmas \ref{integration_lemma}, \ref{integration_lemma_M}, \ref{integration_lemma_M_2}, the arguments for \eqref{f_v_i_v_6}--\eqref{f_v_i_v_7} indeed imply \eqref{f_v_i_v_12_C}.

Using \eqref{f_v_i_v_12_C}, it follows that
\begin{equation}
\label{f_v_i_v_12}
III^{\nu,\kappa,\nu^2,L}=IV^{\nu,\kappa,\nu^2,L}+ C_2(\nu,L,\varepsilon)\,, \qquad |C_2(\nu,L,\varepsilon)| \lesssim_{d,\kappa,v} o_{L}^{\varepsilon}(1)\,,
\end{equation}
where
\begin{multline}
\label{f_v_i_v_12_B}
IV^{\nu,\kappa,\nu^2,L} \deq \frac{1}{|\Lambda_L|} \sum_{n=1}^{\infty} \frac{1}{n!}\, \int_{\Lambda_{L_{\varepsilon}}} \dd {x}_{1}\, \int_{\Lambda_L} \dd x_2\, \cdots \int_{\Lambda_L} \dd x_n
\\
\prod_{j=1}^{n} \left[\nu \sum_{T_j \in \nu \N^*} \frac{\ee^{-\kappa T_j}}{T_j}\, \int {\bb W}^{\infty,T_j}_{x_j,x_j}(\dd \omega_{j})\, 
\ee^{-\mathcal{V}^{\nu,\nu^2,\infty}({\omega}_{j},\omega_{j})/2}\right]
\,\varphi^{\infty}({\omega}_{1},\ldots,{\omega}_{n})\,.
\end{multline}

In the fifth step, we integrate in space over all of $\Z^d$.
We define
\begin{multline}
\label{f_v_i_v_12_D}
V^{\nu,\kappa,\nu^2,L} \deq \frac{1}{|\Lambda_L|} \sum_{n=1}^{\infty} \frac{1}{n!}\, \int_{\Lambda_{L_{\varepsilon}}} \dd {x}_{1}\, \int_{\Z^d} \dd x_2\, \cdots \int_{\Z^d} \dd x_n
\\
\prod_{j=1}^{n} \left[\nu \sum_{T_j \in \nu \N^*} \frac{\ee^{-\kappa T_j}}{T_j}\, \int {\bb W}^{\infty,T_j}_{x_j,x_j}(\dd \omega_{j})\, 
\ee^{-\mathcal{V}^{\nu,\nu^2,\infty}({\omega}_{j},\omega_{j})/2}\right]
\,\varphi^{\infty}({\omega}_{1},\ldots,{\omega}_{n})\,.
\end{multline}
By \eqref{f_v_i_v_12_B}--\eqref{f_v_i_v_12_D} and a telescoping argument, it follows that
\begin{multline}
\label{f_v_i_v_12_E}
\left| IV^{\nu,\kappa,\nu^2,L} -V^{\nu,\kappa,\nu^2,L} \right| \leq  \frac{1}{|\Lambda_L|} \sum_{n=1}^{\infty} \frac{(n-1)}{n!}\, \int_{\Lambda_{L_{\varepsilon}}} \dd {x}_{1}\, \int_{\Z^d \setminus \Lambda_L} \dd x_2\, \int_{\Z^d} \dd x_3\, \cdots \int_{\Z^d} \dd x_n
\\
\prod_{j=1}^{n} \left[\nu \sum_{T_j \in \nu \N^*} \frac{\ee^{-\kappa T_j}}{T_j}\, \int {\bb W}^{\infty,T_j}_{x_j,x_j}(\dd \omega_{j})\, 
\ee^{-\mathcal{V}^{\nu,\nu^2,\infty}({\omega}_{j},\omega_{j})/2}\right]
\,\left|\varphi^{\infty}({\omega}_{1},\ldots,{\omega}_{n})\right|\,.
\end{multline}
In the integral on the right-hand side of \eqref{f_v_i_v_12_E}, we note that
\begin{equation}
\label{f_v_i_v_12_F}
\bigl|{\omega}_{1}(0)-{\omega}_{2}(0)\bigr|=\bigl|{x}_{1}-{x}_{2}\bigr| \geq L \varepsilon\,.
\end{equation}
Using \eqref{f_v_i_v_12_F} in \eqref{f_v_i_v_12_E}, and arguing as for \eqref{f_v_i_v_12_C}, we obtain that
\begin{equation}
\label{f_v_i_v_12_G}
\left| IV^{\nu,\kappa,\nu^2,L} -V^{\nu,\kappa,\nu^2,L} \right| \lesssim_{d,\kappa,v} o_{L}^{\varepsilon}(1)\,.
\end{equation}

In the sixth step, we use translation invariance to conclude the proof.
Namely, noting that the integrand in \eqref{f_v_i_v_12_D} does not depend on $x_1$, we have that
\begin{multline}
\label{f_v_i_v_12_H}
V^{\nu,\kappa,\nu^2,L} = \frac{|\Lambda_{L_{\varepsilon}}|}{|\Lambda_L|}\,\sum_{n=1}^{\infty} \frac{1}{n!}  \int_{\Z^d} \dd x_2\, \cdots \int_{\Z^d} \dd x_n\,\Biggl[\nu \sum_{T_j \in \nu \N^*} \frac{\ee^{-\kappa T_1}}{T_1}\, \int {\bb W}^{\infty,T_1}_{0,0}(\dd {\omega}_{1})\ee^{-\mathcal{V}^{\nu,\infty}({\omega}_{1})}\Biggr]\, 
\\
\times
\prod_{j=2}^{n} \Biggl[\nu \sum_{T_j \in \nu \N^*} \frac{\ee^{-\kappa T_j}}{T_j}\, \int {\bb W}^{\infty,T_j}_{x_j,x_j}(\dd \omega_{j})\, 
\ee^{-\mathcal{V}^{\nu,\nu^2,\infty}({\omega}_{j},\omega_{j})/2}\Biggr]
\,\varphi^{\infty}({\omega}_{1},\ldots,{\omega}_{n})\,.
\end{multline}
By Proposition \ref{operator_norm_bound} (i), we know that $g^{\nu,\kappa,\nu^2,L}$ is bounded uniformly in $\nu \in (0,1/\kappa]$. Using \eqref{f_v_i_v_3}, \eqref{f_v_i_v_6}, \eqref{f_v_i_v_8}, \eqref{f_v_i_v_12}, and \eqref{f_v_i_v_12_G}, we deduce that the same is true for \eqref{f_v_i_v_12_H}.
We now use \eqref{f_v_i_v_3}, \eqref{f_v_i_v_6}, \eqref{f_v_i_v_8}, \eqref{f_v_i_v_12} (combined with \eqref{v-v^L_ell^1}), \eqref{f_v_i_v_12_G}, and \eqref{f_v_i_v_12_H} 
and we conclude that
\begin{multline}
\label{f_v_i_v_conclusion}
\lim_{L \rightarrow \infty} g^{\nu,\kappa,\nu^2,L}=\sum_{n=1}^{\infty} \frac{1}{n!}  \int_{\Z^d} \dd x_2\, \cdots \int_{\Z^d} \dd x_n\,\left[\nu \sum_{T_j \in \nu \N^*} \frac{\ee^{-\kappa T_1}}{T_1}\, \int {\bb W}^{\infty,T_1}_{0,0}(\dd {\omega}_{1})\ee^{-\mathcal{V}^{\nu,\infty}({\omega}_{1})}\right]\, 
\\
\prod_{j=2}^{n} \left[\nu \sum_{T_j \in \nu \N^*} \frac{\ee^{-\kappa T_j}}{T_j}\, \int {\bb W}^{\infty,T_j}_{x_j,x_j}(\dd \omega_{j})\, 
\ee^{-\mathcal{V}^{\nu,\nu^2,\infty}({\omega}_{j},\omega_{j})/2}\right]
\,\varphi^{\infty}({\omega}_{1},\ldots,{\omega}_{n})\,,
\end{multline}
uniformly in $\nu \in (0,1/\kappa]$. This concludes the proof of claim (i) and the result follows.
\end{proof}

Let us recall a general fact about interchanging the order of limits, whose proof we omit.
\begin{lemma} 
\label{convergence_general_fact}
Suppose that $\rho: \N^* \times (0,\infty) \rightarrow \C$ is a function that satisfies the following properties. 
\begin{itemize}
\item[(1)] There exists $\nu_0>0$ such that the limit
\begin{equation}
\label{convergence_general_fact_(1)}
\lim_{L \rightarrow \infty} \rho(L,\nu)=:\rho(\infty,\nu)\,,
\end{equation}
exists uniformly in $\nu \in (0,\nu_0]$.
\item[(2)] For all $L \in \N^* $, there exists $\rho(L,0)$ such that 
\begin{equation}
\label{convergence_general_fact_(2)}
\lim_{\nu \rightarrow 0} \rho(L,\nu)=\rho(L,0)\,.
\end{equation}
\end{itemize}
Then, the following properties hold.
\begin{itemize}
\item[(i)] $\rho(\infty,0) \deq \lim_{L \rightarrow \infty} \rho(L,0)$ exists. 
\item[(ii)] With $\rho(\infty,0)$ given as in (i), we have that
$\lim_{\nu \rightarrow 0}\rho(\infty,\nu)=\rho(\infty,0)$.
\end{itemize}
\end{lemma}

We now have the necessary tools to prove Theorem \ref{Infinite_volume_theorem_1}.

\begin{proof}[Proof of Theorem \ref{Infinite_volume_theorem_1}]
We first prove claim (i). This follows from Theorem \ref{mean_field_convergence} (i), Proposition \ref{MF_cluster_expansion} (i), and Lemma \ref{convergence_general_fact} by taking $\rho(L,\nu)=g^{\nu,\kappa,\nu^2,L}$ for $\nu \in (0,1/\kappa]$ and $\rho(L,0)=g^{\cl,L}$ with $\nu_0=1/\kappa$. 

In order to prove claim (ii), we also apply a suitable modification of Lemma \ref{convergence_general_fact} where all of the convergence is taken in the norm $\|\cdot\|_{L_0,p}$ given by \eqref{L_0_p_norm}. Namely, we first use Proposition \ref{MF_cluster_expansion} (ii) and Theorem \ref{mean_field_convergence} (ii) to note that 
assumptions (1) and (2) of Lemma \ref{convergence_general_fact} hold if we take $\rho(L,\nu)=\nu^p\,\Gamma^{\nu,\kappa,\nu^2,L}_p$, $\rho(L,0)=\Gamma^{\cl,L}_{p}$ and if the convergence in \eqref{convergence_general_fact_(1)}--\eqref{convergence_general_fact_(2)} is interpreted with respect to $\|\cdot\|_{L_0,p}$.
By Lemma \ref{convergence_general_fact} (i), we deduce that the limit 
$\lim_{L \rightarrow \infty} \gamma^{L}_{p}$ exists in $\|\cdot\|_{L_0,p}$. 
Note that, a priori this quantity depends on $L_0$ (since the norm $\|\cdot\|_{L_0,p}$ depends on $L_0$). However, by recalling \eqref{Pi_{L,p}}--\eqref{L_0_p_norm} and by construction, it follows that this limit is independent of $L_0$.
We conclude the result of claim (ii) from Lemma \ref{convergence_general_fact} (ii).
\end{proof}

\begin{remark}
In the noninteracting case $v = 0$, the measure $\mu \equiv \mu^{L}$ given in \eqref{mu_measure_1} is a discrete-time version of Lawler's and Werner's loop soup intensity measure \cite{Lawler_Werner}, to which it converges as $\nu \to 0$.  Indeed, for $v = 0$ the measure $\mu(\dd \omega)$ converges as $\nu \to 0$ to
$\xi(\dd \omega) = \int_0^\infty \dd T \, \frac{\ee^{-\kappa T}}{T} \,  \bb W^{T}(\dd \omega)$,
which is precisely the intensity measure of the loop soup (a Poisson process with intensity $\xi$). 
\end{remark}

\subsection{Infinite-volume limit of the specific relative Gibbs potential and reduced density matrices II: the large-mass regime}
\label{Infinite_volume_limit_large_mass}

In this subsection, we work in the large-mass regime.
Throughout, we assume that the interaction potential on $\Lambda_L$ is given by \eqref{v^{L}} for $v$ as in Assumption \ref{interaction_potential_v_m_infty}.
We recall that we are considering the parameters $\kappa=\kappa_0/\nu$ for fixed $\kappa_0$ and $\lambda=1$ and that the many-body Hamiltonian is given by \eqref{H_n_mass_m}.

Recalling the definition of $\wt{\cal V}^{\nu,1} (\omega) \equiv \wt{\cal V}^{\nu,1,L} (\omega)$ with interaction $v^L$ given by \eqref{V_interaction_1_large_mass_B}, we define the self-interaction
\begin{equation}
\label{tilde_V_m}
\wh{\cal V}^{\nu,1,L}(\omega) \deq \wt{\cal V}^{\nu,1,L} (\omega)+\frac{1}{\nu}\,v^L(0) T(\omega) \,\ind{R=0} \geq 0 \,.
\end{equation}
Furthermore, we modify the definition of the measures \eqref{mu_measure_1}--\eqref{mu_measure_2} according to 
\begin{align}
\label{mu_measure_large_mass_1}
&{\mu}^{L}(\dd \omega) \deq \nu \sum_{T\in \nu \N^*}\frac{\ee^{-\kappa_0 T/\nu}}{T} \,  \bb W^{L,T}(\dd \omega)
\,\ee^{-\wh{\cal V}^{\nu,1,L}(\omega)/2}\,, 
\\
\label{mu_measure_large_mass_2}
&\hat {\mu}^{L}_{y,x}(\dd \omega) \deq \sum_{T \in \nu \N^*} \ee^{-\kappa_0 T/\nu} \, \bb W^{L,T}_{y,x}(\dd \omega)\,\ee^{-\wh{\cal V}^{\nu,1,L}(\omega)/2}\,.
\end{align}
With $v$ and $R$ as in Assumption \ref{interaction_potential_v_m_infty}, we define $v^{(1)}, v^{(2)}: \Z^d \rightarrow [0,\infty]$ by
\begin{equation}
\label{v^{(j)}}
v^{(1)} \deq v \,\ind{|x| <R}\,, \qquad v^{(2)} \deq v-v^{(1)}\,.
\end{equation}
Note that, by Assumption \ref{interaction_potential_v_m_infty} (i), we have that 
\begin{equation}
\label{v^{(2)}_ell^1}
v^{(2)} \in \ell^1(\Z^d)\,.
\end{equation}
With notation as in \eqref{v^{(j)}}, we define $v^{L,(1)}, v^{L,(2)}: \Lambda_L \rightarrow [0,\infty]$ by
\begin{equation}
\label{v^{L,(j)}}
v^{L,(j)}(x) \deq \sum_{k \in (L\Z)^d}v^{(j)}(x+k)\,,\quad j=1,2\,.
\end{equation}
Throughout, we assume that
\begin{equation} 
\label{tilde_kappa}
\kappa_0 > 1\,,\quad \nu \leq 1\,.
\end{equation}
With notation as in \eqref{mu_measure_large_mass_1}--\eqref{v^{(j)}}, and assuming \eqref{tilde_kappa}, we note the following analogue of Lemma \ref{integration_lemma}.
\begin{lemma}
\label{integration_lemma_large_mass}
Let $\omega \in \Omega^{L,T(\omega)}$ with $T(\omega) \in \nu \N^*$, $q \in \N$, and $x \in \Lambda_L$ be given.
Then, the following estimates hold.
\begin{itemize}
\item[(i)] $\int \mu^{L} (\dd \tilde \omega)\,  T(\tilde \omega)^{q} \,|\zeta^L(\omega,\tilde \omega)| \lesssim \frac{T(\omega)}{\kappa_0^{q+1}}\,q!\,\bigl(1+\|v^{(2)}\|_{\ell^1}\bigr)\,\nu^{q-1}\,.$
\item[(ii)] $\int_{\Lambda_L}\dd y\,\int \hat{\mu}^{L}_{y,x} (\dd\tilde \omega) \, T(\tilde \omega)^{q}\,|\zeta^L(\omega,\tilde \omega)| \lesssim \frac{T(\omega)}{\kappa_0^{q+2}}\,(q+1)!\,\bigl(1+\|v^{(2)}\|_{\ell^1}\bigr)\,\nu^{q-1}.$
\item[(iii)] $\int \mu^{L} (\dd \tilde \omega) \,T(\tilde \omega)^{q} \lesssim  \frac{(q-1)!}{\kappa_0^{q}}\,|\Lambda_L|\,\nu^{q}$ if $q\in \N^*$.
\item[(iv)] $\int_{\Lambda_L} \dd y\,\int \hat{\mu}^{L}_{y,x}(\dd \tilde \omega) \,T(\tilde \omega)^{q} \lesssim \frac{q!}{\kappa_0^{q+1}}\,\nu^{q}.$
\end{itemize}
Here we recall the definition \eqref{Ursell_function}. 
\end{lemma}
\begin{proof}
We first prove (i).
We let $\cal V^{\nu,1,L,(2)}(\omega,\tilde \omega)$ be given as in \eqref{V_interaction_1}, where we replace $v \equiv v^L$ by $v^{L,(2)}$ as given in \eqref{v^{L,(j)}} above. 
We define the set $\cal I(\omega,\nu)$ as
\begin{equation}
\label{cal{I}(omega,m)}
\cal I(\omega,\nu) \deq \Biggl\{\tilde \omega \in \bigcup_{\tilde T>0}\Omega^{L,\tilde T}\,,\;
\exists t \in [0,T(\omega)]\; \exists \tilde t \in [0,T(\tilde \omega)]\,,\,\omega( t )=\tilde \omega( \tilde t\,)\,,\,t-\tilde t \in \nu\Z \Biggr\}\,.
\end{equation}
In other words $\tilde \omega \in \cal I(\omega,\nu)$ if and only if $\omega$ and $\tilde \omega$ intersect at times which are equal modulo $\nu\Z$.
By Assumption \ref{interaction_potential_v_m_infty}, the construction of $\cal V^{\nu,1,L,(2)}(\omega,\tilde \omega)$ and \eqref{v^{(j)}}, \eqref{cal{I}(omega,m)}, we have
\begin{equation}
\label{zeta^L_bound}
|\zeta^L(\omega,\tilde \omega)|\leq \ind{\cal I(\omega,\nu)}(\tilde \omega)+
\cal V^{\nu,1,L,(2)}(\omega,\tilde \omega)\,.
\end{equation}
We now estimate the contribution to the left-hand side of (i) coming from each of the expressions in the bound \eqref{zeta^L_bound}. For the first term, we use \eqref{tilde_V_m}--\eqref{mu_measure_large_mass_1} to write
\begin{equation*}
\int \mu^{L} (\dd \tilde \omega)\,  T(\tilde \omega)^{q}\,\ind{\cal I(\omega,\nu)}(\tilde \omega)
\leq \nu\sum_{\tilde T \in \nu\N^*} \ee^{-\kappa_0 \tilde T/\nu}\,\tilde T^{q-1}\,\int \bb W^{L,\tilde T}_{y,x}(\dd \tilde \omega)\,\ind{\cal I(\omega,\nu)}(\tilde \omega)\,,
\end{equation*}
which, recalling \eqref{cal{I}(omega,m)} is
\begin{align}
\notag
\leq \nu\sum_{\tilde T \in \nu\N^*} \ee^{-\kappa_0 \tilde T/\nu}\,\tilde T^{q-1}\,
\sum_{r \in \nu\N} \ind{r<T(\omega)} &\sum_{s \in \nu\N} \ind{s<\tilde T}\,\int_0^{\nu} \dd t
\\
\label{integration_lemma_large_mass_i_1}
&\int_{\Lambda_L}\dd x\, \psi^{L,t+s} \bigl(\omega(t+r)-x\bigr)\,\psi^{L,\tilde T-(t+s)}\bigl(x-\omega(t+r)\bigr)
\,.
\end{align}
By using Lemma \ref{estimates_heat_kernel} (i) and (iii), we note that \eqref{integration_lemma_large_mass_i_1} is
\begin{equation}
\label{integration_lemma_large_mass_i_2}
\leq \sum_{\tilde T \in \nu\N^*} \ee^{-\kappa_0 \tilde T/\nu}\, \tilde T^q\, T(\omega)
\leq \nu^{q} \,T(\omega)\,\sum_{j=1}^{\infty} \ee^{-\kappa_0 j}\,j^q \lesssim  \frac{T(\omega)}{\kappa_0^{q+1}}\,q!\,\nu^{q}\,.
\end{equation}
In order to obtain \eqref{integration_lemma_large_mass_i_2}, we recalled \eqref{tilde_kappa} 
and used Lemma \ref{Riemann_sum_lemma}.

For the second term, we argue argue analogously as in \eqref{integration_lemma_i_1}--\eqref{integration_lemma_i_3a} with $v$ replaced by $v^{L,(2)}$ and deduce that
\begin{equation}
\label{integration_lemma_large_mass_i_3}
\int \mu^{L} (\dd \tilde \omega)\,  T(\tilde \omega)^{q}\,\cal V^{\nu,1,L,(2)}(\omega,\tilde \omega)
\lesssim  \frac{T(\omega)}{\kappa_0^{q+1}}\,q!\,\|v\|_{\ell^1}\,\nu^{q-1}\,.
\end{equation}
In order to deduce \eqref{integration_lemma_large_mass_i_3}, we used the observation that 
\begin{equation}
\label{integration_lemma_large_mass_i_4}
\|v^{L,(2)}\bigr\|_{\ell^1(\Lambda_L)} =\|v^{(2)}\|_{\ell^1(\Z^d)}\,,
\end{equation}
which follows from \eqref{v^{L,(j)}}.
Claim (i) then follows from \eqref{zeta^L_bound}, \eqref{integration_lemma_large_mass_i_2}--\eqref{integration_lemma_large_mass_i_3}. 

We now prove (ii). Let $x \in \Lambda_L$ be fixed. As in the proof of (i), we need to estimate the two terms coming from \eqref{zeta^L_bound}. By \eqref{tilde_V_m}, \eqref{mu_measure_large_mass_2}, the first term is 
\begin{multline}
\label{integration_lemma_large_mass_ii_1}
\leq \sum_{\tilde T \in \nu \N^*} \ee^{-\kappa_0 \tilde T/\nu} \, \tilde T^q\, \sum_{r \in \nu \N} \ind{r<T(\omega)} \sum_{s \in \nu \N} \ind{s<T(\tilde \omega)}\int_0^{\nu}\dd t\,\int_{\Lambda_L}\dd y\,\int \bb W^{L,\tilde T}_{y,x}(\dd \tilde \omega)\,\ind{\omega(t+r)=\tilde \omega(t+s)}
\\
\lesssim  \frac{T(\omega)}{\kappa_0^{q+2}}\,(q+1)!\,\nu^{q}\,.
\end{multline}
In \eqref{integration_lemma_large_mass_ii_1}, we bounded the indicator function by $1$ and argued as in \eqref{integration_lemma_large_mass_i_2}.
By analogous arguments as in \eqref{integration_lemma_ii_1}--\eqref{integration_lemma_ii_1_A}, and recalling  \eqref{integration_lemma_large_mass_i_4}, the second term coming from \eqref{zeta^L_bound} is 
\begin{equation}
\label{integration_lemma_large_mass_ii_2}
\lesssim \frac{T(\omega)}{\kappa_0^{q+2}}\,(q+1)!\,\|v^{(2)}\|_{\ell^1}\,\nu^{q-1}\,.
\end{equation}
Claim (ii) now follows from \eqref{zeta^L_bound}, \eqref{integration_lemma_large_mass_ii_1}--\eqref{integration_lemma_large_mass_ii_2}. 

We now prove (iii), we use \eqref{tilde_V_m}--\eqref{mu_measure_large_mass_1} and argue analogously as in \eqref{integration_lemma_iii_1_a}--\eqref{integration_lemma_iii_1} to deduce that 
\begin{equation*}
\int \mu^{L} (\dd \tilde \omega) \,T(\tilde \omega)^{q} 
\lesssim \frac{(q-1)!}{\kappa_0^{q}}\,|\Lambda_L|\,\nu^{q}\,.
\end{equation*}
Claim (iv) follows by analogous arguments.
\end{proof}
\begin{remark}
\label{cluster_expansion_remark_2} 
In light of Lemma \ref{integration_lemma_large_mass} (i), (iii) and \eqref{tilde_kappa}, the arguments in Remark \ref{cluster_expansion_remark} show that we can apply \cite[Theorem 1]{Ueltschi_2004} in this context as well.
\end{remark}
We can now deduce an analogue of Proposition \ref{operator_norm_bound}.

\begin{proposition} 
\label{operator_norm_bound_proposition_2}
For $\kappa_0$ sufficiently large, we have the following bounds for all $L \in \N^*$.
\begin{itemize}
\item[(i)] The specific relative Gibbs potential \eqref{quantum_free_energy} satisfies
$g^{\nu,\kappa_0/\nu,1,L}=O_{\kappa_0,\|v\|_{\ell^1}}(1)$.
\item[(ii)] For $p \in \N^*$, we have
$\,\bigl\|\Gamma^{\nu,\kappa_0/\nu,1,L}\bigr\|_{\ell^{\infty}_{\b x} \ell^1_{\b y}}=O_{\kappa_0,p,\|v\|_{\ell^1}}(1)$.
\end{itemize}
\end{proposition}
\begin{proof}
The proof is analogous to that of Proposition \ref{operator_norm_bound}. Instead of Lemma \ref{integration_lemma}, we apply Lemma \ref{integration_lemma_large_mass}. We only need to note that the powers of $\nu$ that we obtain by applying the estimates in Lemma \ref{integration_lemma_large_mass} cancel out. In order to do this, we note a general fact about trees. Let $\cal T \in {\fra T}_n$ with a distinguished root $r$ be given. We recall that $\mathrm{V}(\cal T)$ denotes the set of vertices of $\cal T$. For $w \in \mathrm{V}(\cal T)$, we denote by $\cal Q(w)$ the set of \emph{direct descendants} of $w$. By definition, this is the set of all $w' \in \mathrm{V}(\cal T)$ such that the unique path in $\cal T$ connecting $w'$ to $r$ starts with the edge joining $w'$ to $w$. 
By induction on the number of vertices of $\cal T$, we obtain that 
\begin{equation} 
\label{tree_identity}
\sum_{w \in \mathrm{V}(\cal T) \setminus \{r\}}(1-|\cal Q(w)|)=|\cal Q(r)|\,.
\end{equation}
Using \eqref{tree_identity}, we deduce that, when applying Lemma \ref{integration_lemma_large_mass} in the argument of the proof of Proposition \ref{operator_norm_bound}, the powers of $\nu$ in the upper bound cancel out. The claim follows.
\end{proof}
Arguing analogously as for Proposition \ref{operator_norm_bound_corollary}, we can use Proposition \ref{operator_norm_bound_proposition_2} to deduce the following result.

\begin{corollary}
\label{operator_norm_bound_corollary_2}
With assumptions as in Proposition \ref{operator_norm_bound_proposition_2}, and consider $p \in \N^*$.
We can take $L=\infty$ in \eqref{gamma_p_cluster_expansion_claim} and obtain an operator $\Gamma^{\nu,\kappa_0/\nu,1,\infty}$ on $\ell^2(\Z^{d})^{\otimes p}$ which satisfies
$\bigl\|\Gamma^{\nu,\kappa_0/\nu,1,\infty}\bigr\|_{\ell^{\infty}_{\b x} \ell^1_{\b y}}=O_{\kappa_0,p,\|v\|_{\ell^1}}(1)$. 
\end{corollary}

With notation as in \eqref{D^{L,c}}, we note the following analogue of Lemma \ref{integration_lemma_M} for given $c>0$ small.
\begin{lemma}
\label{integration_lemma_M_large_mass}
Let $\omega \in \Omega^{L,T(\omega)}$ with $T(\omega) \in \nu \N^*$ and $q \in \N$ be given. Then, the following estimates hold.
\begin{itemize}
\item[(i)] $\int \mu^{L} (\dd \tilde \omega)\,T(\tilde \omega)^{q} \,|\zeta^{L}(\omega,\tilde \omega)|\,\ind{\cal D^{L}_{c}}(\tilde \omega) \lesssim_{\kappa_0,q,d} o_L\bigl(T(\omega)\,[1+\|v^{(2)}\|_{\ell^1}]\,\nu^{q-1}\bigr)\,.$ 
\item[(ii)] Let $L_0 \in \N^*$ as in \eqref{L_0_choice} and $x \in \Lambda_{L_0}$ be given. Then, we have 
\begin{equation*}
\int_{\Lambda_{L_0}} \dd y\,\int \hat{\mu}^{L}_{y,x} (\dd\tilde \omega) \,  T(\tilde \omega)^{q}\,|\zeta^{L}(\omega,\tilde \omega)|\,\ind{\cal D^{L}_{c}}(\tilde \omega) \lesssim_{\kappa_0,q,d} o_L\Bigl(T(\omega)\,[1+\|v^{(2)}\|_{\ell^{\infty}}+\|v^{(2)}\|_{\ell^1}]\,\nu^{q-1}\Bigr) \,.
\end{equation*}
\item[(iii)] $\frac{1}{|\Lambda_L|}\,\int {\mu}^{L}(\dd \tilde{\omega})\,T(\tilde{\omega})^{q}\,\ind{\mathcal{D}_{c}^{L}}(\tilde{\omega}) \lesssim_{\kappa_0,q,d} o_{L} (\nu^{q})\,.$
\item[(iv)] With $L_0$ as in (ii) and $x \in \Lambda_{L_0}$, we have 
\begin{equation*}
\int_{\Lambda_{L_0}} \dd y\,\int \hat{\mu}^{L}_{y,x}(\dd \tilde \omega) \,T(\tilde \omega)^{q} \,\ind{\cal D^{L}_{c}}(\tilde \omega)\lesssim_{\kappa_0,q,d}\,o_L(\nu^{q})\,.
\end{equation*}
\end{itemize}
\end{lemma}

We also note an analogue of Lemma \ref{integration_lemma_M_2}. Before stating the result, we need to modify some of the notation. With $c>0$ as earlier, we let $\cal V^{\nu,1,L,(2)}_{c}(\omega,\tilde \omega)$ denote the quantity given as in \eqref{V_interaction_1} with interaction potential replaced by 
${v^{L,(2)}_{c}}(x) \deq v^{L,(2)}(x) \,\ind{|x|_L \geq cL}$.
Here, we recall  \eqref{v^{(2)}_ell^1}--\eqref{v^{L,(j)}} to see that, similarly as in \eqref{v^{L}_{c}_estimate}, we have
$\lim_{L \rightarrow \infty} \|{v^{L,(2)}_{c}}\|_{\ell^1(\Lambda_L)}=0$.

We modify \eqref{zeta^{L,c}} and let
\begin{equation}
\label{zeta^{(2),L,c}}
\zeta^{L,(2)}_{c}(\omega,\tilde \omega) \deq \exp\bigl(-\cal V^{\nu,1,L,(2)}_{c}(\omega,\tilde \omega)\bigr)-1\,.
\end{equation}
With this notation, we have the following lemma.
\begin{lemma}
\label{integration_lemma_M_2_large_mass}
With assumptions and notation as in Lemma \ref{integration_lemma_M_large_mass}, the following estimates hold.
\begin{itemize}
\item[(i)] $\int \mu^{L} (\dd \tilde \omega)\, T(\tilde \omega)^{q} \,|\zeta^{L,(2)}_{c}(\omega,\tilde \omega)| \lesssim_{\kappa,q} T(\omega)\,\bigl\|v^{L,(2)}_{c}\bigr\|_{\ell^1(\Lambda_L)}\,\nu^{q-1}\,.$
\item[(ii)] 
$\int_{\Lambda_{L_0}} \dd y\,\int \hat{\mu}^{L}_{y,x} (\dd \tilde \omega) \, T(\tilde \omega)^{q}\,|\zeta^{L,(2)}_{c}(\omega,\tilde \omega)|  \lesssim_{\kappa,q} T(\omega)\,\bigl\|v^{L,(2)}_{c}\bigr\|_{\ell^1(\Lambda_L)}\,\nu^{q-1}$
for all $x \in \Lambda_{L_0}$. 
\end{itemize}
\end{lemma}
We prove Lemmas \ref{integration_lemma_M_large_mass} and \ref{integration_lemma_M_2_large_mass} in Appendix \ref{Appendix C.4}.
Let us note an analogue of Proposition \ref{MF_cluster_expansion}.
\begin{proposition}
\label{large_mass_cluster_expansion}
With assumptions as in Proposition \ref{operator_norm_bound_proposition_2}, the following claims hold.
\begin{itemize}
\item[(i)] The quantity 
\begin{equation}
\label{operator_norm_bound_2_corollary_i}
\lim_{L \rightarrow \infty} g^{\nu,\kappa_0/\nu,1,L} = : g^{\nu,\kappa_0/\nu,1,\infty}
\end{equation}
exists and the convergence in \eqref{operator_norm_bound_2_corollary_i} holds uniformly in $\nu \leq 1$.
\item[(ii)] Let $p \in \N^*$ and $L_0 \in \N^*$ be given. 
Let $\Gamma^{\nu,\kappa_0/\nu,1,\infty}$ be as in Corollary \ref{operator_norm_bound_corollary_2} (ii). Then, we have 
\begin{equation}
\label{gamma_infty_convergence_2}
\Gamma^{\nu,\kappa_0/\nu,1,\infty}=\lim_{L \rightarrow \infty} \Gamma^{\nu,\kappa_0/\nu,1,L}\,.
\end{equation}
The convergence in \eqref{gamma_infty_convergence_2} holds in $\|\cdot\|_{L_0,p}$ given by \eqref{L_0_p_norm} and is uniform in $\nu \leq 1$ and $L_0 \in \N^*$.
\end{itemize}
\end{proposition}

\begin{proof}[Proof of Proposition \ref{large_mass_cluster_expansion}]
The proof is similar to that of Proposition \ref{MF_cluster_expansion}. We just comment on the main differences. As in Proposition \ref{MF_cluster_expansion}, modifications of the proof of (ii) allow us to obtain claim (i). We hence just give the proof of (ii). By Proposition \ref{operator_norm_bound_proposition_2} (ii), Corollary \ref{operator_norm_bound_corollary_2} (ii), and arguing as for \eqref{(ii)_star}, it suffices to prove
\begin{multline}
\label{(ii)_star_2}
\lim_{L \rightarrow \infty} \biggl\|\int \hat \mu^{L}_{y_1,x_1}(\dd {\omega}_{1}) \cdots \hat \mu^{L}_{y_p,x_p}(\dd \omega_p)\,\mu^{L}(\dd \omega_{p+1}) \cdots \mu^{L}(\dd \omega_n)\,\varphi^{L}({\omega}_{1},\ldots,\omega_n)
\\
-\int \hat \mu^{\infty}_{y_1,x_1}(\dd {\omega}_{1}) \cdots \hat \mu^{\infty}_{y_p,x_p}(\dd \omega_p)\,\mu^{\infty}(\dd \omega_{p+1}) \cdots \mu^{\infty}(\dd \omega_n)\,\varphi^{\infty}({\omega}_{1},\ldots,\omega_n)
 \biggr\|_{L_0,p} = 0\,,
\end{multline}
uniformly in $\nu \leq 1$ and $L_0$. 
Note that in \eqref{(ii)_star_2}, the path measures are given by \eqref{mu_measure_large_mass_1}--\eqref{mu_measure_large_mass_2}. Recalling \eqref{A^{L}_B^{L}}, and arguing analogously as in \eqref{(ii)_star_proof1}--\eqref{(ii)_star_proof2}, we have that 
for all $k \in \{1,\ldots,n\}$, 
\begin{equation}
\label{(ii)_star_proof1_1}
\lim_{L \rightarrow \infty} \biggl\|\int \hat \mu^{L}_{y_1,x_1}(\dd {\omega}_{1}) \cdots \hat \mu^{L}_{y_p,x_p}(\dd \omega_p) \,\mu^{L}(\dd \omega_{p+1}) \cdots \mu^{L}(\dd \omega_n)\,\ind{\cal B^{L}}(\omega_k)\,\varphi^{L}({\omega}_{1},\ldots,\omega_n)\biggr\|_{L_0,p}=0\,,
\end{equation}
which follows from the observation that 
\begin{equation}
\label{(ii)_star_proof2_2}
\lim_{L \rightarrow \infty}  \biggl\|\int \hat \mu^{L}_{y_1,x_1}(\dd {\omega}_{1}) \cdots \hat \mu^{L}_{y_p,x_p}(\dd \omega_p)\,\mu^{L}(\dd \omega_{p+1}) \cdots  \mu^{L}(\dd \omega_n)\,\ind{\cal B^{L}}(\omega_k)\,\prod_{\{i,j\} \in \cal T} |\zeta^{L}(\omega_i,\omega_{j})|\biggr\|_{L_0,p}=0\,,
\end{equation}
for a fixed $\cal T \in {\fra T}_n$. The proof of \eqref{(ii)_star_proof2_2} is analogous to that of \eqref{(ii)_star_proof2}, except that we now use Lemmas \ref{integration_lemma_M_large_mass}--\ref{integration_lemma_M_2_large_mass} instead of Lemmas \ref{integration_lemma_M}--\ref{integration_lemma_M_2}.
Similarly, we have that for all $k \in \{1,\ldots,n\}$,
\begin{equation}
\label{(ii)_star_proof3_2}
\lim_{L \rightarrow \infty} \biggl\|\int \hat \mu^{\infty}_{y_1,x_1}(\dd {\omega}_{1}) \cdots \hat \mu^{\infty}_{y_p,x_p}(\dd \omega_p)\,\mu^{\infty}(\dd \omega_{p+1}) \cdots \mu^{\infty}(\dd \omega_n)\,\ind{\cal B^{L}}(\omega_k)\,\varphi^{\infty}({\omega}_{1},\ldots,\omega_n)\biggr\|_{L_0,p}=0\,.
\end{equation}
By using \eqref{(ii)_star_proof1_1}, \eqref{(ii)_star_proof3_2}, and arguing as in the proof of Proposition \ref{MF_cluster_expansion}, the claim follows if we show
\begin{multline}
\label{(ii)_star_proof4_2}
\lim_{L \rightarrow \infty} \biggl\|\int \hat \mu^{L}_{y_1,x_1}(\dd {\omega}_{1}) \cdots \hat \mu^{L}_{y_p,x_p}(\dd \omega_p)\,\mu^{L}(\dd \omega_{p+1}) \cdots \mu^{L}(\dd \omega_n)\,\prod_{k=1}^{n}\ind{\cal A^{L}}(\omega_k)\,\varphi^{L}({\omega}_{1},\ldots,\omega_n)
\\
-\int \hat \mu^{\infty}_{y_1,x_1}(\dd {\omega}_{1}) \cdots \hat \mu^{\infty}_{y_p,x_p}(\dd \omega_p)\,\mu^{\infty}(\dd \omega_{p+1}) \cdots \mu^{\infty}(\dd \omega_n)\,\prod_{k=1}^{n}\ind{\cal A^{L}}(\omega_k)\,\varphi^{\infty}({\omega}_{1},\ldots,\omega_n)
\biggr\|_{\ell^{\infty}_{\b x}\ell^1_{\b y}} = 0\,.
\end{multline}
By arguing as in the proof of \eqref{(ii)_star_proof4}, we note that \eqref{v-v^L_ell^1} still holds in this setting
by \eqref{v^{L}} and Assumption \ref{interaction_potential_v_m_infty} (i).
\end{proof}

We now have the necessary tools to prove Theorem \ref{Infinite_volume_theorem_2}.

\begin{proof}[Proof of Theorem \ref{Infinite_volume_theorem_2}]
The proof is similar to that of Theorem \ref{Infinite_volume_theorem_1}. We combine Theorem \ref{Large_mass_lattice_4}, Proposition \ref{large_mass_cluster_expansion} and Lemma \ref{convergence_general_fact}. 
\end{proof}

\appendix

\section{Derivation of the Symanzik and Ginibre loop representations}
\label{Sym-Gin}

In this appendix we derive the Symanzik and Ginibre loop representations of the classical field theory and the interacting Bose gas, respectively. We shall use the following standard tool.

\begin{lemma}[Feynman-Kac formula] \label{FK_discrete}
For any $V: \Lambda \rightarrow \C$ and $t > 0$ we have
\begin{equation*}
\bigl(\ee^{t (\Delta / 2 - V)}\bigr)_{y,x} = \int \bb W_{y,x}^{t} (\dd \omega) \, \ee^{- \int_0^{t} \dd s \, V(\omega(s))}\,.
\end{equation*}
\end{lemma}

\subsection{The Symanzik representation: proofs of Proposition \ref{Symanzik_representation_theorem} and Corollary \ref{correlation_inequality_theorem}} \label{sec:symanzik_proof}

In this appendix, we give give the proof of Proposition \ref{Symanzik_representation_theorem}. Let us first comment on the main proof strategy. Our starting point is the observation that the weight 
\begin{equation}
\label{exponential_weight}
\ee^{-\frac{1}{2} \int \dd x\, \int \dd y\, |\phi(x)|^2 \,v(x-y)\, |\phi(y)|^2}\,,
\end{equation}
occurring in \eqref{Z_hat} and \eqref{gamma_hat}
is a function of $|\phi|^2 \equiv (|\phi(u)|^2)_{u \in \Lambda}$.
We rewrite \eqref{exponential_weight} using the Hubbard-Stratonovich formula (see \eqref{Hubbard_Stratonovich_formula} below). 
As a result, we obtain integrals over a field $\sigma: \Lambda \rightarrow \R$. After performing a Gaussian integration in the field $\phi$, we can resum the $\sigma$ integration to obtain the result. For similar arguments based on the Hubbard-Stratonovich formula, we refer the reader to \cite[Sections 3-4]{FKSS_2020_1}.
Let us note that arguments based on rewriting \eqref{exponential_weight} using the Fourier transform were applied in \cite[Section 2]{Brydges_Froehlich_Sokal_1983}, \cite[Section 2]{Brydges_Froehlich_Spencer_1982}, and \cite[Section 5]{Froehlich_1982}.
We now prove  Proposition \ref{Symanzik_representation_theorem}.
\begin{proof}[Proof of Proposition \ref{Symanzik_representation_theorem}]
Let us first prove claim (i). We identify $v \col \Lambda \to \R$ with a positive quadratic form 
$f \mapsto \scalar{f}{v f} \deq \int \dd x \int \dd y \, f(x) v(x - y) f(y).$ 
Note that the positivity of the quadratic form follows since $v$ is of positive type.
Let $\mu_{v}$ be a Gaussian measure on $\R^\Lambda$ with covariance $v$, i.e.
\begin{equation}
\label{mu_v}
\int  \mu_{v} (\dd \sigma) \, \sigma(x) \,\sigma(y)=v(x-y)\,.
\end{equation}
The \emph{Hubbard-Stratonovich formula} then states that
\begin{equation}
\label{Hubbard_Stratonovich_formula}
\int \mu_{v} (\dd \sigma) \, \ee^{\ii \langle f, \sigma \rangle} 
= e^{-\frac{1}{2} \langle f , v f \rangle}\,,
\end{equation}
which follows from Lemma \ref{lem:Gaussian}. By \eqref{Hubbard_Stratonovich_formula} with $f=|\phi|^2$ we can rewrite \eqref{Z_hat} as
\begin{equation}
\label{Z_hat_2A}
\cal Z^{\cl}=\int \mu_{(-\Delta/2+\kappa)^{-1}}(\dd \phi)\biggl(\int \mu_{v} (\dd \sigma)\, \ee^{\ii \int \dd x\, \sigma(x) |\phi(x)|^2}\biggr)\,,
\end{equation}
which by using Fubini's theorem and evaluating a Gaussian integral equals
\begin{align}
\notag
&\int \mu_{v} (\dd \sigma)\, \int \mu_{(-\Delta/2+\kappa)^{-1}}(\dd \phi)\,\ee^{\ii \int \dd x\, \sigma(x) |\phi(x)|^2}
\\
\label{Z_hat_2}
&=\int \mu_{v} (\dd \sigma)\,
\det  \bigl(-\Delta/2+\kappa-\ii \sigma\bigr)^{-1} \, \det \bigl(-\Delta/2+\kappa\bigr)
\,.
\end{align}

We note that 
\begin{equation}
\label{**}
\det \bigl(-\Delta/2+\kappa-\ii \sigma\bigr)^{-1}\,\det \bigl(-\Delta/2+\kappa\bigr)
= \exp \Bigl\{-\tr \log\bigl(-\Delta/2+\kappa-\ii \sigma\bigr)+\tr \log \pb{-\Delta/2+\kappa}\Bigr\}\,,
\end{equation}
since the arguments of the logarithm have strictly positive real part. 

We note that for all $a, b \in \C$ of strictly positive real part we have 
\begin{equation}
\label{integral_identity_formula}
\log a - \log b = -\int_{0}^{\infty} \frac{\dd t}{t}\,\bigl(\ee^{-ta}-\ee^{-tb}\bigr)\,.
\end{equation} 
A direct calculation yields
\begin{equation}
\label{integral_identity_formula_proof}
\log a - \log b = - \int_0^\infty \dd t \,\pbb{\frac{1}{t + a} - \frac{1}{t + b}}\,.
\end{equation}
We deduce \eqref{integral_identity_formula} from \eqref{integral_identity_formula_proof}  by noting that for $c \in \C$ with $\re c>0$ and $t \geq 0$ we have
$\frac{1}{t+c}=\int_{0}^{\infty} \dd s\, \ee^{-s(t+c)}$,
and by using Fubini's theorem.
By using \eqref{integral_identity_formula} followed by Lemma \ref{FK_discrete},  the fact that $\tr A = \int \dd u \, A_{u,u}$ and \eqref{W_path_measures}, we can write
\begin{equation}
\label{Lemma1.1_application2A}
-\tr \log \bigl(-\Delta/2+\kappa-\ii \sigma \bigr) + \tr \log \bigl(-\Delta/2+\kappa\bigr)=
\int_{0}^{\infty} \frac{\dd T}{T} \int \bb W^T(\dd \omega) \,\ee^{-\kappa T}\,\pbb{\ee^{\ii \int_0^T \dd t \, \sigma(\omega(t))} - 1}\,.
\end{equation}
From \eqref{Z_hat_2A}--\eqref{Lemma1.1_application2A}, we conclude
\begin{equation}
\label{Z_hat_rewritten}
\cal Z^{\cl}=\int \mu_{v} (\dd \sigma)\,\exp\Biggl\{\int_{0}^{\infty} \frac{\dd T}{T} \,\ee^{-\kappa T} \int \bb W^{T}(\dd \omega)\,\biggl(\ee^{\ii \int_{0}^{T} \dd t\, \sigma (\omega(t))}-1\biggr)\Biggr\}\,.
\end{equation}
In what follows, we fix $\epsilon>0$ and rewrite for fixed $\sigma$ the expression \eqref{Lemma1.1_application2A} as
\begin{equation}
\label{Lemma1.1_application3}
\int \bb L^{\cl,\epsilon}(\dd \omega)\,\ee^{\ii \int_{0}^{T} \dd t\, \sigma (\omega(t))}+K^{\epsilon}
+\int_{0}^{\epsilon} \frac{\dd T}{T}\, \ee^{-\kappa T}\, \int \bb W^{T}(\dd \omega)\,\biggl(\ee^{\ii \int_{0}^{T} \dd t\, \sigma (\omega(t))}-1\biggr)\,,
\end{equation}
where we recall \eqref{L_cl_epsilon} and \eqref{K_epsilon_definition}.
We now show that the third term in \eqref{Lemma1.1_application3} is 
$\epsilon \abs{\Lambda}O(\|\sigma\|_{\infty})$.

Indeed, we obtain this by noting that 
\begin{equation*}
\Bigl|\ee^{\ii \int_{0}^{T} \dd t\, \sigma (\omega(t))}-1\Bigr| \leq T  \|\sigma\|_{\infty} 
\end{equation*}
and using \eqref{heat_kernel_identity} combined with Lemma \ref{estimates_heat_kernel} (i). In particular, we can rewrite \eqref{Lemma1.1_application3} as 
\begin{equation}
\label{Lemma1.1_application4}
\int \bb L^{\cl,\epsilon}(\dd \omega)\,\ee^{\ii \int_{0}^{T} \dd t\, \sigma (\omega(t))}+K^{\epsilon}
+\epsilon \abs{\Lambda}O(\|\sigma\|_{\infty})\,.
\end{equation}
We now need to exponentiate and integrate in $\sigma$.
Before doing so, we analyse \eqref{Lemma1.1_application4} more closely. We first note that, by \eqref{K_epsilon_definition} we have
\begin{equation}
\label{exp_int_sigma_1}
\mathrm{Re} \biggl(\int \bb L^{\cl,\epsilon}(\dd \omega)\,\ee^{\ii \int_{0}^{T} \dd t\, \sigma (\omega(t))}\biggr)+K^{\epsilon}
\leq 0\,.
\end{equation}
Hence, \eqref{exp_int_sigma_1} implies that 
\begin{equation}
\label{exp_int_sigma_2}
\exp\biggl\{\int \bb L^{\cl,\epsilon}(\dd \omega)\,\ee^{\ii \int_{0}^{T} \dd t\, \sigma (\omega(t))}+K^{\epsilon}\biggr\}=O(1)\,,
\end{equation}
uniformly in $\epsilon,\kappa>0$, $\Lambda$ and $\sigma: \Lambda \rightarrow \R$.

We now analyse the third (i.e.\ the error) term in \eqref{Lemma1.1_application4}. Given $C>0$ we show that
\begin{equation}
\label{M_epsilon}
\int \mu_{v} (\dd \sigma)\,{\ee^{C\epsilon  \|\sigma\|_{\infty} }} \rightarrow 1\,\quad\mbox{as}\quad\epsilon \rightarrow 0\,.
\end{equation}
We note that \eqref{M_epsilon} follows from the dominated convergence theorem provided that we show that for $C>0$ we have $\int \mu_{v} (\dd \sigma) \,\ee^{C \|\sigma\|_{\infty} } <\infty$. We prove this by writing 
\begin{equation}
\label{exp_Linfty_bound}
\int \mu_{v} (\dd \sigma) \,\ee^{C \|\sigma\|_{\infty} } \leq \int \dd x \, \int \mu_{v} (\dd \sigma) \,\ee^{C \abs{\sigma(x)}}\,.
\end{equation} 
Note that in \eqref{exp_Linfty_bound}, we used the fact that $\mu_v$ is a positive measure.
We expand the exponential, and use the Cauchy-Schwarz inequality, Wick's theorem (Lemma \ref{lem:Gaussian}), and \eqref{mu_v} to deduce
\begin{equation}
\label{exp_Linfty_bound_application}
\int \mu_{v} (\dd \sigma) \,\ee^{C \abs{\sigma(x)}} \leq \sum_{i=0}^{\infty} \frac{C^i}{i!} \biggl(\int \mu_{v} (\dd \sigma)\, \sigma(x)^{2i}\biggr)^{1/2} =
\sum_{k=0}^{\infty} \frac{C^i}{i!}\,\sqrt{\frac{(2i)!}{i!\,2^i}}\,v(0)^{i/2}<\infty\,,
\end{equation}
as desired.

We now combine \eqref{Z_hat_rewritten}--\eqref{Lemma1.1_application3}, \eqref{Lemma1.1_application4}, \eqref{exp_int_sigma_2}--\eqref{M_epsilon}, and apply an $L^{\infty}(\dd \mu_v)-L^1(\dd \mu_v)$ H\"{o}lder's inequality in $\sigma$, 
to deduce $\cal Z^{\cl} = \lim_{\epsilon \to 0} \tilde{\cal Z}^{\cl,\epsilon}$, where
\begin{equation} \label{Z_E_SYM_rep}
\tilde{\cal Z}^{\cl,\epsilon} \deq \int \mu_{v} (\dd \sigma)\,\exp\biggl\{\int \bb L^{\cl,\epsilon}(\dd \omega)\,\ee^{\ii \int_{0}^{T} \dd t\, \sigma (\omega(t))}+K^{\epsilon}\biggr\}\,.
\end{equation}

What remains, therefore, is to show that 
\begin{equation}
\label{Z'_epsilon=Z_epsilon}
\tilde{\cal Z}^{\cl,\epsilon} = \cal Z^{\cl,\epsilon}\,.
\end{equation}
By expanding the exponential and using Fubini's theorem we have
\begin{equation}
\label{Z_hat_calculation2}
\tilde{\cal Z}^{\cl,\epsilon}
=\sum_{n=0}^{\infty} \frac{1}{n!} \int  \bb L^{\cl,\epsilon} (\dd {\omega}_{1}) \cdots \bb L^{\cl,\epsilon} (\dd \omega_n)  \, \biggl(\int \mu_{v} (\dd \sigma)\, \ee^{\ii \sum_{i=1}^{n} \int_0^{T_i}  \dd t\, \sigma(\omega_i(t)) }\biggr)\,
\exp(K^{\epsilon})\,.
\end{equation}
The $\sigma$-integration can be performed in \eqref{Z_hat_calculation2} by using the Hubbard-Stratonovich formula \eqref{Hubbard_Stratonovich_formula}, noting that $\sum_{i=1}^{n} \int_0^{T_i} \dd t\, \sigma(\omega_i(t)) = \scalar{f}{\sigma}$ with $f(x) \deq \sum_{i = 1}^n \int_0^{T_i} \dd t \, \delta(x - \omega_i(t))$. We therefore obtain \eqref{Z'_epsilon=Z_epsilon} as claimed, and we deduce (i).

The proof of claim (ii) is similar to that of (i). We just outline the main differences. 
We first apply \eqref{Hubbard_Stratonovich_formula} with $f=|\phi|^2$ in \eqref{gamma_hat} followed by Fubini's theorem to deduce that 
\begin{equation}
\label{gamma_hat_2}
(\Gamma_p^{\cl})_{\b x, \b y} = \frac{1}{\cal Z^{\cl}}
\int \mu_{v}(\dd \sigma) \int  \mu_{(-\Delta/2+\kappa)^{-1}} (\dd \phi) \, \ee^{\ii \langle \phi, \sigma \phi\rangle}\,\bar \phi(y_1)\,\cdots\,\bar \phi(y_p)\,\phi(x_1)\,\cdots\,\phi(x_p)\,.
\end{equation}
By \eqref{mu_A_complex} and Lemma \ref{complex_Wick_theorem}, 
we can rewrite \eqref{gamma_hat_2} as
\begin{equation}
\label{gamma_hat_3}
= \frac{1}{\cal Z^{\cl}} \sum_{\pi \in S_p}\,\int \mu_{v}(\dd \sigma) \,\prod_{i=1}^{p}\biggl(\frac{1}{-\Delta/2+\kappa-\ii \sigma}\biggr)_{y_{\pi(i)},x_i}\,
\det \bigl(-\Delta/2+\kappa-\ii \sigma\bigr)^{-1}\,\det \bigl(-\Delta/2+\kappa\bigr)\,.
\end{equation}
Furthermore we have for all $x,y \in \Lambda$
\begin{equation}
\label{gamma_hat_FK}
\biggl(\frac{1}{-\Delta/2+\kappa-\ii \sigma}\biggr)_{y,x}= \int_{0}^{\infty} \dd T \,\Bigl(\ee^{T(\Delta/2-\kappa+\ii \sigma)}\Bigr)_{y,x}
=\int_{0}^{\infty} \dd T\,\ee^{-\kappa T}\,\int \bb W_{y,x}^{T}(\dd \omega)\,\ee^{\ii \int_{0}^{T} \dd t \, \sigma(\omega(t))}\,,
\end{equation}
where in the last equality we used Lemma \ref{FK_discrete}.
We now combine \eqref{gamma_hat_3} and \eqref{gamma_hat_FK} to deduce that
\begin{multline}
\label{gamma_hat_4}
(\Gamma_p^{\cl})_{\b x, \b y}= \frac{1}{\cal Z^{\cl}} \sum_{\pi \in S_p} \int_{(0,\infty)^p} \dd \b T\,\ee^{-\kappa \abs{\b T}}\,\int \bb W^{\b T}_{\pi \b y, \b x}(\dd \b \omega)\,\biggl(\int \mu_{v} (\dd \sigma)\, \ee^{\ii \sum_{j=1}^{p} \int_0^{T_j} \dd t\,  \sigma(\omega_{j}(t))}\biggr)\,
\\
\times \det \bigl(-\Delta/2+\kappa-\ii \sigma\bigr)^{-1} \,\det \bigl(-\Delta/2+\kappa\bigr)\,.
\end{multline}
We rewrite the factor of $\det \bigl(-\Delta/2+\kappa-\ii \sigma\bigl)^{-1}\,\det \bigl(-\Delta/2+\kappa\bigr)$ as in \eqref{**}--\eqref{Lemma1.1_application2A}, recall  \eqref{Lemma1.1_application4}, \eqref{exp_int_sigma_2}--\eqref{M_epsilon}, and proceed analogously as in the remainder of the proof of Proposition \ref{Symanzik_representation_theorem} (i) to rewrite
\eqref{gamma_hat_4} as 
\begin{equation}
\label{gamma_gamma_epsilon}
(\Gamma_{p}^{\cl})_{\b x,\b y}=\lim_{\epsilon \rightarrow 0} (\tilde{\Gamma}_p^{\cl,\epsilon})_{\b x,\b y}\,,
\end{equation}
where for $\epsilon>0$, we let
\begin{multline}
\label{gamma'_epsilon_definition}
(\tilde{\Gamma}_p^{\cl,\epsilon})_{\b x, \b y} \deq \frac{1}{\cal Z^{\cl}} \sum_{\pi \in S_p} \int_{(0,\infty)^p} \dd \b T\,\ee^{-\kappa\abs{\b T}}\,\int \bb W^{\b T}_{\pi \b y,\b x}(\dd \b \omega)\,\sum_{n = 0}^\infty \frac{1}{n!} \int \bb L^{\cl,\epsilon} (\dd \tilde {\omega}_{1}) \cdots \bb L^{\cl,\epsilon} (\dd \tilde \omega_n)
\\
\times
\exp(-V^{\cl}(\b \omega \tilde {\b \omega})) \,\biggl(\int \mu_{v} (\dd \sigma)\, \ee^{\ii \sum_{j=1}^{p} \int_0^{T_j} \sigma(\omega_j(t)) \, \dd t+\ii \sum_{i=1}^{n} \int_0^{T_i} \sigma(\tilde{\omega}_i(t)) \, \dd t}\biggr)\, \exp(K^{\epsilon})
\end{multline}
We can now perform the $\sigma$ integration in \eqref{gamma'_epsilon_definition} analogously as in \eqref{Z'_epsilon=Z_epsilon}--\eqref{Z_hat_calculation2}. Note that we now apply \eqref{Hubbard_Stratonovich_formula} with 
$f= \sum_{j=1}^p \int_0^{T_j} \dd t\, \delta(x-\omega_j(t))+ \sum_{i = 1}^n \int_0^{T_i} \dd t \, \delta(x - \tilde{\omega}_i(t))$.
In particular, recalling \eqref{gamma^epsilon_definition} we obtain that $(\tilde{\Gamma}_p^{\cl,\epsilon})_{\b x, \b y}=(\Gamma_p^{\cl,\epsilon})_{\b x, \b y}$ and we deduce the claim (ii) from \eqref{gamma_gamma_epsilon}.
\end{proof}

We also give the proof of Corollary \ref{correlation_inequality_theorem}.

\begin{proof}[Proof of Corollary \ref{correlation_inequality_theorem}]
We recall \eqref{Gaussian_measure_Phi}, and argue as in \eqref{gamma_hat_2}--\eqref{gamma_hat_FK} with $v$ replaced by $\lambda v$ to rewrite \eqref{complex_correlation1} as
\begin{multline}
\label{Correlation_1}
(\widehat{\Gamma}_{p}^{\cl,\lambda})_{\b x, \b y}=\frac{1}{\pi^{|\Lambda|} \,\det (-\Delta/2+\kappa)^{-1}}\,\sum_{\pi \in S_p} \int_{(0,\infty)^p} \dd \b T\,\ee^{-\kappa \abs{\b T}}\,\int \bb W^{\b T}_{\pi \b y, \b x}(\dd \b \omega)\,\int \prod_{u \in \Lambda} \dd \phi(u)\,\ee^{\scalar{\phi}{(\Delta/2-\kappa) \phi}} 
\\
\times \int \mu_{\lambda v}(\dd \sigma)\,\exp\Biggl[\biggl\langle\ii |\phi|^2+\ii \sum_{j=1}^{p}\tau_{(\cdot)}(\omega_j),\sigma\biggr\rangle\Biggr]\,.
\end{multline}
In \eqref{Correlation_1}, given $u \in \Lambda$ and a path $\omega$, $\tau_u(\omega)$ denotes the \emph{local time}, i.e.\ the amount of time that $\omega$ spends at $u$.
We use \eqref{Hubbard_Stratonovich_formula} with 
\begin{equation*}
f(x)=|\phi(x)|^2+\sum_{j=1}^{p}\tau_{x}(\omega_j)
\end{equation*}
to obtain that
$(\widehat{\Gamma}_{p}^{\cl,\lambda})_{\b x, \b y}=\sum_{\pi \in S_p} \int_{(0,\infty)^p} \dd \b T\,\ee^{-\kappa \abs{\b T}}\,\int \bb W^{\b T}_{\pi \b y, \b x}(\dd \b \omega)\,\int \mu_{(-\Delta/2+\kappa)^{-1}}(\dd \phi)\,\ee^{-\frac{\lambda}{2} \scalar{f}{vf}}$,
from where we deduce \eqref{correlation_inequality_1} since $\scalar{f}{vf} \geq 0$ because $v$ is of positive type.
\end{proof}

\subsection{The Ginibre representation: proof of Proposition \ref{Ginibre_loop_representation}} \label{sec:ginibre_proof}

\begin{proof}[Proof of Proposition \ref{Ginibre_loop_representation}]
We recall \eqref{Hamiltonian_H_n} and apply the Feynman-Kac formula, Lemma \ref{FK_discrete} to obtain that
\begin{equation}
\label{e^{-H_n}}
(\ee^{- H_n^{\nu,\lambda}})_{\b y, \b x} = \int \bb W^{\nu \b 1}_{\b y, \b x}(\dd \b \omega) \exp (-V^{\nu,\lambda}(\b \omega))\,,
\end{equation}
where we write $\b 1 \equiv \b 1_n$ and recall \eqref{b_1} (in the sequel we drop the subscript $n$ in \eqref{b_1}). Recalling \eqref{P_n^+}, we have
\begin{equation}
\label{symmetrization_identity1}
(\ee^{-H_n} P^+_n)_{\b y, \b x} = \frac{1}{n!} \sum_{\pi \in S_n} (\ee^{-H_n})_{\pi \b y, \b x}\,.
\end{equation}
We use  \eqref{e^{-H_n}}--\eqref{symmetrization_identity1} in \eqref{gamma_p_definition} and obtain
\begin{equation}
\label{gamma_p_sum}
(\Gamma_p^{\nu,\kappa,\lambda})_{\b x,\b y} 
= \frac{1}{\Xi^{\nu,\kappa,\lambda}}  \sum_{n = 0}^\infty \frac{\ee^{-\kappa \nu(p+n)}}{n!} \sum_{\pi \in S_{p+n}} \int_{\Lambda^n} \dd \b u \,
\int \bb W^{\nu \b 1}_{\pi(\b y \b u),\b x \b u}(\dd \b \omega) \exp (-V^{\nu,\lambda}(\b \omega))\,.
\end{equation}

Now we perform the first step of the loop integration. We distinguish between two types of paths. The first type are \emph{open paths} with endpoints $x_j$ and $y_{j'}$ for  $j,j' \in\{1, \ldots, p\}$. The \emph{internal points} in these open paths are of the form $u_j$ for some $1 \leq j \leq n$. The second type are \emph{closed paths}, all of whose vertices are of the form $u_j$ for some $1 \leq j \leq n$.  Let us denote by $r$ the number of vertices contained in all of the closed paths.  We give an example in Figure \ref{gibbs_loops_graph_example} below. The first step of the loop integration consists in integrating over the internal vertices of the open paths.

\begin{figure}[!ht]
\begin{center}
\includegraphics[scale=0.5]{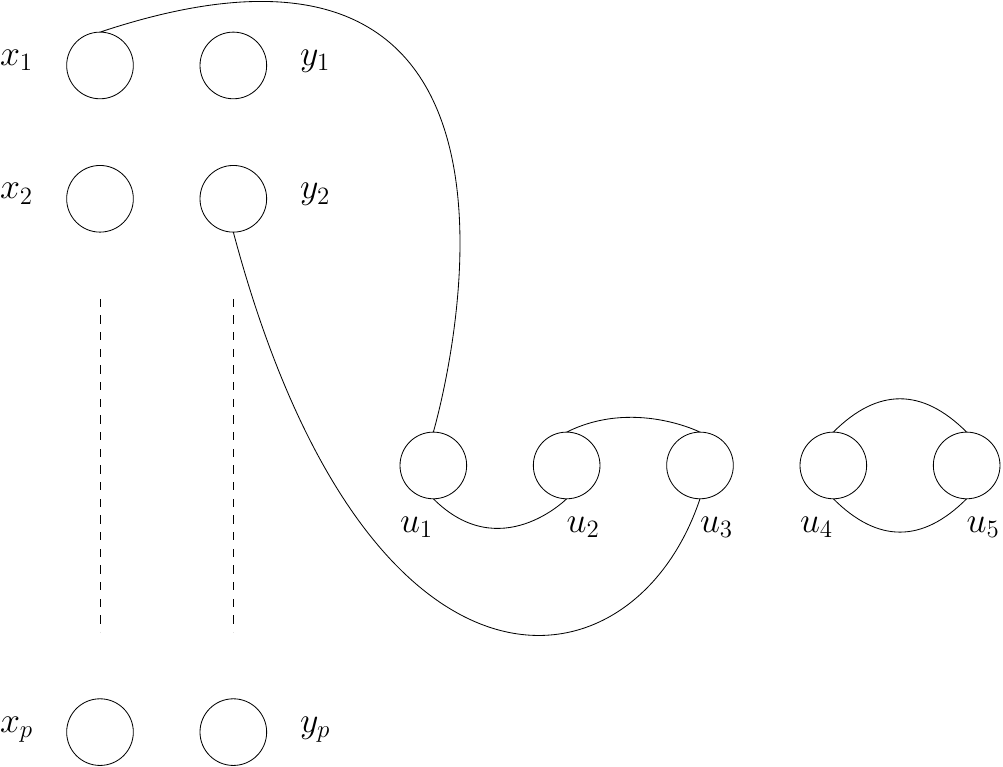}
\end{center}
\caption{In this example, we have $n=5$ and $r=2$. The open path with endpoint $x_1$ has $y_2$ as its other endpoint. It has three internal points $u_1,u_2,u_3$. 
The points $u_4$, $u_5$ belong to a closed path of length two. 
\label{gibbs_loops_graph_example}
}
\end{figure}

In the sequel, we write $k_i$ for the total number of edges in the open path with one endpoint $x_i$ for $i=1,\ldots,p$. Hence, in the example given in Figure \ref{gibbs_loops_graph_example} we have $k_1=4$.
We rewrite \eqref{gamma_p_sum} as
\begin{align}
\notag
&(\Gamma^{\nu,\kappa,\lambda}_p)_{\b x,\b y} = \frac{1}{\Xi^{\nu,\kappa,\lambda}}  \sum_{\b k \in (\N^*)^p} \sum_{n=0}^{\infty} \sum_{r=0}^{\infty} \delta (|\b k|+r-n-p)\,\binom{n}{r}\,(n-r)!
\\
\label{gamma_p_sum_2}
&\times \frac{1}{n!}\,\ee^{-\kappa \nu (|\b k| +r)}\, \sum_{\pi \in S_p} \int \bb W^{\nu \b k}_{\pi \b y, \b x}(\dd \b \omega)\,\int_{\Lambda^r} \dd \b u  \sum_{\sigma \in S_r} \bb W^{\nu \b 1}_{\sigma \b u, \b u}( \dd \tilde{\b \omega}) \,\exp(-V^{\nu,\lambda}(\b \omega \tilde {\b \omega}))\,.
\end{align}
In \eqref{gamma_p_sum_2}, we chose for fixed $r$ the $\binom{n}{r}$ elements of the form $u_j$ that are taken as vertices of closed paths. For fixed $\b k$, the remaining $n-r$ $u$'s can be distributed among the open paths in $(n-r)!$ different ways. Furthermore, we used the presence of the delta function to deduce that $n+p=|\b k| +r$ and to then perform the sum in $n$.
In particular, we conclude that
\begin{align}
\notag
(\Gamma^{\nu,\kappa,\lambda}_p)_{\b x,\b y} = \frac{1}{\Xi^{\nu,\kappa,\lambda}}  \sum_{\b T \in (\nu \N^*)^p} \sum_{\pi \in S_p} &\int \ee^{-\kappa \abs{\b T}} \,\bb W^{\b T}_{\pi \b y, \b x}(\dd \b \omega)
\\
\label{gamma_step1}
& 
 \sum_{r = 0}^\infty \frac{\ee^{-\kappa \nu r}}{r!} \int_{\Lambda^r} \dd \b u \,\sum_{\sigma \in S_r} \int \bb W^{\nu \b 1}_{\sigma \b u, \b u} (\dd \tilde {\b \omega}) \exp (-V^{\nu,\lambda}(\b \omega \tilde {\b \omega}))\,.
\end{align}

Next, we perform the second step of the loop integration, by decomposing $\sigma$ into cycles.
For a permutation $\sigma$ and $k \in \N^*$, denote by $a_k(\sigma)$ the number of cycles of length $k$ in $\sigma$. Write $\b a(\sigma) = (a_1(\sigma), a_2(\sigma), \dots) \in \N^{\N^*}$.
For $\b a = (a_1, a_2, \dots) \in \N^{\N^*}$ we define $\abs{\b a} \deq \sum_{k = 1}^\infty a_k$ (number of cycles) and $r(\b a) \deq \sum_{k = 1}^\infty k a_k$ (number of elements in all cycles). Thus, if $\sigma \in S_r$ then $r(\b a(\sigma)) = r$. Note that the number of permutations $\sigma$ satisfying $\b a(\sigma) = \b a$ is equal to
\begin{equation*}
\frac{r(\b a)!}{\prod_{k = 1}^\infty (k^{a_k} a_k!)}\,.
\end{equation*}
For  $\b \omega \in \Omega^{\b T}$ we let
\begin{align}
\notag
\wt{\Xi}^{\nu,\kappa,\lambda}(\b \omega) &\deq \sum_{r = 0}^\infty \frac{\ee^{-\kappa \nu r}}{r!} \int_{\Lambda^r} \dd \b u \, \sum_{\sigma \in S_r} \int \bb W^{\nu \b 1}_{\sigma \b u, \b u} (\dd \tilde {\b \omega}) \exp (-V^{\nu,\lambda}(\b \omega \tilde {\b \omega}))
\\
\label{tilde_Xi_definition}
&=
\sum_{\b a \in \N^{\N^*}} \prod_{k = 1}^\infty \frac{1}{k^{a_k} a_k!} \int \bb W^{\nu \b \ell(\b a)}(\dd \tilde {\b \omega}) \ee^{-\kappa \nu r(\b a)} \exp (-V^{\nu,\lambda}(\b \omega \tilde {\b \omega}))\,,
\end{align}
where $\b \ell(\b a) \equiv (\ell_1 (\b a), \ldots, \ell_{\abs{\b a}}(\b a)) \in (\N^*)^{\abs{\b a}}$ is an (arbitrary) family of cycle lengths corresponding to $\b a$, i.e.\ satisfying $\sum_{i = 1}^{\abs{\b a}} \ind{\ell_i(\b a) = k} = a_k$ for all $k \in \N^*$.   Here, we recall the notation \eqref{loop_measure_quantum}. Recalling \eqref{L_nu} and using the multinomial identity in \eqref{tilde_Xi_definition}, we find (writing $n = \abs{\b a}$ for the number of cycles) that
\begin{equation}
\label{tilde_Xi}
\wt{\Xi}^{\nu,\kappa,\lambda}(\b \omega) = \sum_{n = 0}^\infty 
\frac{1}{n!}\int \bb L^{\nu,\kappa}(\dd \tilde {\omega}_{1}) \cdots \bb L^{\nu,\kappa}(\dd \tilde \omega_n)
\,\exp(-V^{\nu,\lambda}(\b \omega \tilde{\b \omega}))=\Xi^{\nu,\kappa,\lambda}(\b \omega)\,,
\end{equation}
where we recall the notation from \eqref{Z_nu_omega}.
We recall \eqref{rho_intro} and use the same arguments as above to obtain 
\begin{equation}
\label{tilde_Xi_empty_set}
\wt{\Xi}^{\nu,\kappa,\lambda}(\emptyset) =\Xi^{\nu,\kappa,\lambda}\,.
\end{equation}

The identity \eqref{ginibre_representation} now follows by substituting \eqref{tilde_Xi}--\eqref{tilde_Xi_empty_set} into \eqref{gamma_step1}. The proof of \eqref{Z_nu} is analogous.
\end{proof}

\section{Remarks on Gaussian integrals}
\label{Remarks on Gaussian integrals}

We collect several standard facts about Gaussian integrals.
\begin{lemma} 
\label{lem:Gaussian}
Let $\cal C > 0$ be a positive real $n \times n$ matrix. We define the Gaussian probability measure on $\R^n$ with covariance $\cal C$ through
\begin{equation*}
\mu_{\cal C}(\dd \b x) \deq \frac{1}{\sqrt{(2 \pi)^{n} \det \cal C }} \, \ee^{-\frac{1}{2} \scalar{\b x}{\cal C^{-1} \b x}} \, \dd \b x\,.
\end{equation*}
It has Fourier transform given by
$\int_{\R^n} \mu_{\cal C}(\dd x) \, \ee^{\ii \scalar{\b a}{\b x}} = \ee^{-\frac{1}{2} \scalar{\b a}{\cal C \b a}} $
for all $\b a \in \R^n$. 
\end{lemma}

On $\C^n$, we denote by $\langle z,w \rangle = \sum_{i} \ol{z}_i w_i$ the complex inner product and by $\dd z$ the Lebesgue measure.

\begin{lemma}
\label{complex_Gaussian}
Let $\cal C$ be a complex and symmetric $n \times n$ matrix with $\re \cal C=(\cal C+{\cal C}^*)/2>0$. Then we have
\begin{equation*}
\int_{\C^n} \dd \b z \, \ee^{-\scalar{\b z}{{\cal C}^{-1} \b z}} = \pi^n \det \cal C\,.
\end{equation*}
\end{lemma}
Hence, for $\cal C$ a complex and symmetric $n \times n$ matrix with $\re \cal C>0$, let us define the Gaussian probability measure on $\mathbb{C}^n$ with covariance $\cal C$ as
\begin{equation}
\label{mu_A_complex}
\mu_{\cal C}(\dd \b z) \deq \frac{1}{\pi^{n} \det \cal C} \, \ee^{-\scalar{\b z}{{\cal C}^{-1} \b z}} \, \dd \b z\,.
\end{equation}
\begin{lemma}
\label{complex_Wick_theorem}
Let $\cal C$ be a complex and symmetric $n \times n$ matrix with $\re \cal C>0$ and let $\mu_{\cal C}$ be given as in \eqref{mu_A_complex}. For 
$i_1, \ldots, i_p, j_1, \ldots, j_p \in \{1,\ldots,n\}$, we have
$\int_{\mathbb{C}^n} \mu_{\cal C}(\dd \b z)\,\bar{z}_{j_1}\,\cdots \,\bar{z}_{j_p}\,z_{i_1}\,\cdots z_{i_p}
= \sum_{\pi \in S_p} \prod_{k=1}^{p} {\cal C}_{i_k, j_{\pi(k)}}$.
\end{lemma}

\section{The heat kernel on the lattice}
\label{The heat kernel on the lattice}

We note several useful estimates for the heat kernel $\psi^{L,t}$ on the finite lattice $\Lambda_L$ and on the infinite lattice $\Z^d$.
\begin{lemma} 
\label{estimates_heat_kernel}
\begin{itemize}
\item[(i)] $0 \leq \psi^{L,t}(x) \leq 1$.
\item[(ii)] $\psi^{L,t}(x)=\delta(x)+ O(t)$. Here $\delta$ denotes the Kronecker delta function on $\Lambda_L$.
\item[(iii)] $\int_{\Lambda_L}  \dd x \,\psi^{L,t}(x)=1$.
\end{itemize}
\end{lemma}

For the following estimates, we use Fourier analysis. We denote the dual lattice by $\Lambda_L^* \deq \frac{2 \pi }{L} \Lambda_L$. The heat kernel $\psi^{L,t}$ on $\Lambda_L$ can be written as
\begin{equation}
\label{I(x)_1}
\psi^{L,t}(x) = \frac{1}{L^d} \sum_{\xi \in \Lambda_L^*} \ee^{-t \lambda_\xi} \, \ee^{\ii \xi \cdot x}\,, \qquad \lambda_\xi  \deq \pbb{d-\sum_{j=1}^{d} \cos \xi_j}\,.
\end{equation}
For $L = \infty$, we have
\begin{equation}
\label{I(x)}
\psi^{\infty,t}(x) = \frac{1}{(2 \pi)^d} \int_{[-\pi,\pi)^d} \dd \xi \,  \ee^{-t \lambda_\xi} \, \ee^{\ii \xi \cdot x}\,.
\end{equation}

\begin{lemma}
\label{heat_kernel_estimate_infinite_lattice}
The following estimates hold.
\begin{itemize}
\item[(i)]There exist constants $c_1,c_2>0$ with $c_2$ depending on $d$ such that for every $\delta \in (0,1)$, and $x,y \in \Lambda$ and $t \geq 0$, we have
\begin{equation} 
\label{psi_infty_bound}
\psi^{\infty,t}(x) = O\Bigl(\ee^{c_1 t \delta}\,\ee^{-c_2 \delta |x|}\Bigr)\,.
\end{equation}
\item[(ii)] Given $L \in \N^*$,  $x \in \Lambda_L$, and $t \geq 0$, we have 
\begin{equation} 
\label{psi_L_bound}
\psi^{L,t}(x)= O_d\biggl(\frac{1}{\delta}\,\ee^{c_1 t \delta}\,\ee^{-c_2 \delta |x|_L}\biggr)\,,
\end{equation}
where $c_1,c_2>0$ and $\delta \in (0,1)$ are as in part (i). 
\end{itemize}
\end{lemma}

\begin{proof}
We first prove (i). We can assume without loss of generality, that $|x_1|=\max_{1 \leq i \leq d} |x_i|$. In particular, we have that $|x_1| \geq \frac{1}{\sqrt{d}}|x|$. We then rewrite \eqref{I(x)} as
\begin{equation} 
\label{I(x)_2}
\psi^{\infty,t}(x) =\frac{1}{(2\pi)^d} \int_{-\pi}^{\pi}\dd \xi_2 \cdots \int_{-\pi}^{\pi}\dd \xi_d \, \ee^{-t \bigl[(d-1)-\sum_{j=2}^{d} \cos (\xi_j) \bigr]}\,\ee^{\ii \sum_{j=2}^{d} \xi_j  x_j}\, \int_{-\pi}^{\pi} \dd \xi_1\,\ee^{-t [1-\cos (\xi_1) ]} \, \ee^{\ii \xi_1 x_1}\,.
\end{equation}
By a contour deformation, we can rewrite the $\xi_1$ integral in \eqref{I(x)_2} as 
\begin{equation} 
\label{I(x)_3}
 \int_{-\pi}^{\pi} \dd \xi_1\,\ee^{-t [1-\cos (\xi_1\pm \ii \delta) ]} \, \ee^{\ii (\xi_1\pm \ii \delta) x_1}\,,
\end{equation}
where the sign is taken to be $+$ if $x_1 \geq 0$ and $-$ otherwise. Therefore, since $\cos(\xi_1 \pm  \ii \delta) -1 \leq C \delta$, we deduce that the expression in \eqref{I(x)_3} is
\begin{equation} 
\label{I(x)_4}
=O\Bigl( \ee^{C t \delta} \,\ee^{-\delta |x_1|}\Bigr)= O \biggl( \ee^{C t \delta} \,\ee^{-\frac{\delta |x|}{\sqrt{d}}}\biggr)\,.
\end{equation}
Substituting \eqref{I(x)_3}--\eqref{I(x)_4} into \eqref{I(x)_2}, we deduce (i). 
In order to show (ii), we note that, by periodicity we have for $x \in \Lambda_L$
\begin{equation}
\label{psi_L_bound_proof_1}
\psi^{L,t}(x)=\sum_{k \in (L\Z)^d}\psi^{\infty,t}(x+k)\,.
\end{equation}
Using \eqref{psi_infty_bound} for each term on the right-hand side of \eqref{psi_L_bound_proof_1}, and considering Riemann sums, we deduce \eqref{psi_L_bound}. 
\end{proof}

\section{Proofs of auxiliary claims from Section \ref{The infinite-volume limit}}
\label{Appendix C}
\subsection{Kruskal's algorithm and proof of Lemma \ref{Ursell_function_estimate}}
\label{Appendix C.1}
In this subsection, we give an outline of Kruskal's algorithm, which we then use to prove Lemma \ref{Ursell_function_estimate}.
Kruskal's algorithm \cite{Kruskal_1956} defines a map $\cal K: {\fra G}_n^c \rightarrow {\fra T}_n$ with the property that $\cal K(\cal G) \subset \cal G$ is a spanning tree of $\cal G \in {\fra G}_n^c$. 
For completeness, let us briefly recall the construction of the map $\cal K$. We first order all the edges of the complete graph on $n$ vertices  according to  an arbitrary (strict) linear order $<$. Given $\cal G \in {\fra G}_n^c$, we define the following sequence $(\cal F_k) \equiv (\cal F_k(\cal G))$ of \emph{forests} on $n$ vertices.
\begin{itemize}
\item[(i)] $\cal F_0 \deq \emptyset.$
\item[(ii)] Let $k \in \N$ be given. We find the smallest edge $e_{k+1} \in \cal G \setminus \cal F_{k}$ with the property that $\cal F_{k} \cup \{e_{k+1}\}$ contains no cycles, in which case we let $\cal F_{k+1} \deq \cal F_k \cup \{e_{k+1}\}$. If no such $e_{k+1}$ exists, we let $\cal{F}_{k+1} \deq \cal F_k$ and we terminate the procedure.
\end{itemize}
Given $\cal G \in  {\fra G}_n^c$, there exists $k \in \N$ such that the above procedure terminates at the $k$-th step. 
We then define $\cal K(\cal G) \deq \cal F_{k}.$
We note the following observation about the preimage of any tree under the Kruskal map.

\begin{lemma}
\label{Kruskal_algorithm_corollary}
Let $\cal T \in {\fra T}_n$ be given. Then there exists $M(\cal T) \in {\fra G}_n^c$ containing $\cal T$ such that 
\begin{equation*}
\cal K^{-1}(\cal T)=\{\cal G \in \fra{G}_n^c\,, \cal{T} \subset \cal{G} \subset {M}(\cal T)\}\,.
\end{equation*} 
\end{lemma}
\begin{proof}
We let
\begin{equation}
\label{M(T)}
M(\cal T) \deq \bigcup_{\cal G \in {\fra G}_n^c,\, \cal K(\cal G)=\cal T} \cal G\,.
\end{equation}
We obtain that \eqref{M(T)} satisfies the wanted properties if we show that the following three claims hold.
\begin{itemize}
\item[(i)] $\cal K (\cal T)=\cal T$.
\item[(ii)] Let $\cal G_1, \cal G_2 \in {\fra G}_n^c$ be such that $\cal K(\cal G_1)=\cal K(\cal G_2)=\cal T$. Then $\cal K(\cal G_1 \cup \cal G_2)=\cal T$.
\item[(iii)] Let $\cal G_1, \cal G_2 \in {\fra G}_n^c$ be such that $\cal T \subset \cal G_2 \subset \cal G_1$ and $\cal K(\cal G_1)=\cal T$. Then $\cal K(\cal G_2)=\cal T$.
\end{itemize}

Claims (i) and (iii) follow immediately from the construction of the Kruskal algorithm.
We now prove claim (ii). We argue by contradiction. Assume that $\cal K(\cal G_1 \cup \cal G_2)=\cal T'$ for some $\cal T' \in {\fra T}_n$ with $\cal T' \neq \cal T$. In particular, 
there exists $m \in \N$ such that 
\begin{equation}
\label{Kruskal_algorithm_corollary_proof1}
\cal F_m(\cal G_1 \cup \cal G_2) \neq \cal F_m(\cal G_1)=\cal F_m(\cal G_2)\,.
\end{equation}
Note that $m \geq 2$ since 
$\cal F_1(\cal G_1 \cup \cal G_2) = \cal F_1(\cal G_1)=\cal F_1(\cal G_2)$ consists of the smallest edge in $\cal T$. 
In particular, we have that 
\begin{equation}
\label{Kruskal_algorithm_corollary_proof2}
\cal F_{m-1}(\cal G_1 \cup \cal G_2) = \cal F_{m-1}(\cal G_1)=\cal F_{m-1}(\cal G_2)\,.
\end{equation}
By construction
\begin{equation}
\label{Kruskal_algorithm_corollary_proof3}
\cal F_m(\cal G_1 \cup \cal G_2)=\cal F_{m-1}(\cal G_1 \cup \cal G_2) \cup \{e\}
\end{equation}
for some edge $e \in \cal G_1 \cup \cal G_2$.
If $e \in \cal G_1$, then by the construction of the Kruskal algorithm, as well as \eqref{Kruskal_algorithm_corollary_proof2}--\eqref{Kruskal_algorithm_corollary_proof3}, we get that  $\cal F_m(\cal G_1)=\cal F_{m-1}(\cal G_1) \cup \{e\}=\cal F_{m}(\cal G_1 \cup \cal G_2)$, which contradicts \eqref{Kruskal_algorithm_corollary_proof1}. Analogously, we obtain a contradiction if $e \in \cal G_2$. Claim (ii) then follows.
\end{proof}

We now have the necessary tools to prove Lemma \ref{Ursell_function_estimate}.
\begin{proof}[Proof of Lemma \ref{Ursell_function_estimate}]
We show the claim by applying Kruskal's algorithm in \eqref{Ursell_function} and by resumming the contributions of edges that do not belong to the thus obtained spanning trees. Recalling \eqref{Ursell_function}, we have that
\begin{align}
\notag
&{\varphi}^{L}(\omega_1,\ldots,\omega_n) =\frac{1}{n!} \sum_{\cal T \in {\fra T}_n} \sum_{\cal G \in \cal K^{-1}(\cal T)} \prod_{\{i,j\} \in \cal G} \zeta^L(\omega_i,\omega_j)
\\
\notag
&=\frac{1}{n!} \sum_{\cal T \in {\fra T}_n} \prod_{\{i,j\} \in \cal T} \zeta^L(\omega_i,\omega_j) \sum_{\cal G \in \cal K^{-1}(\cal T)} \prod_{\{i,j\} \in \cal G \setminus \cal T} \Bigl(\ee^{-\cal{V}^{\nu,\lambda,L}(\omega_i,\omega_j)}-1\Bigr)
\\
\label{Ursell_function_estimate_proof}
&=
\frac{1}{n!} \sum_{\cal T \in {\fra T}_n} \prod_{\{i,j\} \in \cal T} \zeta^L(\omega_i,\omega_j)\,
{\ee^{-\sum_{\{i,j\} \in M(\cal T) \setminus \cal T} \cal{V}^{\nu,\lambda,L}(\omega_i,\omega_j)}}\,.
\end{align}
Note that in the last line we applied Lemma \ref{Kruskal_algorithm_corollary}.
The claim follows from \eqref{Ursell_function_estimate_proof} by recalling  \eqref{V_nonnegative}.
\end{proof}
\begin{remark}
\label{resummation_remark}
A similar method to bound the Ursell function has been applied in a more general context in \cite[Theorem 3.1]{Brydges_1984}.
\end{remark}

\subsection{Proof of Lemma \ref{Riemann_sum_lemma}}
\label{Appendix C.2}

In this subsection, we prove Lemma \ref{Riemann_sum_lemma}. 
\begin{proof}[Proof of Lemma \ref{Riemann_sum_lemma}]
Let us first consider the case when $q \geq 1$.
We estimate the expression on the left-hand side of \eqref{integration_lemma_i_3a_B} by analysing Riemann sums. We note that the function $f(t)=\ee^{-\kappa t}\, t^{q}$ is increasing on $[0,\frac{q}{k}]$ and decreasing on $[\frac{q}{k},\infty)$.
We choose $\ell_0 \in \N^*$ such that 
$\ell_0 \nu \leq \frac{q}{\kappa} < (\ell_0+1)\nu$.
By construction of $\ell_0$, we have that
\begin{equation}
\label{Riemann_sum_integration_lemma_1}
\nu \sum_{\substack{\tilde T \in \nu \N^* \\ \tilde T \leq (\ell_0-1)\nu}} \ee^{-\kappa \tilde T}\,\tilde T^{q}+\nu \sum_{\substack{\tilde T \in \nu \N^* \\ \tilde T \geq (\ell_0+2)\nu}} \ee^{-\kappa \tilde T}\,\tilde T^{q} \leq \int_0^{\infty} \dd t\,\ee^{-\kappa t}\, t^{q} = \frac{q!}{\kappa^{q+1}}\,.
\end{equation}
We now show that the bound in \eqref{Riemann_sum_integration_lemma_1} also holds for the terms with $\ell \in \{\ell_0,\ell_0+1\}$. We note that for $\ell \in \{\ell_0,\ell_0+1\}$ 
\begin{equation}
\label{Riemann_sum_integration_lemma_3}
\ee^{-\kappa \nu \ell} \lesssim \ee^{-q}\,,\quad (\nu \ell)^{q} \leq \biggl(\frac{q+1}{\kappa}\biggr)^{q}
\leq \ee\,\biggl(\frac{q}{\kappa}\biggr)^{q}\,.
\end{equation}
Using \eqref{Riemann_sum_integration_lemma_3}, applying Stirling's formula, and recalling \eqref{nu_small}, we deduce indeed that
$\nu\,\ee^{-\kappa \nu \ell}\,(\nu \ell)^{q} \lesssim \frac{q!}{\kappa^{q+1}}$.
For $q=0$, we also get the bound \eqref{integration_lemma_i_3a_B}, but the proof is simplified since the function $f(t)=\ee^{-\kappa t}$ is decreasing on $[0,\infty)$ and we can estimate the Riemann sum by the integral. 
\end{proof}

\subsection{Proofs of Lemmas \ref{integration_lemma_M} and \ref{integration_lemma_M_2}}
\label{Appendix C.3}
In this subsection, we prove Lemmas \ref{integration_lemma_M} and \ref{integration_lemma_M_2}, which were used in the proof of Proposition \ref{MF_cluster_expansion}.
Before proving Lemma \ref{integration_lemma_M}, we note the following fact that relates the path measures \eqref{W_path_measures} $\bb{W}^{L,T}_{y,x}$ on the box $\Lambda_L$ and the corresponding measures $\bb{W}^{\infty,T}_{y,x}$ on the whole lattice.

\begin{lemma}
\label{W_pi_L}
We define by $\pi_L: \Z^d \rightarrow \Lambda_L$ the canonical projection map
\begin{equation}
\label{pi_L}
\pi_L(x+k) \deq x \quad \text{for} \quad x \in \Lambda_L\,, k \in (L\Z)^d\,.
\end{equation}
With notation as in \eqref{pi_L}, we have
\begin{equation*}
\bb{W}^{L,T}_{y,x}(\dd \omega)=\sum_{k \in (L\Z)^d} \left(\pi_L\right)_{\sharp} \bb{W}^{\infty,T}_{y+k,x}(\dd \omega)
\end{equation*}
for all $T>0$ and $x,y \in \Lambda_L$.
\end{lemma}

\begin{proof}[Proof of Lemma \ref{W_pi_L}]
Given $m \in \N$, a continuous function $f: (\Lambda_L)^m \rightarrow \mathbb{C}$, and times $0<t_1<\cdots<t_m<T$, we have
\begin{multline}
\label{D.9}
\int_{\Omega^{L,T}} \sum_{k \in (L\mathbb{Z})^d} \bigl(\pi_{L} \bigr)_{\sharp} \bb{W}^{\infty,T}_{y+{k},x}(\dd \omega)\, f(\omega(t_1), \ldots, \omega(t_{m})) 
\\
=
\sum_{k \in (L\Z)^d} \int_{\Omega^{\infty,T}}  {\bb W}^{\infty,T}_{y+k,x}(\dd \tilde{\omega})\,  (f \circ \pi_L) (\tilde{\omega}(t_1), \ldots, \tilde{\omega}(t_m))\,.
\end{multline}
We rewrite \eqref{D.9} as
\begin{multline}
\label{D.10}
\sum_{k \in (L\Z)^d} \int_{\Z^d} \dd x_{1} \cdots \int_{\mathbb{Z}^d} \dd x_m \, \psi^{\infty,t_{1}}(x_{1}-x)\, \psi^{\infty,t_{2}-t_{1}}(x_{2}-x_{1})\, \cdots \psi^{\infty,T-t_{m}}(y+k-x_{m})\, 
\\
\times
f(\pi_{L} x_{1}, \ldots,\pi_L x_{m})
\\
=\sum_{k,k_{1},\ldots,k_{m} \in (L\mathbb{Z})^d} \int_{\Lambda_{L} + k_{1}} \dd x_{1} \cdots \int_{\Lambda_{L}+k_{m}} \dd x_{m} \,\psi^{\infty,t_{1}}(x_{1}-x)\, \psi^{\infty,t_{2}-t_{1}}(x_{2}-x_{1}) \times \cdots\\
\times
\cdots  \psi^{\infty,t_{m}-t_{m-1}} (x_{m}-x_{m-1})\,\psi^{\infty,T-t_{m}}(y+k-x_{m})\,f(\pi_{L} x_{1}, \ldots, \pi_{L} x_{m})
\\
=\sum_{k,k_{1},\ldots,k_{m} \in (L\Z)^d} \int_{\Lambda_{L}} \dd x_{1} \cdots \int_{\Lambda_{L}} \dd x_{m} \,\psi^{\infty,t_{1}}(x_{1}-x+k_{1})\, \psi^{\infty,t_{2}-t_{1}}(x_{2}-x_{1}+k_{2}-k_{1}) \times \cdots\\
\times
\cdots  \psi^{\infty,t_{m}-t_{m-1}} (x_{m}-x_{m-1}+k_{m}-k_{m-1})\,\psi^{\infty,T-t_{m}}(y-x_{m}+k-k_{m})\,f(x_{1}, \ldots,x_{m})
\,.
\end{multline}
We recall \eqref{psi_L_bound_proof_1}, and then we sum in $k$, followed by $k_{m},k_{m-1}, \ldots, k_{1}$ to deduce that 
\begin{multline}
\label{D.11}
\eqref{D.10}=\int_{\Lambda_{L}} \mathrm{d} x_{1}\, \cdots \int_{\Lambda_{L}} \mathrm{d} x_{m}\, \psi^{L,t_{1}}(x_{1}-{x})\, 
\psi^{L,t_{2}-t_1}(x_2-x_1)\, \cdots\, \psi^{L,t_{m}-t_{m-1}}(x_{m}-x_{m-1})\, 
\\
\times 
\psi^{L,T-t_{m}}(y-x_{m})\, f(x_{1}, \ldots, x_{m})
=\int_{\Omega^{L,T}} {\bb W}^{L,T}(\dd \omega) \, f(\omega(t_{1}),\ldots,\omega(t_{m}))\,. 
\end{multline}
The claim follows from \eqref{D.11}.
\end{proof}

\begin{proof}[Proof of  Lemma \ref{integration_lemma_M}]
We first prove (i). The argument is based on that of the proof of Lemma \ref{integration_lemma} (i). Instead of \eqref{integration_lemma_i_1}, we need to estimate 
\begin{multline} 
\label{integration_lemma_i_1_M}
\int \mu^{L} (\dd \tilde \omega)\, T(\tilde \omega)^{q} \, \cal{V}^{\nu,\nu^2,L}(\omega,\tilde \omega) \,\ind{\cal D^{L}_{c}}(\tilde \omega) \leq \nu^2 \sum_{\tilde T \in \nu\N^*} \ee^{-\kappa \tilde T}\,\tilde{T}^{q-1}\,  
\\
\times \sum_{r \in \nu \N} \ind{r<T(\omega)}\sum_{s\in \nu \N}\ind{s<\tilde T}\,
\int_0^{\nu} \dd t\, \biggl\{ \int_{\Lambda_L}  \dd x\,\int \bb W^{L,\tilde T}_{x,x}(\dd \tilde \omega)\,v^{L}\pb{\omega(t + r) - \tilde \omega(t+s)} \,\ind{\cal D^{L}_{c}}(\tilde \omega)\biggr\}\,.
\end{multline}
By arguing as for \eqref{integration_lemma_i_2_a}--\eqref{integration_lemma_i_2}, we can rewrite the expression in the curly brackets in \eqref{integration_lemma_i_1_M} as
\begin{equation}
\label{integration_lemma_i_2_M}
\int_{\Lambda_L}  \dd z\,v^{L}(y-z)\, \biggl[\int \bb W^{L,\tilde T}_{z,z} (\dd \tilde \omega)\,\ind{\cal D^{L}_{c}}(\tilde \omega)\biggr]\,.
\end{equation}
We now estimate the expression in square brackets in \eqref{integration_lemma_i_2_M} for a fixed $z \in \Lambda_L$. Let us recall \eqref{D^{L,c}} and note that by the triangle inequality, we have
\begin{equation}
\label{D^{L}_{c}_inclusion_1}
\cal{D}^{L}_{c} \subset \overline{\cal{D}}^{L}_{c}(z)\,,
\end{equation}
where
\begin{equation}
\label{overline_D^{L}_{c}(z)_definition}
\overline{\cal{D}}^{L}_{c}(z):=\left\{\omega \in \bigcup_{T>0} \Omega^{L,T}\,,\,|\omega(s)-z|_L \geq \frac{cL}{2} \,\, \mbox{for some} \,\, s\in [0,T(\omega)]\right\}\,.
\end{equation}
Furthermore, we define
\begin{equation}
\label{D^{L}_{c}(z)_definition}
\cal{D}^{L}_{c}(z):=\left\{\omega \in \bigcup_{T>0} \Omega^{\infty,T}\,,\,|\omega(s)-z| \geq \frac{cL}{2} \,\, \mbox{for some} \,\, s\in [0,T(\omega)]\right\}\,.
\end{equation}
From \eqref{pi_L}, \eqref{overline_D^{L}_{c}(z)_definition}--\eqref{D^{L}_{c}(z)_definition}, and the definition of the periodic norm $|\cdot|_L$, it follows that 
\begin{equation}
\label{D^{L}_{c}_inclusion_2}
\overline{\cal{D}}^L_c(z) \subset \pi_L \circ \cal{D}^L_c(z)\,.
\end{equation}
We now use Lemma \ref{W_pi_L}, \eqref{D^{L}_{c}_inclusion_1}, and \eqref{D^{L}_{c}_inclusion_2} to estimate 
\begin{multline}
\label{5.67*}
\int \bb{W}^{L,\tilde{T}}_{z,z}(\mathrm{d} \tilde{\omega})\,\mathbf{1}_{\cal{D}^{L}_{c}}(\tilde{\omega}) \leq 
\int \bb{W}^{\infty,\tilde{T}}_{z,z}(\mathrm{d} \tilde{\omega})\,\mathbf{1}_{\cal{D}^{L}_{c}(z)}(\tilde{\omega})+
\sum_{k \in (L\mathbb{Z})^d \setminus \{0\}} \int \bb{W}^{\infty,\tilde{T}}_{z+k,z}(\mathrm{d} \tilde{\omega})\,\mathbf{1}_{\cal{D}^{L}_{c}(z)}(\tilde \omega)
\\
\leq 
\int\bb{W}^{\infty,\tilde{T}}_{z,z}(\mathrm{d} \tilde \omega)\,\mathbf{1}_{\cal{D}^{L}_{c}(z)}(\tilde \omega)+ \sum_{k \in (L\mathbb{Z})^d \setminus \{0\}} \psi^{\infty,\tilde T}
(k) \eqd I+II\,.
\end{multline}
By Definition \ref{o_L_definition}, \eqref{D^{L}_{c}(z)_definition}, \eqref{I(x)} and the dominated convergence theorem, the contribution $I$ from \eqref{5.67*} satisfies
\begin{equation}
\label{5.67*_I}
I=\int\bb{W}^{\infty,\tilde{T}}_{z,z}(\mathrm{d} \tilde \omega)\,\mathbf{1}_{\cal{D}^{L}_{c}(z)}(\tilde \omega)=\int\bb{W}^{\infty,\tilde{T}}_{0,0}(\mathrm{d} \tilde \omega)\,\mathbf{1}_{\cal{D}^{L}_{c}(0)}(\tilde \omega)=o_L \bigl(\psi^{\infty,\tilde{T}}(0)\bigr)=o_L(1)\,.
\end{equation}
By Lemma \ref{heat_kernel_estimate_infinite_lattice} (i),  the contribution $II$ from \eqref{5.67*} satisfies
\begin{equation}
\label{5.67*_II}
II \lesssim  \sum_{k \in (L\mathbb{Z})^d \setminus \{0\}} 
\ee^{\kappa \tilde{T}/2}\,\ee^{-C|k|} \lesssim \ee^{\kappa \tilde{T}/2}\, \ee^{-CL}\,.
\end{equation}
We now substitute \eqref{5.67*}--\eqref{5.67*_II} into \eqref{integration_lemma_i_1_M}--\eqref{integration_lemma_i_2_M} and argue as in the proof of Lemma \ref{integration_lemma} (i) to obtain claim (i).

We now show (ii). 
Recalling \eqref{v^{L}} and Assumption \ref{interaction_potential_v^{L}} (i), (iii), and (iv), we obtain that 
for $x \in \Lambda_L$
\begin{equation*}
0 \leq v^L(x) \lesssim_d \|v\|_{{\ell^{\infty}}} + \sum_{k \in (L\Z)^d \setminus \{0\}} v(k) \lesssim \|v\|_{{\ell^{\infty}}} + \|v\|_{\ell^{1}}\,.
\end{equation*}
Hence, we have
\begin{equation}
\label{v^L_l_infty}
\|v^L\|_{\ell^{\infty}(\Lambda_L)} \lesssim_d \|v\|_{{\ell^{\infty}}} + \|v\|_{\ell^{1}}\,,
\end{equation}
uniformly in $L$.

Arguing similarly as in \eqref{integration_lemma_ii_1}
 and using \eqref{D^{L}_{c}_inclusion_1}, \eqref{v^L_l_infty}, we obtain 

\begin{multline}
\label{integration_lemma_ii_1_M}
\nu \int_{\Lambda_{L_0}} \dd y\,\int \hat{\mu}^{L}_{y,x} (\dd\tilde \omega) \, T(\tilde \omega)^{q}\,\cal{V}^{\nu,\nu^2,L}(\omega,\tilde \omega)\,\ind{\cal D^{L}_{c}}(\tilde \omega)
\\
\leq \nu^2 \int_{\Lambda_{L_0}}  \dd y\, \sum_{\tilde T \in \nu \N^*} \ee^{-\kappa \tilde T}\,\tilde T^{q} \sum_{r \in \nu \N} \ind{r<T(\omega)} \sum_{s\in \nu \N} \ind{s<\tilde T}\int \bb W^{L,\tilde T}_{y,x}(\dd \tilde \omega)\,\ind{\overline{\cal D}^{L}_{c}(x)}(\tilde \omega)
\\
\times
\int_0^{\nu} \dd t\,v^{L}\pb{\omega(t + r) - \tilde \omega(t+s)}
\\
\lesssim_d \nu \sum_{\tilde T \in \nu \N^*} \ee^{-\kappa \tilde T}\,\tilde T^{q+1}\,T(\omega)\,
\bigg\{\int_{\Lambda_{L_0}} \dd y\, \int \bb W^{L,\tilde T}_{y,x}(\dd \tilde \omega)\,\ind{\overline{\cal D}^{L}_{c}(x)}(\tilde \omega)\biggr\}\,\bigl(\|v\|_{{\ell^{\infty}}} + \|v\|_{\ell^{1}}\bigr)
\,.
\end{multline}
We now estimate the expression in curly brackets in \eqref{integration_lemma_ii_1_M} by using Lemma \ref{W_pi_L}, \eqref{D^{L}_{c}_inclusion_2}, and by arguing as for \eqref{5.67*} above. In particular, we obtain 
\begin{multline}
\label{5.72*}
\int_{\Lambda_{L_0}} \dd y\, \int \bb W^{L,\tilde T}_{y,x}(\dd \tilde \omega)\,\ind{\overline{\cal D}^{L}_{c}(x)}(\tilde \omega)
\\
\leq \int_{\Lambda_{L_0}} \dd y\, \int \bb W^{\infty,\tilde T}_{y,x}(\dd \tilde \omega)\,\ind{\cal D^{L}_{c}(x)}(\tilde \omega)+\sum_{k \in (L \Z)^d \setminus \{0\}} \int_{\Lambda_{L_0}} \dd y\, \int \bb{W}^{\infty,\tilde{T}}_{y+k,x}(\dd \tilde{\omega}) \eqd \widetilde{I}+\widetilde{II}\,.
\end{multline}
We note that the contribution $\widetilde{I}$ to \eqref{5.72*} satisfies
\begin{equation}
\label{5.73*}
\widetilde{I}=\int_{\Lambda_{L_0}} \dd y \, \int \bb{W}^{\infty,\tilde{T}}_{y,x}(\mathrm{d} \tilde{\omega})\,\ind{\cal D^{L}_{c}(x)}(\tilde{\omega})=o_L(1)\,,
\end{equation}
uniformly in $x \in \Lambda_{L_{0}}$.
Namely, by translation invariance, we have 
\begin{multline}
\label{5.73*_A}
\int_{\Lambda_{L_0}} \dd y \, \int \bb{W}^{\infty,\tilde{T}}_{y,x}(\mathrm{d} \tilde{\omega})\,\ind{\cal D^{L}_{c}(x)}(\tilde{\omega})=\int_{\Lambda_{L_0}} \dd y \, \int \bb{W}^{\infty,\tilde{T}}_{y-x,0}(\mathrm{d} \tilde{\omega})\,\ind{\cal D^{L}_{c}(0)}(\tilde{\omega})
\\
\leq \int_{\Z^d} \dd y \, \int \bb{W}^{\infty,\tilde{T}}_{y,0}(\mathrm{d} \tilde{\omega})\,\ind{\cal D^{L}_{c}(0)}(\tilde{\omega})
\,.
\end{multline}
Moreover
\begin{equation}
\label{5.73*_B}
\int_{\Z^d} \dd y \, \int \bb{W}^{\infty,\tilde{T}}_{y,0}(\mathrm{d} \tilde{\omega}) = \int_{\Z^d} \dd y\,\psi^{\infty,\tilde{T}}(y)=1\,.
\end{equation}
Hence \eqref{5.73*_A}--\eqref{5.73*_B} and the dominated convergence theorem imply that \eqref{5.73*} holds uniformly in $x \in \Lambda_{L_{0}}$.

We now estimate the contribution $\widetilde{II}$ to \eqref{5.72*}. By recalling \eqref{L_0_choice} and by using Lemma \ref{heat_kernel_estimate_infinite_lattice} (ii), we have 
\begin{multline}
\label{5.75*}
\widetilde{II}=\sum_{k \in (L\Z)^d \setminus \{0\}} \int_{\Lambda_{L_0}} \dd y \, \psi^{\infty,\tilde{T}}(y+k-x) 
\leq \int_{\Z^d} \dd y \, \ind{|y| \gtrsim L}\,\psi^{\infty,\tilde{T}}(y) \lesssim 
\int_{\Z^d} \dd y \, \ind{|y| \gtrsim L}\,\ee^{\kappa \tilde{T}/2} \,\ee^{-C|y|}
\\
\lesssim \ee^{\kappa \tilde{T}/2} \,\ee^{-CL}\,.
\end{multline}
We now substitute \eqref{5.72*}, \eqref{5.73*}, and \eqref{5.75*} into \eqref{integration_lemma_ii_1_M}  and argue as in the proof of Lemma \ref{integration_lemma} (ii) to obtain claim (ii).

The proof of claim (iii) is analogous to that of Lemma \ref{integration_lemma} (iii). Namely, instead of \eqref{integration_lemma_iii_1_a}, we consider 
\begin{equation}
\label{integration_lemma_iii_1_a}
\int \mu^{L} (\dd \tilde \omega) \, T(\tilde \omega)^{q}\,\ind{\cal D^{L}_{c}}(\tilde \omega)
 \leq  \nu\, \sum_{\tilde T \in \nu \N^*} \ee^{-\kappa \tilde T}\,\tilde T^{q-1}  \int_{\Lambda_L}  \dd x \,\int \bb W^{L,\tilde T}_{x,x}(\dd \tilde \omega)\,\ind{\cal D^{L}_{c}}(\tilde \omega)\,.
\end{equation}
We get the wanted bound by arguing as for \eqref{integration_lemma_iii_1}.

We now prove claim (iv).
Using \eqref{mu_measure_2} and \eqref{D^{L}_{c}_inclusion_1}, we have
\begin{equation}
\label{5.77*}
\nu\,\int_{\Lambda_{L_0}} \dd y\,\int \hat{\mu}^{L}_{y,x}(\dd \tilde \omega) \,T(\tilde \omega)^{q} \,\ind{\cal D^{L}_{c}}(\tilde \omega)
\leq \nu \sum_{\tilde{T} \in \nu \N^*} \ee^{-\kappa \tilde{T}}\,\tilde{T}^q\, \biggl\{\int_{\Lambda_{L_0}} \dd y \, \int \bb{W}^{\infty,\tilde{T}}_{y,x}(\mathrm{d} \tilde{\omega})\,\ind{\cal D^{L}_{c}(x)}(\tilde{\omega})\biggr\}\,.
\end{equation}
We note that the quantity in curly brackets in \eqref{5.77*} is the one estimated in \eqref{5.72*} above. Therefore, the estimate of claim (iii) follows from the proof of claim (ii). 
\end{proof}

\begin{proof}[Proof of Lemma \ref{integration_lemma_M_2}]
By recalling \eqref{zeta^{L,c}} and arguing analogously as in the proof of Lemma \ref{integration_lemma} (i) and (ii), we deduce the following estimates.
\begin{itemize}
\item[(1)] $\int \mu^{L} (\dd \tilde \omega)\,  T(\tilde \omega)^{q} \,|\zeta^{L}_{c}(\omega,\tilde \omega)| \lesssim \frac{T(\omega)}{\kappa^{q+1}}\,q!\,\|v^{L}_{c}\|_{\ell^1(\Lambda_L)}\,.$
\item[(2)] $\int_{\Lambda_{L_0}} \dd y\,\int \hat{\mu}^{L}_{y,x} (\dd \tilde \omega) \, T(\tilde \omega)^{q}\,|\zeta^{L}_{c}(\omega,\tilde \omega)| \lesssim \frac{T(\omega)}{\kappa^{q+2}}\,(q+1)!\,\|v^{L}_{c}\|_{\ell^1(\Lambda_L)}$ for all $x \in \Lambda_{L_0}$. 
\end{itemize}
The claim now follows.
\end{proof}

\subsection{Proofs of Lemmas \ref{integration_lemma_M_large_mass} and  \ref{integration_lemma_M_2_large_mass}}
\label{Appendix C.4}

In this subsection, we prove Lemmas \ref{integration_lemma_M_large_mass} and \ref{integration_lemma_M_2_large_mass}. 

Before proceeding with the proof of Lemma \ref{integration_lemma_M_large_mass}, we note the following result about the continuous-time random walk on $\Lambda_L$.

\begin{lemma}
\label{Lemma_5.21*} 
Recalling  \eqref{D^{L,c}}, we have that, uniformly in $z \in \Lambda_L$
\begin{equation}
\label{Lemma_5.21*_bound} 
\int \bb W^{L,\tilde{T}}_{z,z}(\dd \tilde{\omega})\,\ind{\cal D^{L}_{c}}(\tilde \omega) \lesssim_{\kappa_0,d} \frac{1}{L^{2d}}\,\ee^{\frac{\kappa_0 \tilde{T}}{2\nu}}\,\bigl(1+\tilde{T}^{\frac{3d}{2}+1}\bigr)\,.
\end{equation}
\end{lemma}

\begin{proof}[Proof of Lemma \ref{Lemma_5.21*}]
By using the triangle inequality and Lemma \ref{W_pi_L}, it suffices to estimate (with a different choice of $c$) the quantity
\begin{multline}
\label{Lemma_5.21*_1}
\int \bb W^{L,\tilde{T}}_{z,z}(\dd \tilde{\omega})\,\ind{\mathrm{\sup}_{0 \leq t \leq \tilde{T}}\, |\tilde{\omega}(t)|_L \geq cL}(\tilde \omega)
=\sum_{k \in (L\Z)^d} \int {\bb W}^{\infty,\tilde{T}}_{z+k,z}(\dd \omega)\,
\,\ind{\mathrm{\sup}_{0 \leq t \leq \tilde{T}}\, |\pi_L \circ\, \omega(t)|_L \geq cL}(\omega)
\\
\leq 
\sum_{k \in (L\Z)^d} \int {\bb W}^{\infty,\tilde{T}}_{z+k,z}(\dd \omega)\,
\,\ind{\mathrm{\sup}_{0 \leq t \leq \tilde{T}}\, |\omega(t)| \geq cL}(\omega)\,.
\end{multline}
Here, we recalled \eqref{pi_L} and the construction of $|\cdot|_L$.

By arguing as in \eqref{5.67*_II} and by dropping the indicator function, we have that 
\begin{equation}
\label{Lemma_5.21*_1_B}
\sum_{k \in (L\Z)^d \setminus \{0\}} \int {\bb W}^{\infty,\tilde{T}}_{z+k,z}(\dd \omega)\,
\,\ind{\mathrm{\sup}_{0 \leq t \leq \tilde{T}}\, |\omega(t)| \geq cL}(\omega)
\lesssim  \ee^{\frac{\kappa_0 \tilde{T}}{2\nu}}\,\ee^{-\frac{C \kappa_0 L}{\nu}} \lesssim_{\kappa_0,d} \frac{1}{L^{2d}}\,\ee^{\frac{\kappa_0 \tilde{T}}{2\nu}}\,.
\end{equation}
Let us write the leading term in \eqref{Lemma_5.21*_1} as
\begin{multline}
\label{Lemma_5.21*_2}
\int {\bb W}^{\infty,\tilde{T}}_{z,z}(\dd \omega)\,
\,\ind{\mathrm{\sup}_{0 \leq t \leq \tilde{T}}\, |\omega(t)| \geq cL}(\omega)
=\psi^{0,\tilde{T}}(0)\,\mathbb{P}^{\infty,\tilde{T}}_{z,z}\Biggl(\mathop{\mathrm{\sup}}_{0 \leq t \leq \tilde{T}}\, |\omega(t)| \geq cL \Biggr)
\\
\leq \mathbb{P}^{\infty,\tilde{T}}_{z,z}\Biggl(\mathop{\mathrm{\sup}}_{0 \leq t \leq \tilde{T}}\, |\omega(t)| \geq cL \Biggr)\,.
\end{multline}
We rewrite \eqref{Lemma_5.21*_2} as 
\begin{multline}
\label{Lemma_5.21*_3}
\frac{1}{\mathbb{P}^{\infty}_{z}\bigl(\omega(\tilde{T})=z\bigr)}\,\mathbb{P}^{\infty}_{z}\Biggl(\mathop{\mathrm{\sup}}_{0 \leq t \leq \tilde{T}}\, |\omega(t)| \geq cL \, \bigcap \, 
\{\omega(\tilde{T})=z\}\Biggr) 
\\
\lesssim (1+\tilde{T}^{d/2})\,\mathbb{P}^{\infty}_{z}\Biggl(\mathop{\mathrm{\sup}}_{0 \leq t \leq \tilde{T}}\, |\omega(t)| \geq cL \, \bigcap \, 
\{\omega(\tilde{T})=z\}\Biggr)
\leq (1+\tilde{T}^{d/2})\,\mathbb{P}^{\infty}_{z}\Biggl(\mathop{\mathrm{\sup}}_{0 \leq t \leq \tilde{T}}\, |\omega(t)| \geq cL\Biggr)\,.
\end{multline}
In \eqref{Lemma_5.21*_3}, we used the observation that for $T \gtrsim 1$, we have
\begin{equation*}
\mathbb{P}^{\infty}_{z}\bigl(\omega(\tilde{T})=z\bigr)=\psi^{\infty,\tilde{T}}(0) \sim \frac{1}{\tilde{T}^{d/2}}\,,
\end{equation*}
which follows from \eqref{I(x)_1}--\eqref{I(x)} by stationary phase arguments. For $T \ll 1$, we have $\psi^{\infty,\tilde{T}}(0) \gtrsim 1$ from \eqref{I(x)_1}--\eqref{I(x)}.

Let us for now consider the case when $d=1$. Then, the continuous time random walk $\omega$ can be written as
\begin{equation}
\label{Lemma_5.21*_4}
\omega(t)=\sum_{n \geq 0} \ind{T_n \leq t <T_{n+1}}\,X_n\,,
\end{equation}
where $(X_n)_{n \geq 0}$ is a simple random walk on $\Z$ and $0=T_0<T_1<T_2<\cdots$ are random variables with $\mathrm{exp}(1)$ increments.
We let 
\begin{equation}
\label{Lemma_5.21*_5}
N(t):=\sum_{n \geq 0}\ind{T_n<t}\,.
\end{equation}
By \eqref{Lemma_5.21*_4}--\eqref{Lemma_5.21*_5}, we have, by using a union bound that
\begin{multline}
\label{Lemma_5.21*_6}
\mathbb{P}^{\infty}_{z}\Biggl(\mathop{\mathrm{\sup}}_{0 \leq t \leq \tilde{T}}\, |\omega(t)| \geq cL\Biggr)=\mathbb{P}^{\infty}_{z}\Biggl(\mathop{\mathrm{\sup}}_{0 \leq n \leq N(T)}\, |X_n| \geq cL\Biggr)
\\
=\sum_{m=0}^{\infty} \mathbb{P}^{\infty}_{z}\Biggl(\mathop{\mathrm{\sup}}_{0 \leq n \leq m}\, |X_n| \geq cL,\,N(T)=m\Biggr)
= \sum_{m=0}^{\infty} \mathbb{P}^{\infty}_{z}\Biggl(\mathop{\mathrm{\sup}}_{0 \leq n \leq m}\, |X_n| \geq cL\Biggr)\,\ee^{-T}\,\frac{T^m}{m!}
\\
\leq 
\sum_{m=0}^{\infty} \sum_{n=0}^{m }\mathbb{P}^{\infty}_{z}\bigl( |X_n| \geq cL\bigr)\,\ee^{-T}\,\frac{T^m}{m!} \leq \sum_{m=0}^{\infty} \sum_{n=0}^{m}\ee^{-\frac{CL^2}{n}}\,\ee^{-T}\,\frac{T^m}{m!} 
\\
\leq \sum_{m=0}^{\infty} m\,\ee^{-\frac{CL^2}{m}}\,\ee^{-T}\,\frac{T^m}{m!}\,.
\end{multline}
By considering coordinates of $\omega$ and by suitably changing the constants, the estimate 
\eqref{Lemma_5.21*_6} holds for every dimension $d$.
Let us note that, for $K \in \N^*$, we have
\begin{multline}
\label{Lemma_5.21*_7}
\eqref{Lemma_5.21*_6} \lesssim_K \sum_{m=0}^{K} \frac{m^{K+1}}{L^{2K}}\,\ee^{-T}\,\frac{T^m}{m!}=\frac{1}{L^{2k}}\,\Biggl(\sum_{m=0}^{K} \ee^{-T}\, \frac{m^{K+1} T^m}{m!}+\sum_{m=K+1}^{\infty} \ee^{-T}\, \frac{m^{K+1} T^m}{m!}\Biggr)
\\
\lesssim_K 
\frac{1}{L^{2K}}\,\Biggl((1+T)^K+T^{K+1}\,\sum_{m=K+1}^{\infty} \ee^{-T}\,\frac{T^{m-k-1}}{(m-k-1)!}\Biggr) \lesssim \frac{1}{L^{2K}}\,(1+\tilde{T}^{K+1})\,.
\end{multline}
We now deduce \eqref{Lemma_5.21*_bound} from
\eqref{Lemma_5.21*_1}--\eqref{Lemma_5.21*_3}, and \eqref{Lemma_5.21*_6}--\eqref{Lemma_5.21*_7}, where in the latter we set $K=d$.
\end{proof}

\begin{proof}[Proof of Lemma \ref{integration_lemma_M_large_mass}]
The proof is similar to that of Lemma \ref{integration_lemma_M}. The main difference is that we need to analyse the case when there is a hard core. 
Let us first prove (i). We estimate the two terms that come from \eqref{zeta^L_bound}.
For the first term, we use \eqref{tilde_V_m}--\eqref{mu_measure_large_mass_1}, bound the indicator function $\ind{\cal I(\omega,\nu)}(\tilde \omega)$ by $1$ and estimate
\begin{multline*}
\int \mu^{L} (\dd \tilde \omega)\,  T(\tilde \omega)^{q}\,\ind{\cal I(\omega,\nu)}(\tilde \omega)\,\ind{\cal D^{L}_{c}}(\tilde \omega) 
\\
\leq \nu\sum_{\tilde T \in \nu\N^*} \ee^{-\kappa_0 \tilde T/\nu}\,\tilde T^{q-1}\,\int_{0}^{\nu}\dd t\,\sum_{r \in \nu\N} \ind{r<T(\omega)} \sum_{s \in \nu \N} \ind{s<\tilde T}
\int_{\Lambda_L} \dd z\, \int \bb W^{L,\tilde{T}}_{z,z}(\dd \tilde{\omega})\,\ind{\cal D^{L}_{c}}(\tilde \omega)\,,
\end{multline*}
which by Lemma \ref{Lemma_5.21*}, and by arguing as in \eqref{integration_lemma_large_mass_i_2} is 
\begin{equation}
\label{integration_lemma_M_large_mass_i_1}
\lesssim_{\kappa_0,d} \frac{1}{L^d} \sum_{\tilde{T} \in \nu \N^*} \ee^{-\frac{\kappa_0 \tilde{T}}{2\nu}}\,\tilde{T}^q\,T(\omega) (1+\tilde{T}^{3d/2+1}) \lesssim_{\kappa_0,q,d} \frac{1}{L^d}\,
T(\omega) \,\nu^{q}\,.
\end{equation}
We now estimate the second term coming from \eqref{zeta^L_bound}, i.e.\
\begin{equation}
\label{integration_lemma_M_large_mass_i_6}
\frac{1}{2}\,\int \mu^{L} (\dd \tilde \omega)\,  T(\tilde \omega)^{q}\,\cal V^{\nu,1,L,(2)}(\omega,\tilde \omega)\,\ind{\cal D^{L}_{c}}(\tilde \omega)\,.
\end{equation}

Arguing analogously as in the proof of Lemma \ref{integration_lemma_M} (i), we reduce to estimating \eqref{integration_lemma_i_1_M}
--\eqref{integration_lemma_i_2_M} with $v^L$ replaced by $v^{L,(2)}$. Arguing analogously as in \eqref{5.67*}--\eqref{5.67*_II}, we get that this quantity is 
\begin{equation}
\label{integration_lemma_M_large_mass_i_7_A}
=o_L\Bigl(\ee^{\frac{\kappa_0 \tilde T}{2\nu}}\,\|v^{(2)}\|_{\ell^1}\Bigr)\,.
\end{equation}
In particular, from \eqref{integration_lemma_M_large_mass_i_7_A}, we deduce that
\begin{equation}
\label{integration_lemma_M_large_mass_i_7}
\eqref{integration_lemma_M_large_mass_i_6} \lesssim o_L \Biggl( \frac{1}{\nu}\,\sum_{\tilde T \in \nu\N^*} \ee^{-\frac{\kappa_0 \tilde T}{2\nu}}\,\tilde T^{q}\,T(\omega)\,\|v^{(2)}\|_{\ell^1} \Biggr)\lesssim o_L \Biggl( \frac{T(\omega)}{\kappa_0^{q+3}}\,(q+2)!\,\nu^{q+1}\,\|v^{(2)}\|_{\ell^1}\Biggr)\,.
\end{equation}
Claim (i) follows from \eqref{zeta^L_bound}, \eqref{integration_lemma_M_large_mass_i_1}, and \eqref{integration_lemma_M_large_mass_i_7}. 

The proof of claim (ii) is similar. As in (i), we have to study the two terms that come from \eqref{zeta^L_bound}. By \eqref{tilde_V_m}, \eqref{mu_measure_large_mass_2}, and arguing as for \eqref{integration_lemma_ii_1_M} the first term is
\begin{equation}
\label{integration_lemma_M_large_mass_ii_1}
\leq \sum_{\tilde T \in \nu\N^*} \ee^{-\kappa_0 \tilde T/\nu}\,\tilde T^q\, \int_{0}^{\nu}\dd t\,\sum_{r \in \nu\N} \ind{r<T(\omega)} \sum_{s \in \nu \N} \ind{s<\tilde T}
\bigg\{\int_{\Lambda_{L_0}} \dd y\, \int \bb W^{L,\tilde T}_{y,x}(\dd \tilde \omega)\,\ind{\cal D^{L}_{c}}(\tilde \omega)\biggr\}\,.
\end{equation}
We estimate the quantity in curly brackets in \eqref{integration_lemma_M_large_mass_ii_1} as in \eqref{5.72*}--\eqref{5.75*} and deduce that 
\begin{equation}
\label{integration_lemma_M_large_mass_ii_1_B}
\eqref{integration_lemma_M_large_mass_ii_1} =
 \sum_{\tilde T \in \nu\N^*} \ee^{-\kappa_0 \tilde T/\nu}\,\tilde T^q\, \int_{0}^{\nu}\dd t\,\sum_{r \in \nu\N} \ind{r<T(\omega)} \sum_{s \in \nu \N} \ind{s<\tilde T} \,o_L\Bigl(\ee^{\frac{\kappa_0 \tilde{T}}{2\nu}}\Bigr) \lesssim_{\kappa_0,d,q} o_L(T(\omega) \nu^{q-1})\,.
\end{equation}
The second term coming from \eqref{zeta^L_bound} is 
\begin{align}
\notag
\leq \frac{2}{\nu}\,\sum_{\tilde T \in \nu\N^*} \ee^{-\kappa_0 \tilde T/\nu}&\,\tilde T^q\, \int_{0}^{\nu}\dd t\,\sum_{r \in \nu\N} \ind{r<T(\omega)} \sum_{s \in \nu \N} \ind{s<\tilde T}
\\
\label{integration_lemma_M_large_mass_ii_3}
&\int_{\Lambda_{L_0}} \dd y\,\int \bb W^{L,\tilde T}_{y,x}(\dd \tilde \omega)\,v^{L,(2)}\bigl(\omega(t+r)-\tilde \omega (t+s)\bigr)\,\ind{\cal D^{L}_{c}}(\tilde \omega)\,.
\end{align}
We now argue analogously as in the proof of Lemma \ref{integration_lemma_M} (ii) to deduce that 
\begin{equation}
\label{integration_lemma_M_large_mass_ii_4} 
\eqref{integration_lemma_M_large_mass_ii_3} \lesssim_{\kappa_0,q,d} o_L\Bigl(T(\omega)\,[\|v^{(2)}\|_{\ell^{\infty}}+\|v^{(2)}\|_{\ell^1}]\,\nu^{q-1}\Bigr) \,.
\end{equation}
Claim (ii) follows from \eqref{integration_lemma_M_large_mass_ii_1_B} and \eqref{integration_lemma_M_large_mass_ii_4}.
Claim (iii) follows by analogous arguments as in the proof of Lemma \ref{integration_lemma_M}  (iii).
\end{proof}

\begin{proof}[Proof of Lemma \ref{integration_lemma_M_2_large_mass}]
We use the estimate $|\zeta^{L,(2)}_{c}(\omega,\tilde \omega)| \leq \cal V^{\nu,1,L,(2)}_{c}(\omega,\tilde \omega)$, which follows from \eqref {zeta^{(2),L,c}} and the nonnegativity of $\cal V^{\nu,1,L,(2)}_{c}(\omega,\tilde \omega)$. We obtain (i) by arguing as in \eqref{integration_lemma_large_mass_i_3} above to deduce that 
\begin{equation*}
\int \mu^{L} (\dd \tilde \omega)\,  T(\tilde \omega)^{q}\,\cal V^{\nu,1,L,(2)}_{c}(\omega,\tilde \omega)
\lesssim  \frac{T(\omega)}{\tilde{\kappa}^{q+1}}\,q!\,\|v^{L,(2)}_{c}\|_{\ell^1(\Lambda_L)}\,\nu^{q-1}\,.
\end{equation*}
Note that the first term in the upper bound \eqref{zeta^L_bound} does not appear by construction of $\cal V^{\nu,1,L,(2)}_{c}(\omega,\tilde \omega)$.
Likewise, we obtain (ii) by arguing as in \eqref{integration_lemma_large_mass_ii_2} to deduce that
\begin{equation*}
\int_{\Lambda_L}\dd y\,\int \hat{\mu}^{L}_{y,x} (\dd\tilde \omega) \, T(\tilde \omega)^{q}
\,\cal V^{\nu,1,L,(2)}_{c}(\omega,\tilde \omega)
\lesssim \frac{T(\omega)}{\tilde{\kappa}^{q+2}}\,(q+1)!\,\bigl\|v^{L,(2)}_{c}\bigr\|_{\ell^1(\Lambda_L)}\,\nu^{q-1}\,. \qedhere
\end{equation*}
\end{proof}

\bigskip

\noindent
J\"urg Fr\"ohlich, ETH Z\"urich, Institute for Theoretical Physics, \href{mailto:juerg@phys.ethz.ch}{juerg@phys.ethz.ch}.
\\[0.3em]
Antti Knowles, University of Geneva, Section of Mathematics, \href{mailto:antti.knowles@unige.ch}{antti.knowles@unige.ch}.
\\[0.3em]
Benjamin Schlein, University of Z\"urich, Institute of Mathematics, \href{mailto:benjamin.schlein@math.uzh.ch}{benjamin.schlein@math.uzh.ch}.
\\[0.3em]
Vedran Sohinger, University of Warwick, Mathematics Institute, \href{mailto:V.Sohinger@warwick.ac.uk}{V.Sohinger@warwick.ac.uk}.

\bigskip

\paragraph{Acknowledgements}
The authors thank David Brydges and Daniel Ueltschi for numerous very useful discussions of different technical problems that appeared in our work on this paper.  They thank the referee for their helpful feedback. They also thank Spyridon Garouniatis and Grega Saksida for comments on the file. AK gratefully acknowledges the support of the European Research Council through the RandMat grant and of the Swiss National Science Foundation through the NCCR SwissMAP. BS gratefully acknowledges partial support from the NCCR SwissMAP, from the Swiss National Science Foundation through the Grant ``Dynamical and energetic properties of Bose-Einstein condensates'' and from the European Research Council through the ERC-AdG CLaQS. VS acknowledges support of the EPSRC New Investigator Award grant EP/T027975/1.

\end{document}